\documentclass[twocolumn]{article}
\pdfoutput=1

\usepackage{authblk}
\usepackage{amsfonts}
\usepackage{amsmath}
\usepackage{amssymb}
\usepackage{xfrac}
\usepackage{graphicx}
\usepackage{booktabs,array}
\usepackage[backend=biber,style=phys]{biblatex}
\usepackage{comment}
\usepackage[skip=2pt,font=footnotesize]{caption}
\usepackage{titlesec}
\usepackage[flushleft]{threeparttable}
\usepackage{color}
\usepackage[normalem]{ulem}
\usepackage{tablefootnote}
\usepackage{hyperref}

\newcolumntype{L}[1]{>{\raggedright\arraybackslash}m{#1}}
\newcolumntype{C}[1]{>{\centering\arraybackslash}m{#1}}
\newcolumntype{R}[1]{>{\raggedleft\arraybackslash}m{#1}}

\titleformat*{\section}{\normalsize\bfseries\rmfamily}
\titleformat*{\subsection}{\normalsize\bfseries\rmfamily}
\titleformat*{\subsubsection}{\normalsize\bfseries\rmfamily}

\graphicspath{{./Figures/},{./Workflows/}}
\DeclareGraphicsExtensions{.eps,.pdf,.jpg,.png}
\title{Application of Gaussian process regression to plasma turbulent transport model validation via integrated modelling}
\author[1]{A.~Ho}
\author[1]{J.~Citrin}
\author[2]{F.~Auriemma}
\author[3]{C.~Bourdelle}
\author[4]{F.~J.~Casson}
\author[4]{Hyun-Tae~Kim}
\author[5]{P.~Manas}
\author[4]{G.~Szepesi}
\author[6]{H.~Weisen}
\author[*]{JET~Contributors}
\affil[1]{DIFFER - Dutch Institute for Fundamental Energy Research, Eindhoven, the Netherlands}
\affil[2]{Consorzio RFX, Associazione EURATOM-ENEA sulla Fusione, Padova, Italy}
\affil[3]{CEA, IRFM, Saint-Paul-lez-Durance, France}
\affil[4]{CCFE, Culham Science Centre, Abingdon, UK}
\affil[5]{Max-Planck-Institut f{\"u}r Plasmaphysik, Garching, Germany}
\affil[6]{Swiss Plasma Center, EPFL, Lausanne, Switzerland}
\affil[*]{See the author list of ``X. Litaudon et al 2017 Nucl. Fusion 57 102001"}

\date{}

\begin{document}

\maketitle

\begin{abstract}
This paper outlines an approach towards improved rigour in tokamak turbulence transport model validation within integrated modelling. Gaussian process regression (GPR) techniques were applied for profile fitting during the preparation of integrated modelling simulations allowing for rigourous sensitivity tests of prescribed initial and boundary conditions as both fit and derivative uncertainties are provided. This was demonstrated by a JETTO integrated modelling simulation of the JET ITER-like-wall H-mode baseline discharge \#92436 with the QuaLiKiz quasilinear turbulent transport model, which is the subject of extrapolation towards a deuterium-tritium plasma. The simulation simultaneously evaluates the time evolution of heat, particle, and momentum fluxes over $\sim\!10$ confinement times, with a simulation boundary condition at  $\rho_{\text{tor}} = 0.85$. Routine inclusion of momentum transport prediction in multi-channel flux-driven transport modelling is not standard and is facilitated here by recent developments within the QuaLiKiz model. Excellent agreement was achieved between the fitted and simulated profiles for $n_e$, $T_e$, $T_i$, and $\Omega_{\text{tor}}$ within $2\sigma$, but the simulation underpredicts the mid-radius $T_i$ and overpredicts the core $n_e$ and $T_e$ profiles for this discharge. Despite this, it was shown that this approach is capable of deriving reasonable inputs, including derivative quantities, to tokamak models from experimental data. Furthermore, multiple figures-of-merit were defined to quantitatively assess the agreement of integrated modelling predictions to experimental data within the GPR profile fitting framework.
\end{abstract}

\section{Introduction}
\label{sec:Introduction}

To have confidence in the predictions of any given model in unproven conditions, it must first be rigourously tested to show it behaves as expected and to understand its range of validity, through a process called \emph{model verification and validation (V\&V)}~\cite{aVnV-Greenwald}. However, with a complex non-linear system, such as a tokamak plasma device, the interpretation of experimental data and the verification and validation of any resulting models becomes increasingly difficult~\cite{aVnV-Holland}, though no less important.

This paper outlines a rigourous and automatable approach to data processing and profile fitting, through the use of \emph{Gaussian process regression (GPR)} techniques~\cite{bGP-Rasmussen,bGP-Bishop}, and the consequent improvements to V\&V within the field of nuclear fusion research, as applied to the JETTO transport code~\cite{aJINTRAC-Romanelli,tJETTO-Cenacchi}, coupled with the QuaLiKiz quasilinear turbulent transport code~\cite{aQLK-Citrin,aQLK-Bourdelle}. While Bayesian techniques have been applied to experimental data from tokamaks before~\cite{aBayesian-Irishkin,aTomography-Svensson,aBayesian-Verdoolaege,aBayesian-Fischer}, including the implementation of GPR techniques~\cite{aGP-Chilenski}, the novelty of this work lies in its extension into validation efforts for integrated models. The simulated results and distributions from Monte Carlo studies are compared against the GPR fit distributions, themselves being determined based on experimental data from the JET tokamak. Specifically, this paper applies it to JET discharge \#92436, a JET-ILW baseline discharge with plasma parameters $B_T = 2.77$~T, $I_p = 2.98$~MA, and 33~MW of input heating power, 28~MW from neutral beam injection (NBI) and 5~MW from ion cyclotron (IC) heating. This discharge was selected as it recorded the highest experimental D-D neutron rate to date at JET and is also the subject of extrapolation towards a D-T plasma. In order to proceed with the proposed V\&V procedure, the experimental data needs to first be processed into the appropriate inputs for the code.

For the combined JETTO + QuaLiKiz transport code, the primary quantities under investigation are the simultaneous time evolution of the following plasma kinetic profiles:
\begin{itemize}
	\itemsep 0mm
	\item main ion density, $n_i$,
	\item electron temperature, $T_e$,
	\item ion temperature, $T_i$,
	\item and toroidal flow angular frequency, $\Omega_{\text{tor}}$.
\end{itemize}
Note that the current density, $j$, and densities of the primary impurity ions, $n_{\text{imp}}$, within the discharge are also self-consistently evolved in time for completeness, although these results are not validated in this study due to the lack of experimental measurements of these quantities. However, JETTO combines the main ion and impurity ion densities, along with their respective charges, to form the electron density profile, $n_e$, for which experimental data exists. The profiles are evaluated by the simulation over a sufficient number of confinement times to reach steady state, due to the high sensitivity of the simulation on these quantities at the simulation boundary, especially on the ratio $T_i/T_e$~\cite{aTiTe-Linder}. These inputs are typically expressed on the \emph{square-root normalized toroidal flux} coordinate, or simply \emph{toroidal rho}, $\rho_{\text{tor}}$, defined as follows:
\begingroup\makeatletter\def\f@size{9}\check@mathfonts
\def\maketag@@@#1{\hbox{\m@th\normalsize\normalfont#1}}%
\begin{equation}
	\label{eq:NormalizedToroidalFluxCoordinate}
	\rho_{\text{tor}} = \sqrt{\frac{\psi_{\text{tor}}}{\psi_{\text{tor, LCFS}}}}
\end{equation}\endgroup
where $\psi_{\text{tor}}$ is the \emph{toroidal magnetic flux} associated with the radial point in the plasma and LCFS is the \emph{last-closed-flux-surface}. The advantage of using GPR for profile fitting is the estimation of both the fit uncertainty based on the measurement uncertainties, as well as the analytical calculation of the fit derivative and its uncertainty as well. This additional information allows for statistically rigourous sensitivity studies regarding the impact of the boundary conditions of the simulation, set at $\rho_{\text{tor}} = 0.85$ within the simulations performed in this study, as well as an improved measure of the agreement of and confidence in the transport model when compared to experimental data.

As the quantities of interest, $n_e,T_e,T_i,\Omega_{\text{tor}}$, are also inputs in the calculation of the heat sources, $Q_e,Q_i$, the particle sources, $S_i$, and the fast ion population quantities, $n_{\text{fast}},E_{\text{fast}}$, from the various plasma heating devices, it is in principle possible to propagate these uncertainties through these calculations as well. However, this particular application is outside the scope of this study. Additional measurements, such as the effective ion charge, $Z_{\text{eff}}$, neutron rate, $N$, total diamagnetic energy, $W_{\text{tot}}$, total radiated power, $P_{\text{rad}}$ and normalized internal inductance, $l_i$, were used to adjust and filter the measurement data and constrain the fits further, via an automatised data processing algorithm.

Section~\ref{sec:JETData} outlines the specific measurement data in the JET database used to generate these profiles, as well as briefly discussing the pre-processing steps required in order to automate this procedure, highlighting the generality inherent in the GPR techniques for profile fitting. Section~\ref{sec:ValidationMetrics} introduces the novel figures-of-merit (FOM) used in this study, discussing their application and interpretation. Section~\ref{sec:IntegratedModellingResults} discusses the sensitivity and consequent validation studies performed, highlights their implications, and showcases the statistical rigour and validation metrics made possible with GPR profile fitting. Finally, a summary and future outlook is provided in Section~\ref{sec:Conclusions}.

\section{JET data extraction and processing}
\label{sec:JETData}

In general, the raw measurement signals from the diagnostic devices have already been converted into the physical quantities and profiles. Although this process accrues its own errors, it is assumed that the reported measurement uncertainties associated with the processed data sufficiently capture these effects and that they do not warrant further discussion. To that effect, this paper refers to these converted measurement signals as the \emph{raw data}. This section discusses the GPR fitting algorithm and the pre-processing done to the raw data at JET.

\subsection{Data processing and profile fitting}
\label{subsec:DataProcessing}

\begin{figure*}[tb]
	\centering
	\includegraphics[scale=0.8,trim={0.5cm 0 0 0},clip]{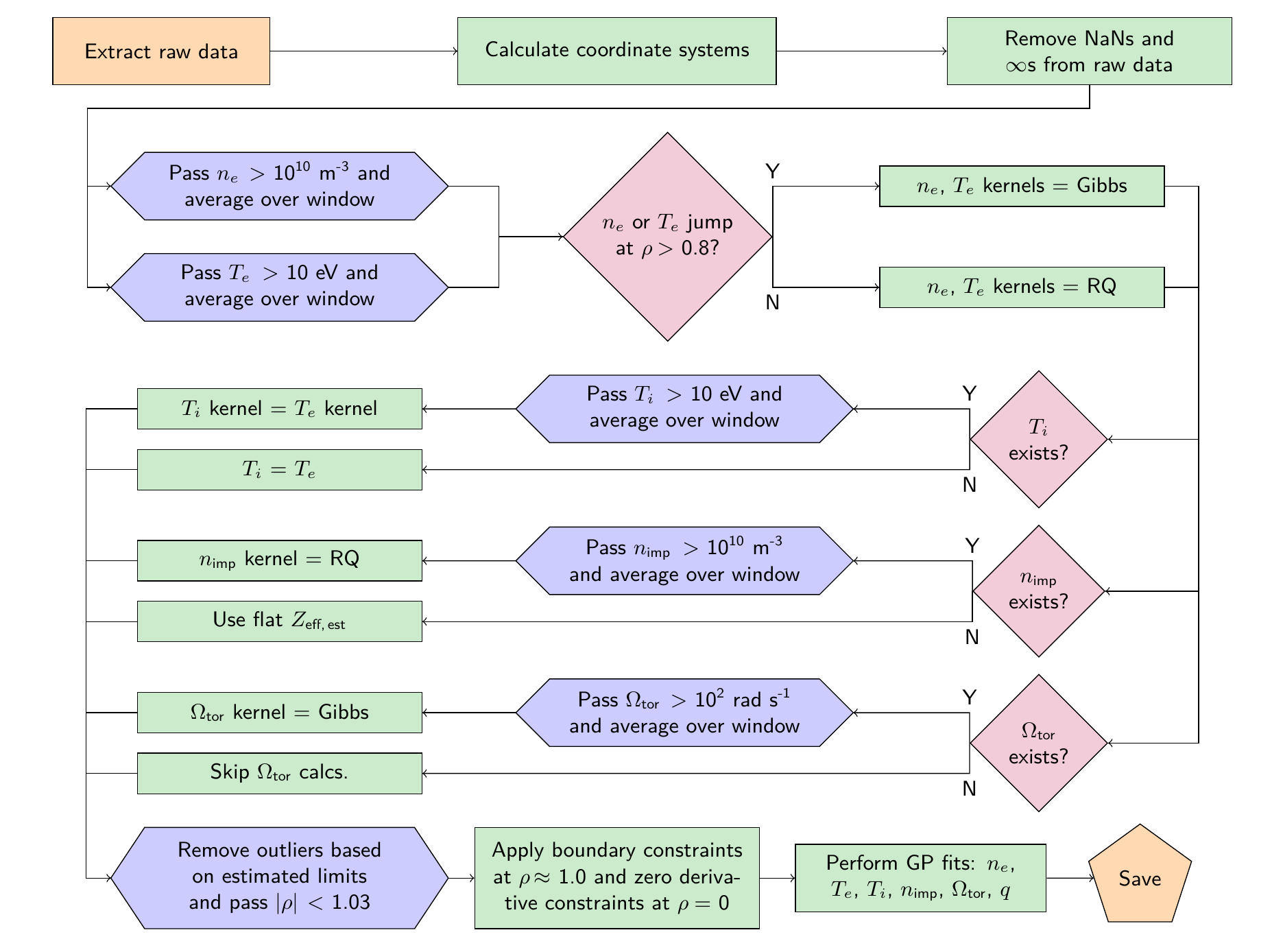}
	\caption{Workflow diagram of the filters and consistency checks applied to the raw data in order to ensure an adequate level of data quality for the fitting algorithm to produce useful profiles.}
	\label{fig:DataProcessingWorkflow}
\end{figure*}

Table~\ref{tbl:JETDiagnosticExtraction} shows the primary data fields in the public JET experimental data storage system, and their corresponding diagnostic devices, from which the raw profile data was derived from. Due to the potential presence of erroneous data as a result of drifting diagnostic calibrations or data processing faults within this database, a workflow was devised to filter out any non-physical data points in a broad and generic fashion. These filters were designed to be generic in nature such that they may be applied to any appropriate raw data, regardless of the origin of that data, such that the resulting workflow can be used for large-scale automation of data extraction and profile fitting. This level of automation was desired to support the collection of a JET 1D profile database, for purpose of sampling QuaLiKiz inputs for creating neural network training sets to extend previous work~\cite{aProof-Citrin}. Figure~\ref{fig:DataProcessingWorkflow} shows the developed workflow and outlines the various criteria determined to produce sufficiently reasonable profile fits from a wide variety of discharges. 

\begin{table}[tb]
	\centering
	\begin{threeparttable}
		\caption{List of diagnostics used for most crucial physical quantities}\vspace{2mm}
		\begin{tabular}{l|l}
			\toprule[1.5pt]
			Quantity & Diagnostic Type \\ 
			\midrule
			$n_e$ & Thomson scattering \\ 
			$T_e$ & Thomson scattering \\ 
			& Electron cyclotron emission \\ 
			$T_i$ & Charge exchange spectroscopy \\ 
			$\Omega_{\text{tor}}$ & Charge exchange spectroscopy \\ 
			\bottomrule[1.5pt]
		\end{tabular}
		\label{tbl:JETDiagnosticExtraction}
	\end{threeparttable}
\end{table}

The basic requirements of the workflow are the existence of electron density, $n_e$, and electron temperature, $T_e$, measurements, along with the magnetic geometry / equilibrium in order to define the radial coordinate systems on which the profiles are to be mapped. If additional diagnostics or post-processed results are available, such as the ion temperature, $T_i$, toroidal angular frequency, $\Omega_{\text{tor}}$, and impurity densities, $n_\text{imp}$, measurements, these are also extracted. A number of basic data filters are then applied to the data, such as the removal of corrupted or missing data and non-physical values\footnote[1]{\label{foot:RmajShift}Due to the identified discrepancies in the equilibrium reconstruction for this discharge, the mid-plane major radius vector of the equilibrium was shifted by $\sim 5$~cm inwards before remapping the raw data to $\rho_{\text{tor}}$ and applying the GPR. This shift was calculated by applying a least-squares $\tanh$ fit to the temperature pedestal data and setting the separatrix at the location where $T_e = 100$~eV, as is understood to be the $T_e$ boundary condition for H-mode plasmas from scrape-off layer modelling results.}.

Then, a number of data points are added to the raw data to enforce a small positive boundary constraint at the separatrix, due to known behaviours from scrape-off layer modelling, and a zero derivative constraint is applied at the magnetic axis, due to the assumed symmetry across the magnetic axis. Note that the boundary value constraint for the $\Omega_{\text{tor}}$ profile is a linearly extrapolated value as the rotation profile can switch signs in the separatrix region. Finally, the data is organized into a standardised format and the appropriate kernels are selected according to approximate properties of the filtered profile data. This standardised structure is passed to the newly-created ``GPR1D'' Python package\footnote[2]{\label{foot:GPR1D}An open-source one-dimensional GPR algorithm, available on GitLab at \texttt{https://gitlab.com/aaronkho/GPR1D.git}} to perform the GPR fits.

The GPR algorithm is derived by applying Bayesian statistical principles to the mathematics of regression modelling, with the assumption that the input noise is described by Gaussian probability distribution functions~\cite{bGP-Rasmussen,aGP-Chilenski}. The resulting algorithm effectively performs the regression using an infinite set of basis functions, represented by a careful choice of the \emph{covariance function}, or \emph{kernel function}, of the underlying model. This theoretically gives it the capability of a universal function estimator. However, as this section will discuss, the exact details of its implementation and usage often enforce some constraints on this capability.

The algorithm calculates the mean fit profile and its error through the following predictive equations~\cite{bGP-Rasmussen,ahGP-Kersting}:
\begingroup\makeatletter\def\f@size{9}\check@mathfonts
\def\maketag@@@#1{\hbox{\m@th\normalsize\normalfont#1}}%
\begin{equation}
\label{eq:GPRPredictiveModel}
	\begin{aligned}
	K\!\left(X,X'\right) &= k\!\left(x,x',\mathbf{\theta}\right) |_{x=X,x'=X'} \\
	R\!\left(X,X'\right) &= r\!\left(x\right) r\!\left(x'\right) \delta\!\left(x - x'\right) |_{x=X,x'=X'} \\
	Y_* &= K\!\left(X_*,X\right) L^{-1} \, Y \\
	\sigma_{Y_*}^2 &= L_* - K\!\left(X_*,X\right) L^{-1} K\!\left(X,X_*\right)
	\end{aligned}
\end{equation}\endgroup
where $\theta$ represents the \emph{hyperparameters} of the chosen kernel function, $k$, the lower-case and upper-case variables denote continuous functions and discrete data points, respectively, $\left(X,Y\right)$ represents the input data points, $\left(X_*,Y_*\right)$ represents the points at which the model is evaluated, and the short-hand $K=K\!\left(X,X\right)$, $R=R\!\left(X,X\right)$, $K_*=K\!\left(X_*,X_*\right)$, $R_*=R\!\left(X_*,X_*\right)$, $L=K+R$, and $L_*=K_*+R_*$ was used to improve readability. Due to the numerical details of the GPR implementation, it is assumed and advised to select the prediction points such that $X_* \notin X$. The resulting regression fit distribution for each point is then Gaussian-distributed, or normally distributed, by the definition of the GPR procedure, which is a probability distribution described by the following expression:
\begingroup\makeatletter\def\f@size{9}\check@mathfonts
\def\maketag@@@#1{\hbox{\m@th\normalsize\normalfont#1}}%
\begin{equation}
\label{eq:NormalDistribution}
	\mathcal{N}\!\left(Y_*,\sigma_{Y_*}^2\right) \equiv \frac{1}{\sqrt{2 \pi \sigma_{Y_*}^2}} \exp\!\left(-\frac{\left(y - Y_*\right)^2}{2 \sigma_{Y_*}^2}\right)
\end{equation}\endgroup
where $y$ is the variable coordinate corresponding to the input data, $Y$.

Within the GPR framework, the hyperparameters of the chosen kernel, $\mathbf{\theta}$, act as the free variables which can be adjusted to fine-tune the resulting fit. One optimization technique for these hyperparameters maximizes the \emph{log-marginal-likelihood (LML)}, $\ln p\!\left(Y|X,\mathbf{\theta}\right)$, through the use of its derivative with respect to each of the hyperparameters, $\theta_j$, which is expressed as follows~\cite{bGP-Rasmussen}:
\begingroup\makeatletter\def\f@size{9}\check@mathfonts
\def\maketag@@@#1{\hbox{\m@th\normalsize\normalfont#1}}%
\begin{equation}
\label{eq:LogMarginalLikelihoodDerivative}
	\frac{\partial \ln{p\!\left(Y|X,\mathbf{\theta}\right)}}{\partial \theta_j} = \frac{1}{2} Y^T L^{-1} \frac{\partial K}{\partial \theta_j} L^{-1} Y - \frac{\lambda}{2} \text{tr}\!\left(L^{-1} \frac{\partial K}{\partial \theta_j}\right)
\end{equation}\endgroup
where the hyperparameter-dependence is also present in $L \equiv L\!\left(\mathbf{\theta}\right)$ and $\lambda$ is the \emph{regularization parameter}, used to control the degree of complexity in the model. This optimization scheme assumes that an analytical form, or a sufficiently accurate numerical approximation, exists for the derivative of the kernel function with respect to these hyperparameters and attempts to find the model that maximizes the probability for observing the experimental data. However, it provides no guarantee that the chosen model accurately depicts the physical process which produced the input data.

The desired optimized solution will then be the combination of hyperparameters, $\mathbf{\theta}$, which satisfy the following criteria:
\begingroup\makeatletter\def\f@size{9}\check@mathfonts
\def\maketag@@@#1{\hbox{\m@th\normalsize\normalfont#1}}%
\begin{equation}
\label{eq:MaximizationCriteria}
	\nabla_{\!\theta} \ln{p\!\left(Y|X\right)} = \mathbf{0}
\end{equation}\endgroup
However, since Equation~\eqref{eq:LogMarginalLikelihoodDerivative} typically forms a non-linear system of equations for the set of $\theta$, it is difficult to calculate the solution explicitly. Thus, an iterative method, such as a gradient-based optimization algorithm, is used to find the closest desired solution.

The primary advantage of the GPR technique, over other common fitting techniques such as spline fitting, is that it provides statistically rigourous uncertainties of the fit, including the fit derivatives, based on the input measurement errors. More information about the theory and terminology behind the GPR can be found in Appendix~\ref{app:GaussianProcess} and in Ref.~\cite{bGP-Rasmussen}.

A common issue with current fitting practices is that, when a pedestal is present within the profile, it is normally difficult to accurately fit both the core and pedestal regions. This is typically due to the dramatic difference in properties between these two neighbouring regions. The GPR methodology offers a potential solution to this without modifying the radial coordinate space through the use of a \emph{Gibbs kernel}, mathematically as such~\cite{bGP-Rasmussen}:
\begingroup\makeatletter\def\f@size{9}\check@mathfonts
\def\maketag@@@#1{\hbox{\m@th\normalsize\normalfont#1}}%
\begin{equation}
\label{eq:GibbsKernel}
	k\!\left(x,x'\right) = \sigma^2 \sqrt{\frac{2l\!\left(x\right)l\!\left(x'\right)}{l^2\!\left(x\right) + l^2\!\left(x'\right)}} \exp{\left(\frac{\left(x - x'\right)^2}{l^2\!\left(x\right) + l^2\!\left(x'\right)}\right)}
\end{equation}\endgroup
where $l\!\left(x\right)$, known as a \emph{warping function}, is chosen based on the desired behaviour of the length scale of the fit. The hyperparameters of this kernel are $\mathbf{\theta}=\left\lbrace\sigma,\mathbf{\Theta}\right\rbrace$, where $\mathbf{\Theta}$ represents the set of additional hyperparameters introduced by the chosen warping function. In order to obtain the required behaviour of the length scale for capturing the pedestal, an inverted Gaussian warping function was selected, expressed as follows:
\begingroup\makeatletter\def\f@size{9}\check@mathfonts
\def\maketag@@@#1{\hbox{\m@th\normalsize\normalfont#1}}%
\begin{equation}
\label{eq:InverseGaussianWarpingFunction}
	l\!\left(x\right) = l_0 - l_1 \exp{\left(\frac{\left(x - \mu \right)^2}{2\sigma_l^2}\right)}
\end{equation}\endgroup
where the additional hyperparameters are $\mathbf{\Theta}=\left\lbrace l_0,l_1,\mu,\sigma_l \right\rbrace$. The ability to avoid the modification of the coordinate space is desired, as it would introduce an interpretation bias in the data which is inconsistent towards the Bayesian treatment and eliminates the need for complex data processing algorithms to handle this in an automated fashion. When a large jump in the data is detected within $\rho>0.8$, for any of the kinetic profiles, which is indicative of the formation of an H-mode pedestal, the workflow attempts to fit the data using this Gibbs kernel.

When a potential pedestal is not detected in the profile data, the Gibbs kernel is entirely replaced with the more basic \emph{rational quadratic (RQ) kernel}, as it is generally more stable and yields sufficiently reasonable results in these cases. The RQ kernel can be expressed mathematically as follows~\cite{bGP-Rasmussen}:
\begingroup\makeatletter\def\f@size{9}\check@mathfonts
\def\maketag@@@#1{\hbox{\m@th\normalsize\normalfont#1}}%
\begin{equation}
\label{eq:RQKernel}
	k\!\left(x,x'\right) = \sigma^2 \left(1 + \frac{\left(x - x'\right)^2}{2 \alpha l^2}\right)^{-\alpha}
\end{equation}\endgroup
where the hyperparameters are given by $\mathbf{\theta}=\left\lbrace\sigma,\alpha,l\right\rbrace$.

The noise function required by the GPR algorithm, $r\!\left(x\right)$, is generated by applying a separate GPR on the measurement uncertainties, also using the RQ kernel. Typically, a large regularization parameter ($\lambda \geq 10$) is applied to the fitting of the measurement errors, due to the overprediction of the fit derivative uncertainties as a direct result of steep gradients in the noise function.

The profile fitting routine is designed to execute the GPR a number of times per physical quantity, first using a defined initial guess and afterwards with a number of randomized guesses within set boundaries, in order to ensure that the hyperparameter optimization routine does not fall into a local maximum. If a pedestal is detected in the profile data and all the attempts to perform a fit using a Gibbs kernel fail, then the workflow reverts to using the RQ kernel into order to provide a reasonable estimate, albeit with a lower fit quality. In practice, this is a rare occurance since the algorithm is more likely to \emph{overfit} if a reasonable solution cannot be found, wherein the algorithm converges on a solution which does not model the underlying smooth structure but the radial variation due to noise instead.

Overfitting is a common problem with GPR fitting, especially if the initial guess is too far from the optimal spot, and typically results in a poor quality profile fit which varies erratically in the radial coordinate. To remedy this, the regularization parameter of the fit itself can also be increased, which then applies a greater penalty to erratic fits. Another solution to this problem within this profile fitting routine is to increase the number of random hyperparameter restarts, which can significantly increase the amount of time required to fit each profile, and may reduce the efficacy of the algorithm in capturing the behaviour of the pedestal.

\subsection{Application of data extraction process}
\label{subsec:ExtractionApplication}

\begin{figure}[tb]
	\centering
	\includegraphics[scale=0.3]{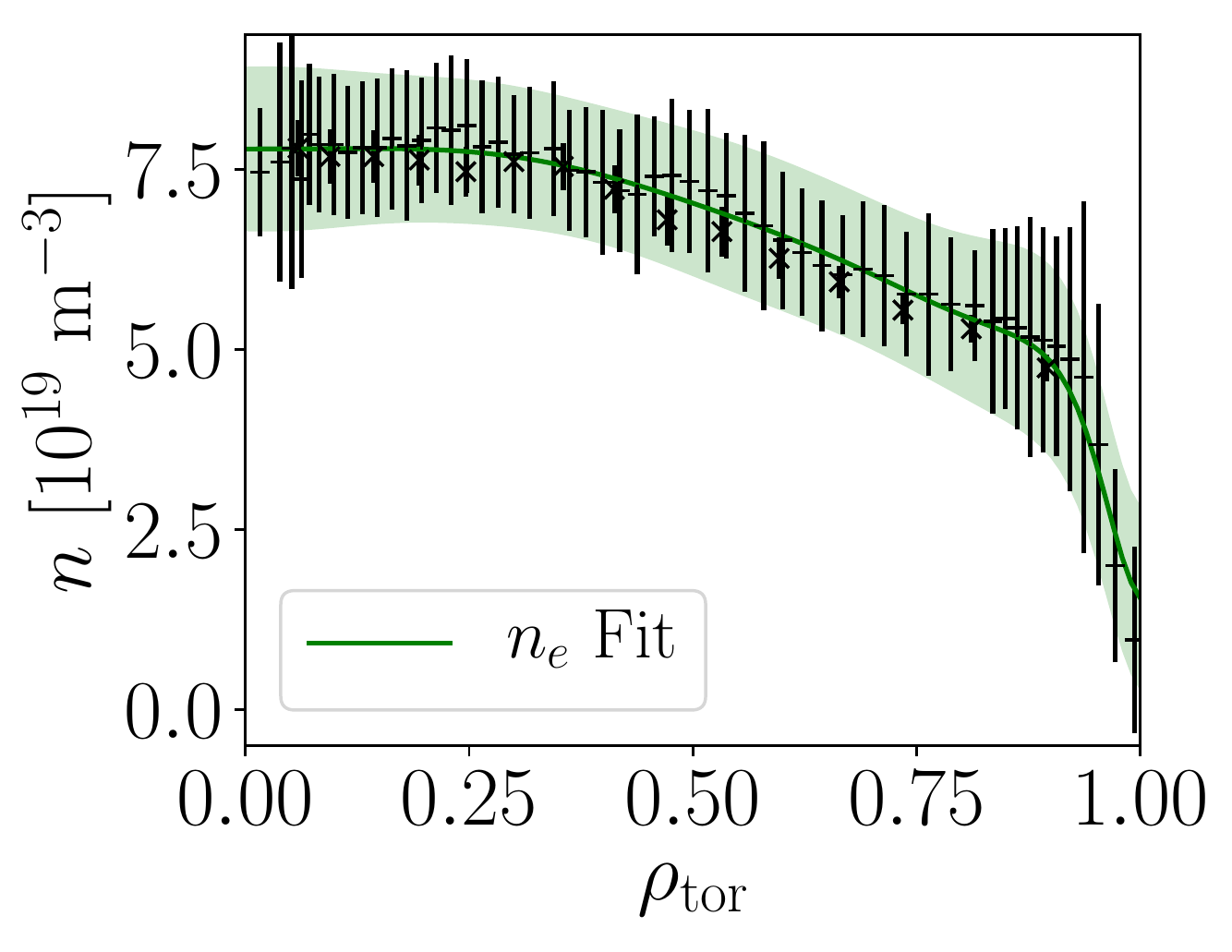}%
	\includegraphics[scale=0.3]{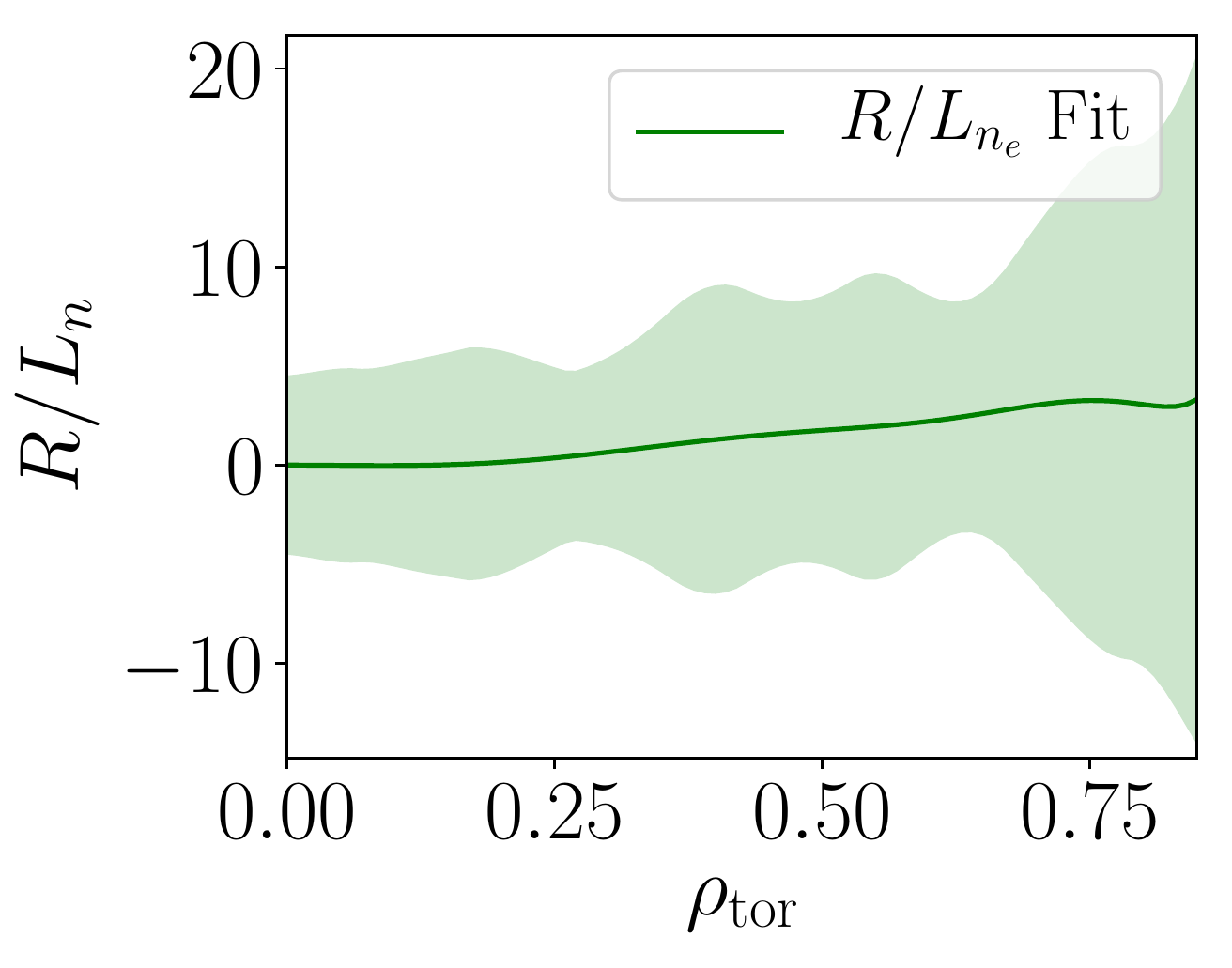}
	\caption{Profiles for JET \#92436, averaged over a 0.5~s time window, from 9.75~s -- 10.25~s, with $\sim\!50$~ms data resolution. Left: Electron (green line) and ion (red line) densities with experimental data (black points), showing good fitting of pedestal shoulder for use as the boundary condition for core transport modelling, $n_i$ estimated using $n_e, Z_{\text{eff}}$. Right: Normalized logarithmic gradients for $n_e$ and $n_i$, only shown for $\rho_{\text{tor}} \le 0.8$. All errors are depicted as $\pm 2 \sigma$, corresponding to a confidence interval of $\sim$95\% within Gaussian statistics.}
	\label{fig:DensitySampleOutput}
\end{figure}

\begin{figure}[tb]
	\centering
	\includegraphics[scale=0.305]{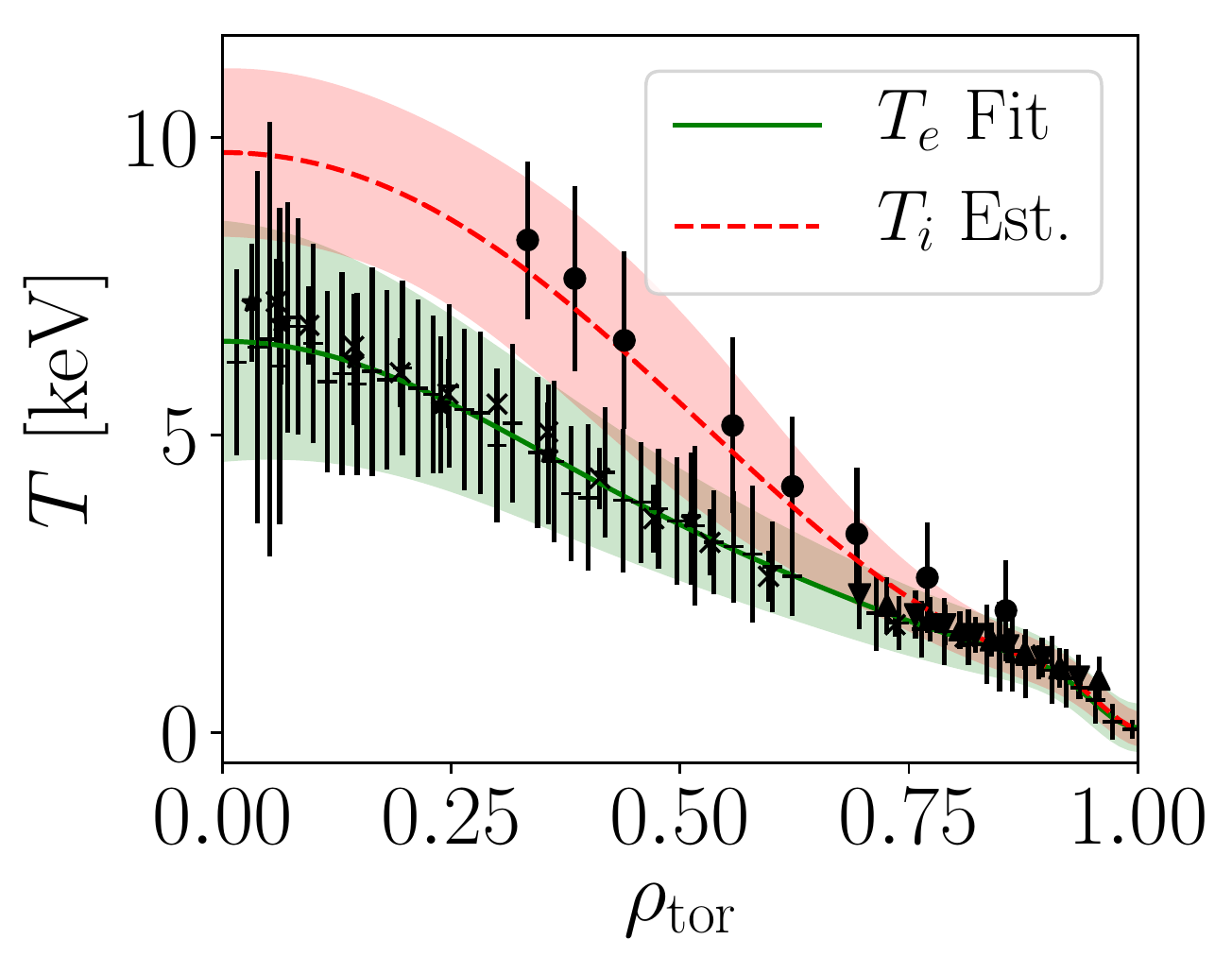}%
	\includegraphics[scale=0.305]{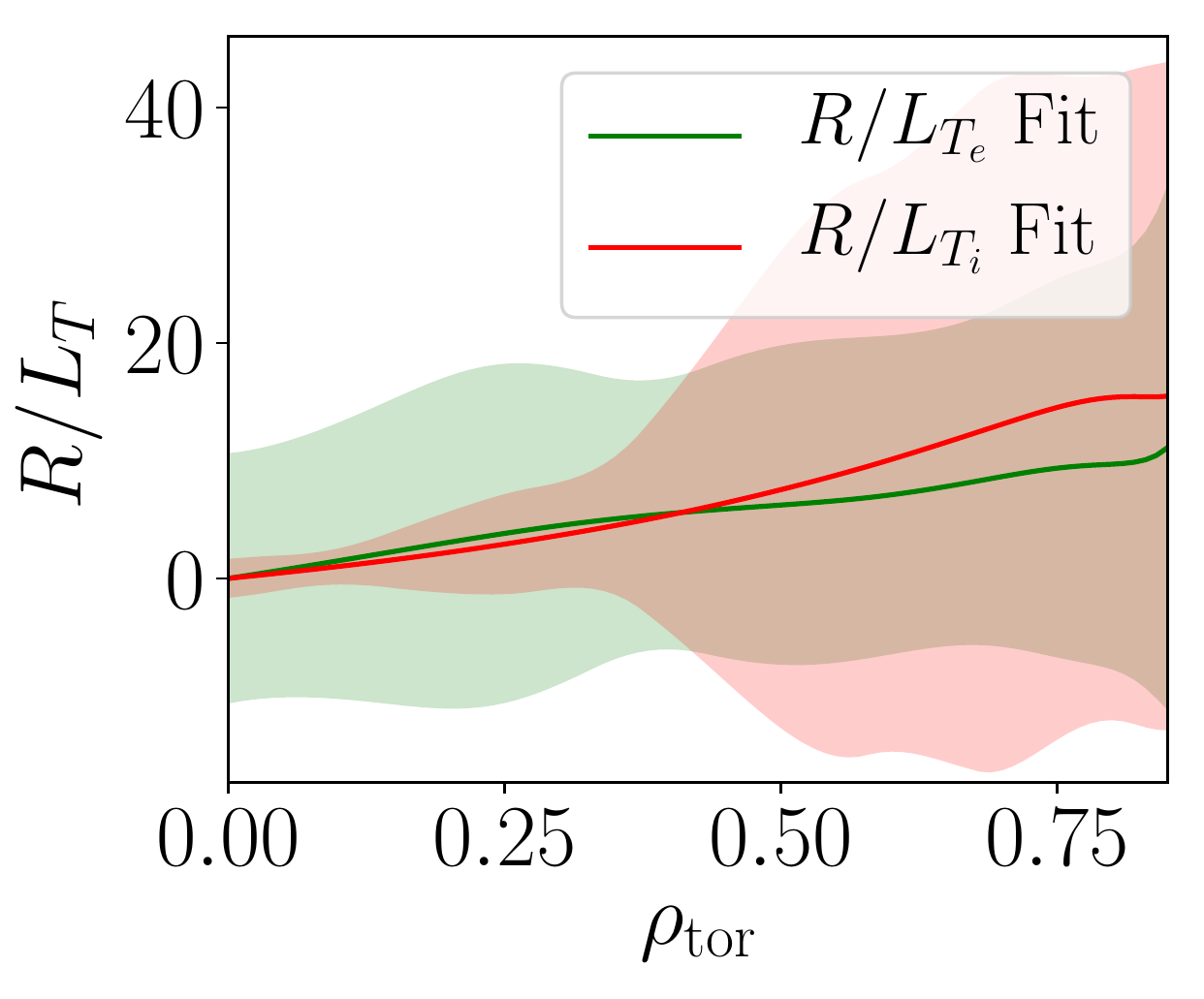}
	\caption{Profiles for JET \#92436, averaged over a 0.5~s time window, from 9.75~s -- 10.25~s, with $\sim\!50$~ms data resolution. Left: Electron and ion temperatures with experimental data, with assumption that $T_i = T_{\text{imp}}$. Impurity ion temperature measurements concatenated over three charge exchange diagnostics: one core diagnostic and two edge diagnostics. Right: Normalized logarithmic gradients for $T_e$ and $T_i$, only shown for $\rho_{\text{tor}} \le 0.8$. All errors are depicted as $\pm 2 \sigma$, corresponding to a confidence interval of $\sim$95\% within Gaussian statistics.}
	\label{fig:TemperatureSampleOutput}
\end{figure}

\begin{figure}[tb]
	\centering
	\includegraphics[scale=0.305]{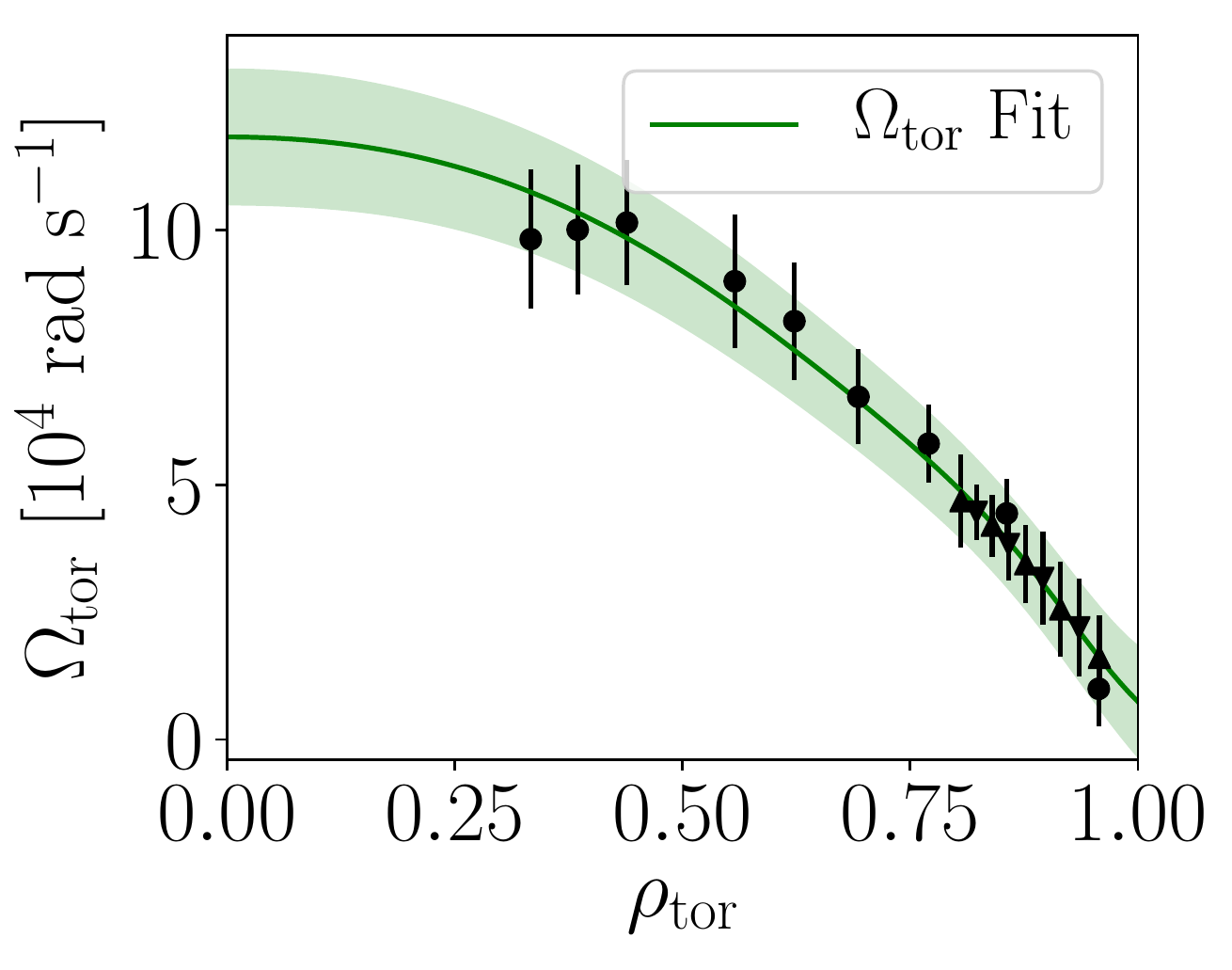}%
	\includegraphics[scale=0.305]{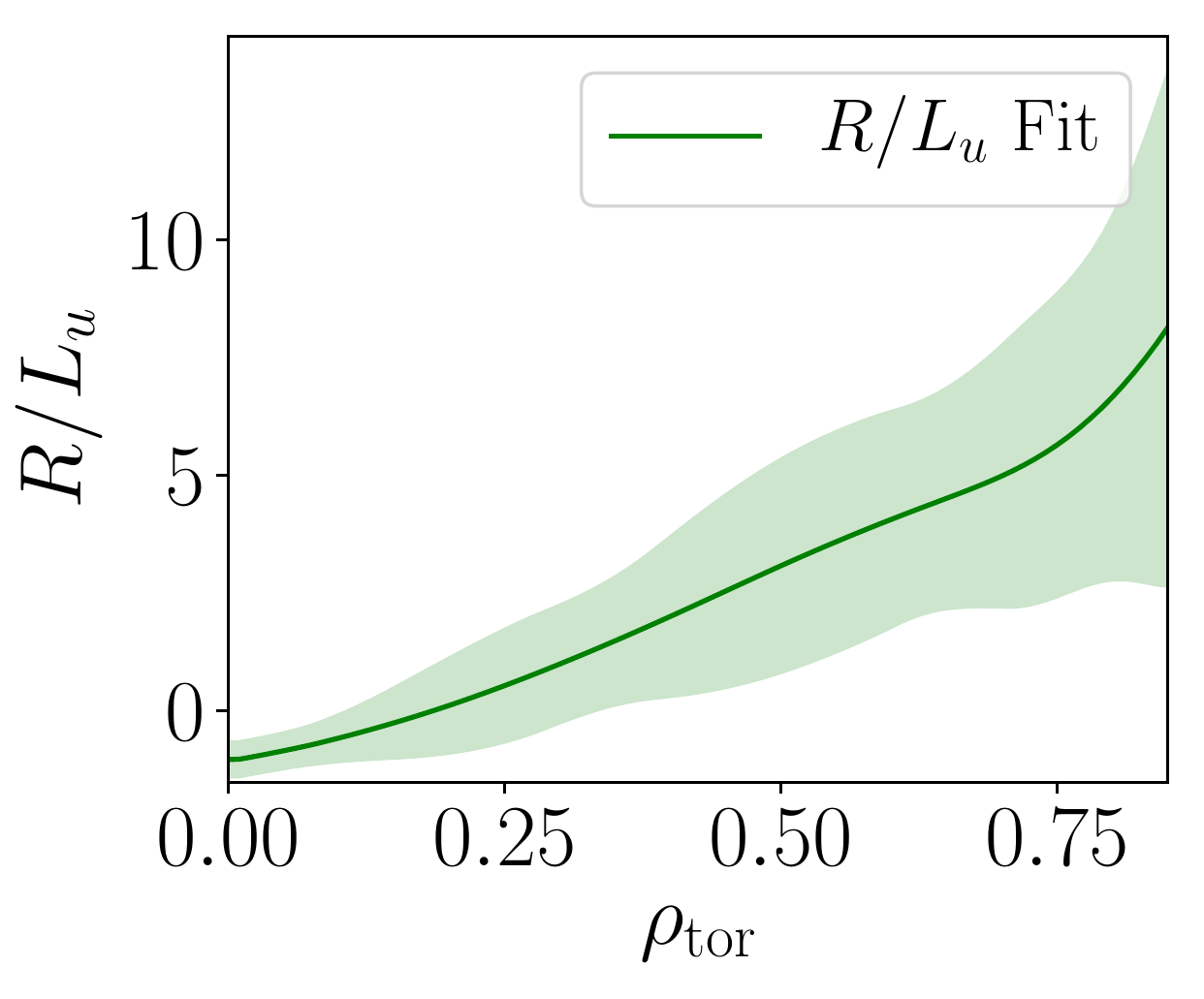}
	\caption{Profiles for JET \#92436, averaged over a 0.5~s time window, from 9.75~s -- 10.25~s, with $\sim\!50$~ms data resolution. Left: Toroidal flow angular frequency with experimental data. Toroidal flow measurements concatenated over three charge exchange diagnostics: one core diagnostic and two edge diagnostics. Right: Normalized toroidal flow shear. All errors are depicted as $\pm 2 \sigma$, corresponding to a confidence interval of $\sim$95\% within Gaussian statistics.}
	\label{fig:ToroidalFlowSampleOutput}
\end{figure}

A demonstration of the GPR profile fitting routine was performed on JET-ILW discharge \#92436, a high-power H-mode baseline scenario plasma with $B_T = 2.73$~T, $I_p = 2.98$~MA, and 28~MW neutral beam injection (NBI) and 5~MW ion cyclotron (IC) auxiliary heating applied. This discharge is of particular interest as it is the JET-ILW baseline with the highest measured neutron rate to date at JET and is the subject of extrapolation towards a D-T plasma. The input data used to generate these fits were averaged over a 0.5~s time window, specifically from 9.75~s -- 10.25~s, after all the discussed filters were applied to the raw data. The results of these fits are shown in Figures~\ref{fig:DensitySampleOutput}, \ref{fig:TemperatureSampleOutput} and \ref{fig:ToroidalFlowSampleOutput}, showing reasonable fits and error estimates, including good performance over the pedestal region and even in the absence of inner core data, as seen in the $T_i$ fit.

With the selection of an appropriate kernel and optimizer, a single profile fit can be performed in $\sim$10~s on a single processor. This means that a single discharge time window can be processed in 1 -- 3~minutes, depending on the amount of raw data available. A significant portion of this time is spent extracting data from the storage system and performing the pre-processing required such that the fit procedure can be automated. While this demonstration is limited to a single discharge, profiles from 13000 time windows from over 2000 different JET discharges have been processed for the aforementioned purpose of sampling QuaLiKiz inputs for creating neural network training sets. This will be the subject of a future publication.

\section{Validation metrics}
\label{sec:ValidationMetrics}

To improve the validation efforts of plasma kinetic profiles predicted by these complex transport codes, \emph{validation metrics} need to be developed from which the model results can be compared to experimental data. These metrics typically require a degree of generality such that they can be applied across a large range of foreseen scenarios, but also retain enough information as to provide a proper quantification of any agreement between the model and experiment~\cite{aVnV-Greenwald}. Although many different validation metrics have been previously formulated for fusion data~\cite{aVnV-Greenwald,aVnV-Holland}, their applicability is normally problem-dependent. This section outlines two \emph{figure-of-merits (FOM)}, which are intended to be applied specifically to the comparison of kinetic profiles with Gaussian-distributed uncertainties, such as those provided by GPR fitting. The first is for comparing profiles without any known uncertainty estimate to the GPR fit uncertainties and the second is for comparing other profiles with Gaussian-distributed uncertainties to the GPR fit uncertainties.

\subsection{Point-distribution comparisons}
\label{subsec:ValidationMetricPointDist}

In the cases where an estimate of the output uncertainty is unavailable, an analysis of statistical agreement can be performed by evaluating the probability density function of the experimental distribution at the point of the simulation output value, as a function of space. As the GPR technique necessarily provides fit uncertainties with a normal distribution, the degree of trust that can be placed on these output profiles given the input distributions can be calculated simply by evaluating a modified form of Equation~\eqref{eq:NormalDistribution}, which is as follows:
\begingroup\makeatletter\def\f@size{9}\check@mathfonts
\def\maketag@@@#1{\hbox{\m@th\normalsize\normalfont#1}}%
\begin{equation}
\label{eq:PointDistributionSignificance}
	S \equiv \exp\!\left(-\frac{\ln\!\left(2\right)}{2} \frac{\left(y_o - \mu_i\right)^2}{2 \sigma_i^2}\right)
\end{equation}\endgroup
where $y_o$ is the simulation output value, and $\mu_i$ and $\sigma_i$ are the mean and standard deviation of the experimental fit distribution, respectively. The normalising prefactor of Equation~\eqref{eq:NormalDistribution} was removed such that $S=1$ indicates a perfect match, and the additional factor in the exponent was added such that $S=0.5$ when the simulation output is at the $\pm 2 \sigma$ boundary of the GPR uncertainty. In principle, an evaluation of the agreement between simulation and experimental fit via this methodology does not strictly require the assumption of Gaussian-distributed uncertainties. Thus, its application is not exclusive to the GPR fit methodology but is demonstrated in this paper using the GPR fit uncertainties.

\subsection{Distribution-distribution comparisons}
\label{subsec:ValidationMetricDistDist}

On the other hand, if an estimate of the output uncertainty is provided and assumed to be Gaussian-distributed and denoted as $\mathcal{N}\!\left(\mu_o,\sigma_o^2\right)$, as defined in Equation~\eqref{eq:NormalDistribution}, a comparison can be made between the input and output distributions based solely on their statistical properties, $\left(\mu_i,\sigma_i^2\right)$ and $\left(\mu_o,\sigma_o^2\right)$, respectively. The proposed FOM developed in this work, denoted as $M$, accounts both the difference of the distribution means in relation to their widths and the ratio of the distribution widths in relation to their means, expressed as follows:
\begingroup\makeatletter\def\f@size{9}\check@mathfonts
\def\maketag@@@#1{\hbox{\m@th\normalsize\normalfont#1}}%
\begin{equation}
\label{eq:ProposedGaussianMetric}
\begin{aligned}
	M &= A \, P_i \, P_o \\
	&= \exp\!\left(-\frac{\left(\mu_i - \mu_o\right)^2}{2 \left(\sigma_i^2 + \sigma_o^2\right)} - \frac{\left(3 \sigma_i\right)^2}{\mu_i^2} - \frac{\left(3 \sigma_o\right)^2}{\mu_o^2}\right)
\end{aligned}
\end{equation}\endgroup
where $A$ represents the \emph{accuracy} of the output distribution compared to the input distribution, and $P_i$ and $P_o$ represent the individual \emph{precision} of the input and output distributions, respectively. These components can then be expanded further into the desired terms as a function of $\mu$ and $\sigma$ using logical considerations.

The accuracy component can be derived by calculating the area under the product of the two distributions, expressed as follows:
\begingroup\makeatletter\def\f@size{9}\check@mathfonts
\def\maketag@@@#1{\hbox{\m@th\normalsize\normalfont#1}}%
\begin{equation}
\label{eq:GaussianAccuracyMetric}
	A = \exp\!\left(-\frac{\left(\mu_i - \mu_o\right)^2}{2 \left(\sigma_i^2 + \sigma_o^2\right)}\right)
\end{equation}\endgroup
By examining Equation~\eqref{eq:GaussianAccuracyMetric} intuitively, it compares the absolute difference between their means, $\left|\mu_i - \mu_o\right|$, while taking into account the combined width of their distributions, $\sigma_i^2 + \sigma_o^2$, making it a suitable measure of accuracy. The negative exponent ensures that $A \in \left(0,1\right]$, with a value of unity meaning a perfect match between the two means and a value of zero meaning no statistical overlap between the two distributions, corresponding to no overlap of the $3 \sigma$ boundaries of the two distributions. Equation~\eqref{eq:GaussianAccuracyMetric} is similar to the metric proposed by P. Ricci~\cite{aVnV-Ricci}, except with the inclusion of the negative exponential operator and the $1/2$ factor. In order to provide a rule-of-thumb for the interpretation of $A$, it is generally noted that distribution pairs with $A \gtrsim 0.8$ typically have their mean values lie within the $\pm\;1\sigma$ range of the distribution of the other.

However, in Equation~\eqref{eq:GaussianAccuracyMetric}, $A \rightarrow 1$ as $\sigma_i \rightarrow \infty$ or $\sigma_o \rightarrow \infty$, which is undesired behaviour as it would award high scores to distributions that are too dispersed to be meaningful. The precision component is then introduced to provide a penalty for this behaviour based simply on the ratio of the distribution width to its mean, as follows:
\begingroup\makeatletter\def\f@size{9}\check@mathfonts
\def\maketag@@@#1{\hbox{\m@th\normalsize\normalfont#1}}%
\begin{equation}
\label{eq:GaussianPrecisionMetric}
	P = \exp\!\left(-\frac{\left(3 \sigma\right)^2}{\mu^2}\right)
\end{equation}\endgroup
where $3 \sigma$ was chosen as the reference width as $\sim\!99.7$\% of the distribution lies between $\left[\mu - 3\sigma,\mu + 3\sigma\right]$ in Gaussian statistics, effectively meaning $99.9$\% of the probability distribution lies on one side of zero when $\left|\mu\right| = 3 \sigma$. Similarly to Equation~\eqref{eq:GaussianAccuracyMetric}, the negative exponent ensures that $P \in \left(0,1\right]$, with a value of unity meaning the quantity in question is perfectly known and a value of zero meaning that the given distribution can be regarded as meaningless due to its width in comparison to its mean. This particular method for qualifying the distribution width is only useful for quantities which have a \emph{non-zero expected value}, as $P = 0$ when $\mu = 0$. In order to provide a rule-of-thumb for this component, it is generally noted that distributions with $P \gtrsim 0.7$ have a relative error of $\sigma/\mu \lesssim 0.2$. When applying this to fusion profiles, which generally have strictly positive values, regions with $P < 0.3$ can effectively be regarded as meaningless without additional information. Due to the dependence of this parameter on the absolute value of the distribution mean itself, it is only suitable for comparing distributions of quantities which do not fluctuate around zero, as is generally the case for fusion profile quantities. In that sense, extra care should be taken when applying this to toroidal rotation profiles, as they can cross zero in certain plasma regimes.

By considering the heuristic statements for each individual component in Equation~\eqref{eq:ProposedGaussianMetric}, a good rule-of-thumb for the proposed FOM is that a value of $M \lesssim 0.1$ implies the two distributions do not match at all, ie. inaccurate, or the data is inconclusive, ie. imprecise. On the other hand, a value of $M \gtrsim 0.9$ implies an excellent match between the two distributions, meaning they are both precise and accurate in comparison to each other. A pair of distributions whose means lie within $\pm\;2\sigma$ of each other, each having a relative error of $\sim\!10$\%, yields $M \simeq 0.5$, which indicates a reasonable match for fusion data. The constant factors added in Equations~\eqref{eq:GaussianAccuracyMetric} and \eqref{eq:GaussianPrecisionMetric} were chosen such that this interpretation is consistent with that described for the point-distribution metric discussed in Section~\ref{subsec:ValidationMetricPointDist}.

The performance and suitability of the proposed FOM can be determined by comparing it to other known statistical distance tests which evaluate the agreement between two distributions. The chosen tests are the validation metric proposed by P. Ricci~\cite{aVnV-Ricci} and the \emph{Kullbeck-Leibler (K-L) divergence} test~\cite{aKLDiv-Perez,aKLDiv-Kullback} for continuous probability distributions, specifically the Gaussian distribution in this case. These tests were modified with a negative exponential operator to simplify their comparison with the proposed FOM in Equation~\eqref{eq:ProposedGaussianMetric} and were computed as follows:
\begingroup\makeatletter\def\f@size{9}\check@mathfonts
\def\maketag@@@#1{\hbox{\m@th\normalsize\normalfont#1}}%
\begin{equation}
\label{eq:ModifiedMetricsComparison}
	\begin{gathered}
	M_{\text{Ricci}} = \exp\!\left(-\frac{\left(\mu_i - \mu_o\right)^2}{\left(\sigma_i^2 + \sigma_o^2\right)}\right) \\
	M_{\text{K-L c.}} = \exp\!\left(-\int_{-\infty}^{\infty} p_o\!\left(y\right) \ln\!\left(\frac{p_o\!\left(y\right)}{p_i\!\left(y\right)}\right) \, \text{d}y \right)
	\end{gathered}
\end{equation}\endgroup
where $N$ is the number of bins in the discrete probability histogram, $p_o\!\left(y\right) \sim \mathcal{N}\!\left(\mu_o,\sigma_o\right)$ is the na{\"i}ve Gaussian envelope computed from the Monte Carlo results, $p_o\!\left(y_j\right)$ is the histogram of the Monte Carlo results, $p_i\!\left(y\right) \sim \mathcal{N}\!\left(\mu_i,\sigma_i\right)$ is the GPR fit distribution and $p_i\!\left(y_j\right)$ is calculated from the GPR fit distribution as such:
\begingroup\makeatletter\def\f@size{9}\check@mathfonts
\def\maketag@@@#1{\hbox{\m@th\normalsize\normalfont#1}}%
\begin{equation}
\label{eq:GPRDiscreteProbabilityDistribution}
p_i\!\left(y_j\right) = \int_{y_{j,\text{lower}}}^{y_{j,\text{upper}}} p_i\!\left(y\right) \, \text{d}y
\end{equation}\endgroup

Depending on the application of these profiles, the spatially-resolved FOM, $M$, can be reduced even further into a single number via its integration with respect to an appropriate quantity. For example, integrating it with respect to volume, $V$, ie. $\int M \text{d}V / \int \text{d}V$, can provide a rough estimate of similarity of the profiles for applications involving volumetric considerations, such as total plasma energy, $W_p$, or neutron rate, $R_n$. Although the uncertainty of the numerical integration can be made small by selecting an appropriate algorithm, the uncertainty of the multiplied quantity provides an additional source of error which could significantly influence the interpretation of this single number. As such, this paper does not comment further on the use of an integrated FOM, as it is highly dependent on the application.

As a final note, both the proposed figures-of-merit discussed in this paper for evaluating the agreement and trustworthiness between simulation inputs and outputs, found in Equations~\eqref{eq:PointDistributionSignificance} and \eqref{eq:ProposedGaussianMetric}, should not be confused with other statistically meaningful quantities, ie. probability, likelihood, etc. Although the derivations of these metrics are based on statistical principles, they are intended only to provide a simple, quantitative, but still inherently heuristic measure of agreement between the experimental fits and the simulation output while simultaneously incorporating any knowledge on the experimental and simulation uncertainties.

\section{Integrated modelling results}
\label{sec:IntegratedModellingResults}

The macroscopic transport phenomena within fusion plasmas are governed by a system of coupled differential equations, which must be solved self-consistently in order to determine the time evolution of the system. The one-dimensional energy transport equation for a given species, $s$, in cylindrical geometry is provided below as an example of one such equation:
\begingroup\makeatletter\def\f@size{9}\check@mathfonts
\def\maketag@@@#1{\hbox{\m@th\normalsize\normalfont#1}}%
\begin{equation}
\label{eq:HeatTransportEquation}
	\frac{3}{2} \frac{\partial \left(n_s T_s\right)}{\partial t} + \frac{1}{r} \frac{\partial \left(r q_s\right)}{\partial r} = Q_s\!\left(r,t\right)
\end{equation}\endgroup
where $n_s$ and $T_s$ are the density and temperature, respectively, $q_s$ represents the \emph{heat transport flux} within the plasma, and $Q_s$ represents the \emph{heat source} of the plasma. This equation along with the mass transport, momentum transport and current diffusion equations form the basic equations of a plasma transport simulation code.

\begin{figure}[tb]
	\centering
	\includegraphics[scale=0.92,trim={0.5cm 0 0 0},clip]{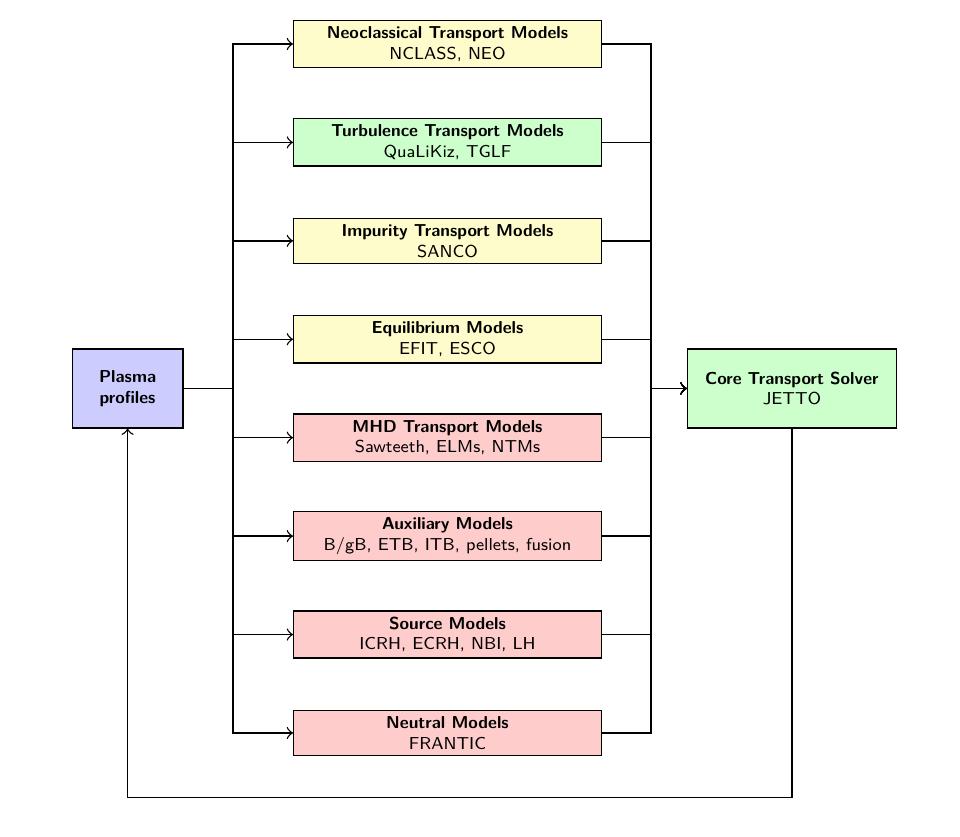}
	\caption{Workflow diagram of the JETTO integrated model, showing the coupled calculation of several smaller models and integrating their results in the core transport solver. The green modules are the primary focus of this study, while the yellow modules were used but not involved in the statistical study and the red modules were not used.}
	\label{fig:IntegratedModelWorkflow}
\end{figure}

Due to the complexity of these equations, they are typically solved numerically and in an iterative manner, requiring that the spatial and temporal coordinates be divided into discrete points, or \emph{grids}, to make its computation viable. Due to the effective timescales of the most influential plasma physics phenomena, the temporal grid typically has small intervals, typically on the order of $10^{-5}$~--~$10^{-3}$~s, to resolve and investigate their behaviour and underlying mechanisms. Such an approach effectively linearises the system of transport equations, allowing the separation of plasma transport processes into distinct sets of linearised equations which can be solved individually, using results from other models as inputs if necessary. The amalgamation and interconnection of these separate components back into a larger simulation suite is called an \emph{integrated model}. The plasma transport simulation code used in this study, JETTO, is an example of such an integrated model, schematically depicted in Figure~\ref{fig:IntegratedModelWorkflow}, and QuaLiKiz is the module used for the calculation of the turbulent transport fluxes, known to be the dominant contributor to transport fluxes, ie. $q_s$ within Equation~\eqref{eq:HeatTransportEquation}, within the core region of tokamak plasmas. This section discusses the settings used in this integrated modelling exercise and the results of the simulation.

\subsection{Nominal simulation settings for JET \#92436}
\label{subsec:NominalSettings}

Once processed by the GPR1D tool, the resulting kinetic profiles, along with their associated derivatives, can be used to calculate the input quantities needed by various plasma models. The uncertainties of the fit calculated by the GPR technique, which are themselves derived from the measurement uncertainties, allow for rigourous model V\&V efforts, described further in Sections~\ref{subsec:ValidationMetricDistDist} and \ref{subsec:ValidationMetricPointDist}.

The fitted kinetic profiles were given as inputs to an interpretive TRANSP~\cite{aTRANSP-Hawryluk} calculation, along with the experimental input heating parameters, in order to determine the associated particle, heat, and momentum source profiles. Then, both the fitted kinetic profiles and calculated source profiles were used to define the initial and boundary conditions for the JETTO + QuaLiKiz integrated model, which then self-consistently evaluates the particle, heat, and momentum flux within the chosen steady-state time window, then consequently the kinetic profiles themselves. Due to the focus on steady-state solutions, it is assumed that only the boundary condition, set at $\rho_{\text{tor}}=0.85$ for the simulation of this discharge, will have a significant impact on the simulation results. Although the initial condition can affect steady-state solutions through the switching of plasma turbulence regimes, the variations in the initial condition required for such effects to be important are typically much larger than the uncertainties given by the GPR profile fits. Thus, any initial condition dependencies can be safely neglected in this study. As the TRANSP calculation also provides the fast ion density and energy density profiles, it was decided to also include them inside the quasilinear turbulent transport model, QuaLiKiz, as additional Maxwellian-distributed ions species. This implementation only accounts for the linear component of the fast ion species as the non-linear saturation rules for the fast ion turbulence contributions are still under study. For reference purposes, a list of other relevant settings and parameters for the simulation can be found in Appendix~\ref{app:BaseSimulationSettings}.

A number of modifications were made to the input profiles in order to increase the self-consistency of the simulation. Some complications are foreseen in incorporating modifications of this nature into any automatised version of the proposed verification workflow, due to their reliance on additional signals and consequent analysis, but the modifications themselves are still presented here for completeness.

\begin{figure}[tb]
	\centering
	\includegraphics[scale=0.375]{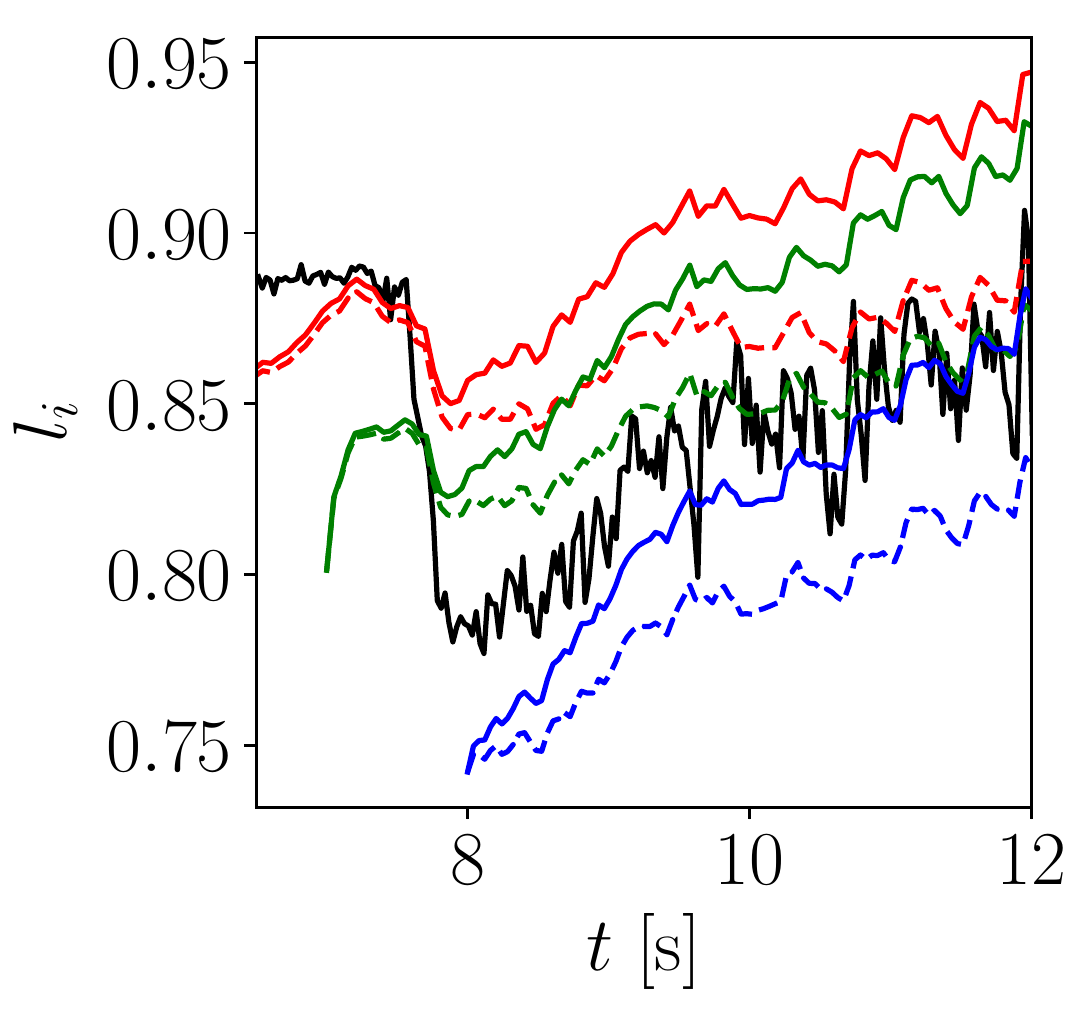}%
	\includegraphics[scale=0.375]{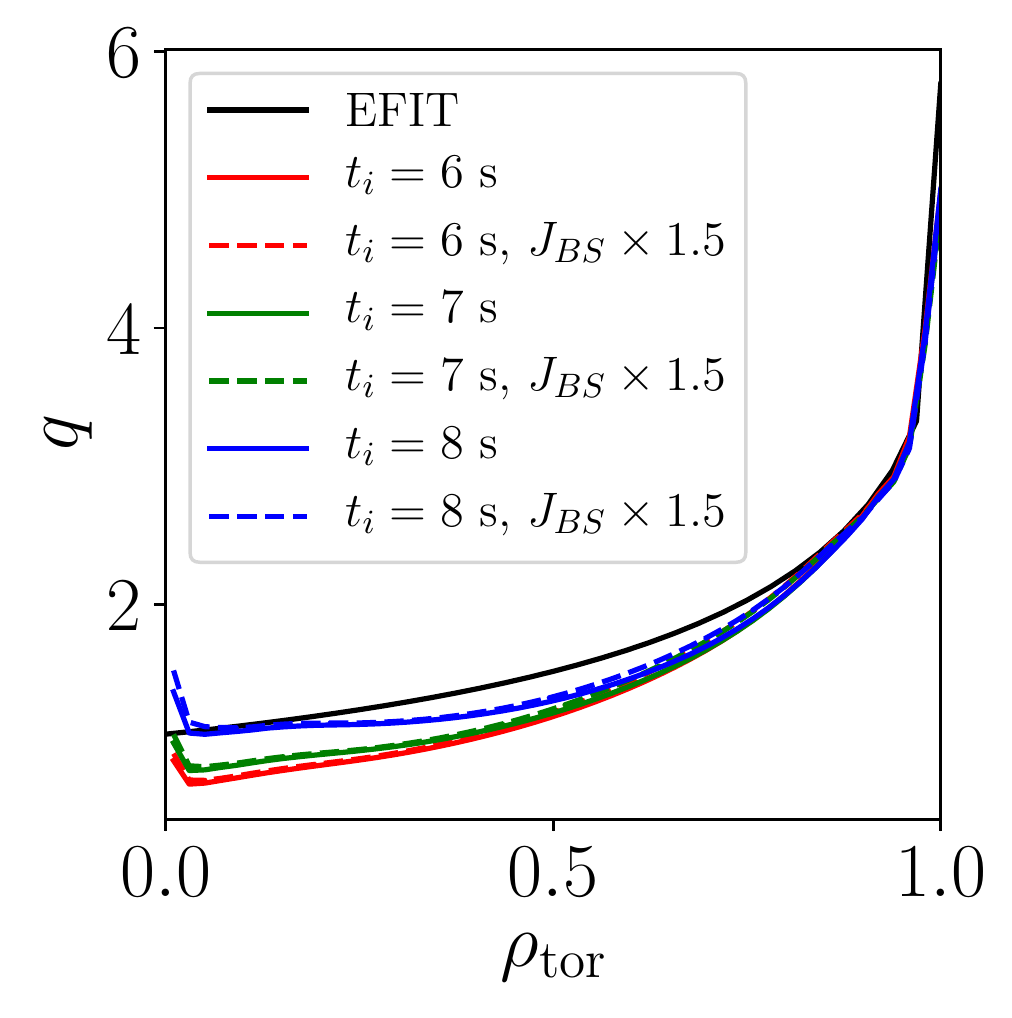}
	\caption{Comparison of the simulated $l_i$ time traces from the various interpretive JETTO runs against the measured $l_i$ signal taken from EFIT/LI3D.}
	\label{fig:InternalInductanceStudiesJET92436}
\end{figure}

Firstly, the input safety factor, $q$, profile was not prescribed using the standard equilibrium fitting (EFIT) routine at JET, but was instead calculated from a separate interpretive JETTO simulation with the fitted profiles as inputs. Within the framework of integrated modelling, this is justified by the knowledge that the current profile evolves on a slower time scale than the kinetic profiles. Thus, to avoid non-physical feedback loops in the time evolution of the kinetic profiles due to excessively far initial conditions, the chosen $q$ profile was such that the time evolution of the simulated internal inductance, $l_i$, had a reasonable match with the measured values at the desired simulation start time, 10~s. Figure~\ref{fig:InternalInductanceStudiesJET92436} shows the comparison study of the $l_i$ traces and their corresponding $q$ profiles. The $q$ profile chosen to be used as the base setting was the $t = 8$~s option.

\begin{figure}[tb]
	\centering
	\includegraphics[scale=0.45]{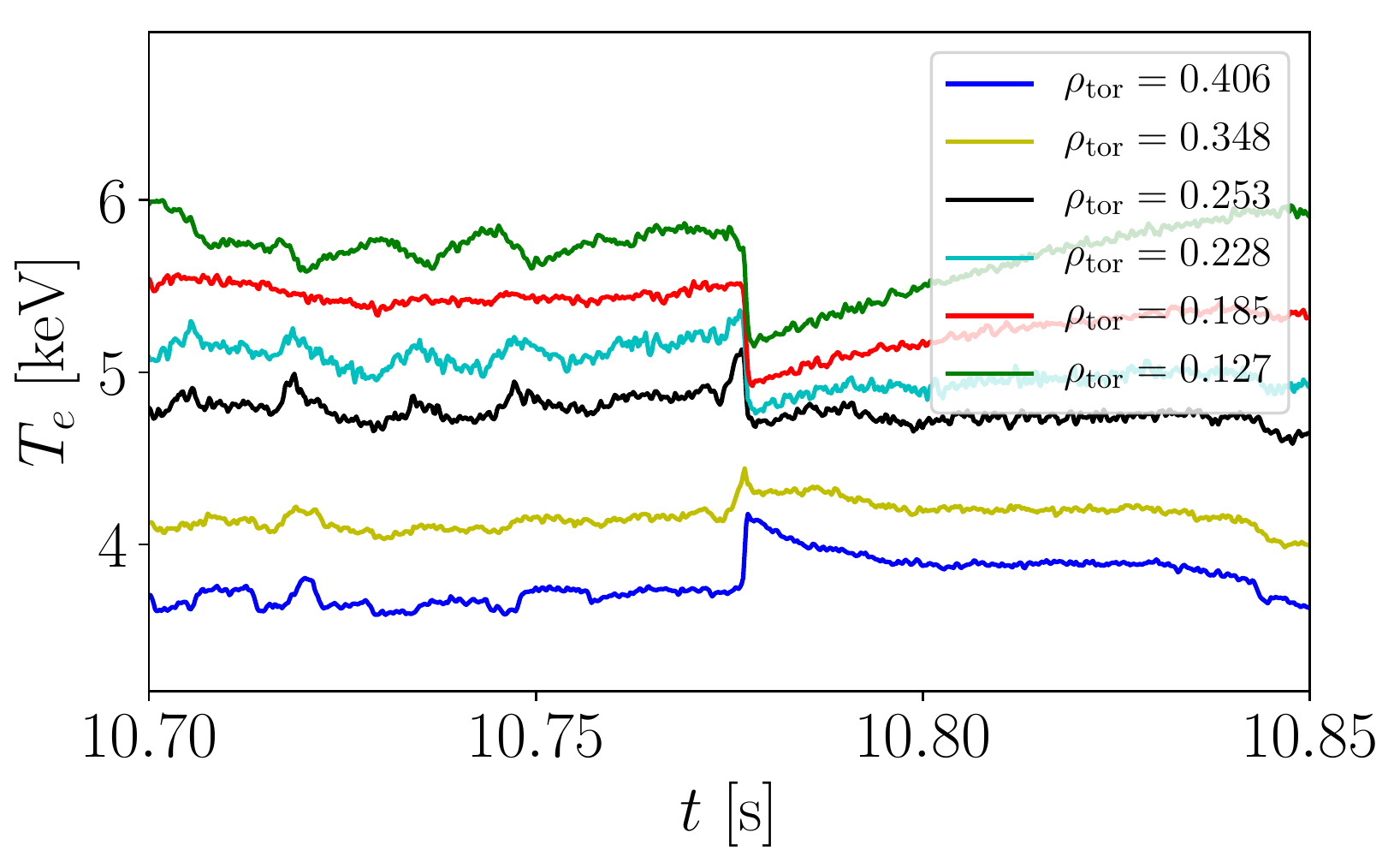}
	\caption{Time traces of the core ECE radiometer channels, showing the signature behaviour of MHD activity in the core and the inversion radius, $\rho_{\text{tor,inv}} \approx 0.25$, of the sawtooth instability located at $t\approx10.77$. The sawtooth behaviour continues throughout the discharge at a frequency of $\sim1$~Hz, with the smaller sawtooth-like behaviour before the crash begin indicative of additional MHD instabilities.}
	\label{fig:CoreMHDActivityStudiesJET92436}
\end{figure}

Secondly, an inspection of the time-resolved temperature measurements from the ECE heterodyne radiometer revealed the presence of MHD behaviour, as can be seen in Figure~\ref{fig:CoreMHDActivityStudiesJET92436}. The presence of sawteeth behaviour is visible throughout the discharge at a frequency of $\sim\!1$~Hz but is only plotted around the crash at 10.77~s for clarity. From these measurements, the inversion radius of the sawtooth crash was estimated to be located at $\rho_{\text{tor,inv}} \simeq 0.25$, but the presence of similar behaviour with a different inversion radius before sawtooth crash is indicative of additional MHD behaviour. As the explicit modelling of all the complex MHD behaviour within the inner core is not necessary for the validation of the turbulent transport model, an ad-hoc emulation of the expected transport of this phenomena was implemented in the model instead. As such, the $q$ profile was further modified with a linear multiplication, such that $q=1$ at the observed inversion radius, and the diffusion coefficients in the simulation were manually increased in the central core region, ie. $\rho_{\text{tor}} < \rho_{\text{tor,inv}}$, as a proxy for sawtooth-induced transport in this region. Both the electron and ion thermal diffusion coefficients, $\chi_e$ and $\chi_i$ respectively, were modified according to a Gaussian-shaped function, centered on $\rho_{\text{tor}} = 0.0$ with a height of $1.0$ $\text{m}^2$ $\text{s}^{-1}$ and a standard deviation of $0.15$ in the toroidal rho coordinates, such that the additional contribution is sufficiently reduced but non-zero at the inversion radius. The density diffusion coefficient, $D$, was also modified using an identical function except with a height of $2.0$ $\text{m}^2$ $\text{s}^{-1}$.

Thirdly, based on the presence of nickel within the vessel during this discharge observed via spectroscopy and a suspected population of beryllium due to the first wall materials and tungsten from the divertor tiles, the impurity transport module, SANCO~\cite{aJINTRAC-Romanelli}, was employed to self-consistently evolve the impurity density profiles. The boundary conditions of the impurity densities were chosen such that the simulation converged on a 0.8\% beryllium (Be) and 0.07\% nickel (Ni) impurity ion composition within the inner core, with both species computed with variable ionisation levels and the percent composition referenced to the measured electron density, $n_e$. The constraint applied was that the beryllium concentration should be between 0.5\% and 1.0\%, as given by the acceptable impurity concentration for the observed IC heat deposition. The tungsten concentration in the plasma was adjusted such that the total radiated power of the simulation was within 10\% of the measured value, leading to a core tungsten density around 0.004\% of $n_e$. The computed radiated power profile from SANCO was then used self-consistently in the simulation.

Finally, since QuaLiKiz is an electrostatic code, an ad-hoc emulation of electromagnetic (EM) $\beta$-stabilisation of ITG turbulence was added as a non-standard option. However, as JET \#92436 is a baseline discharge with a reasonably low $\beta_N$, it is anticipated that these effects will play a minor role in the predicted profiles of the discharge. The effect of this stabilisation mechanism is shown to be important for high performance hybrid scenarios with $\beta_N > 2.5$ and significant fast ion populations~\cite{aNonLinearFI-Citrin,aEMStabilisation-Citrin,aGKSuppress-Doerk,aFastBeta-Garcia}, where:
\begingroup\makeatletter\def\f@size{9}\check@mathfonts
\def\maketag@@@#1{\hbox{\m@th\normalsize\normalfont#1}}%
\begin{equation}
\label{eq:NormalizedPlasmaBeta}
	\beta_N = \beta \frac{a B_T}{I_p} \simeq \frac{2 \mu_0 a}{B_T I_p} \left\langle \sum_s n_s T_s \right\rangle_V
\end{equation}\endgroup
where $\left\langle\;\right\rangle_V$ denotes a volume-average operation. For JET \#92436, with $\beta_{\text{N,th}} = 1.88$ and $\beta_{\text{N,tot}} = 2.11$, the impact of the EM stabilisation is not expected to be significant but the option is retained in the simulation for completeness. The ad-hoc implementation applies a numerical reduction of the normalised ion temperature gradient, $R/L_{T_i}$, input to QuaLiKiz by the ratio of the local thermal energy density over the local total energy density, $W_{\text{th}}/W_{\text{tot}}$\footnote[3]{\label{foot:Wfast}The implementation used in this paper has since been determined to overestimate $W_{\text{fast}}$ by a factor of 2.25. While the effect of this ad-hoc factor on the $T_i$ simulation results is significant, it is expected to have a negligible effect on the conclusions made by this study.}. This ad-hoc correction does not imply that fast ions are solely responsible for the EM-stabilisation effect, but rather acknowledges that the expected level of stabilisation, including contributions from the thermal component, $\beta_{N,\text{th}}$, is strongly correlated with the fast ion content in discharges with substantially high NBI and IC auxiliary heating powers.

\begin{figure}[tb]
	\centering
	\hspace{1mm}\includegraphics[scale=0.275]{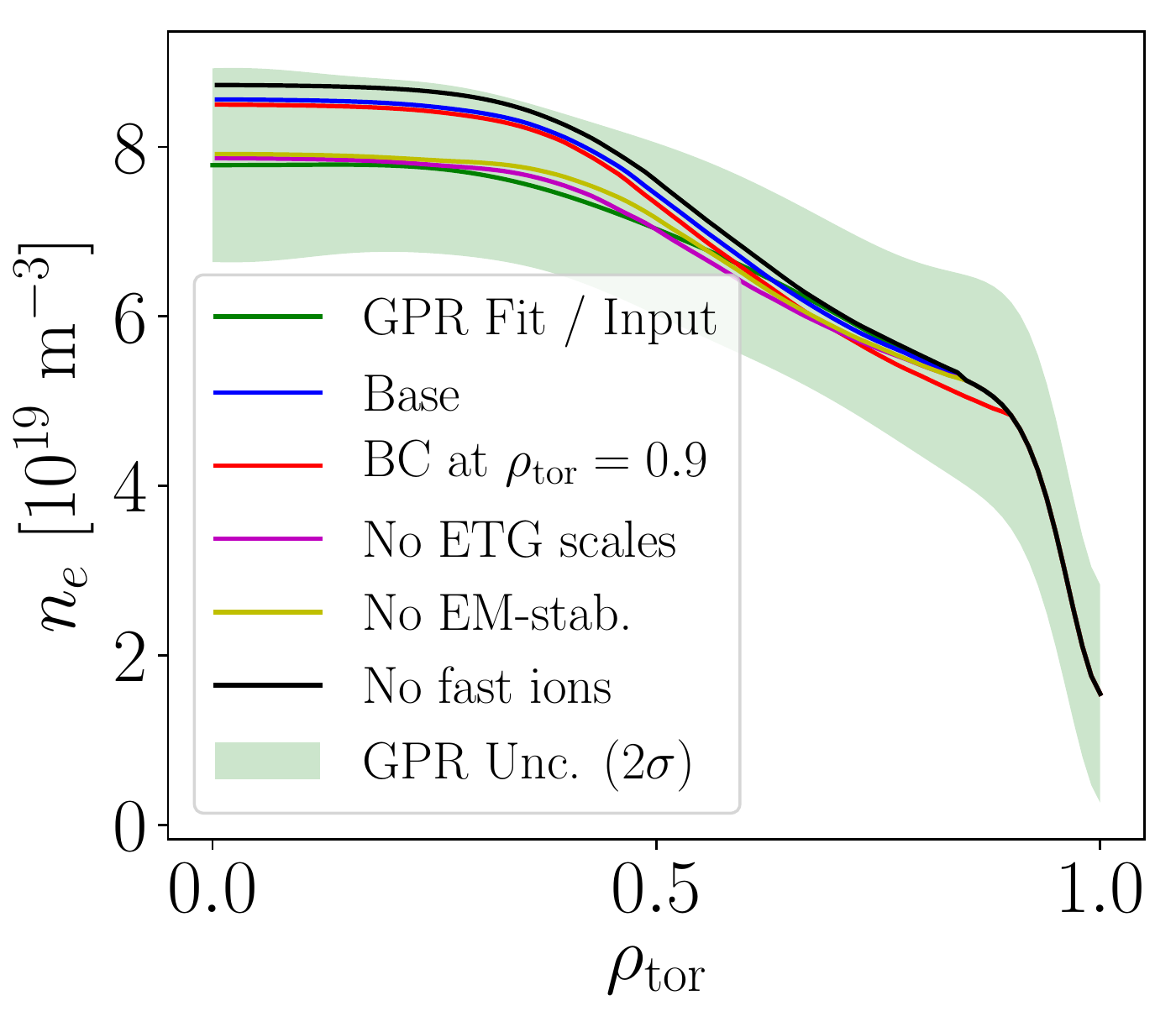}%
	\hspace{1mm}\includegraphics[scale=0.275]{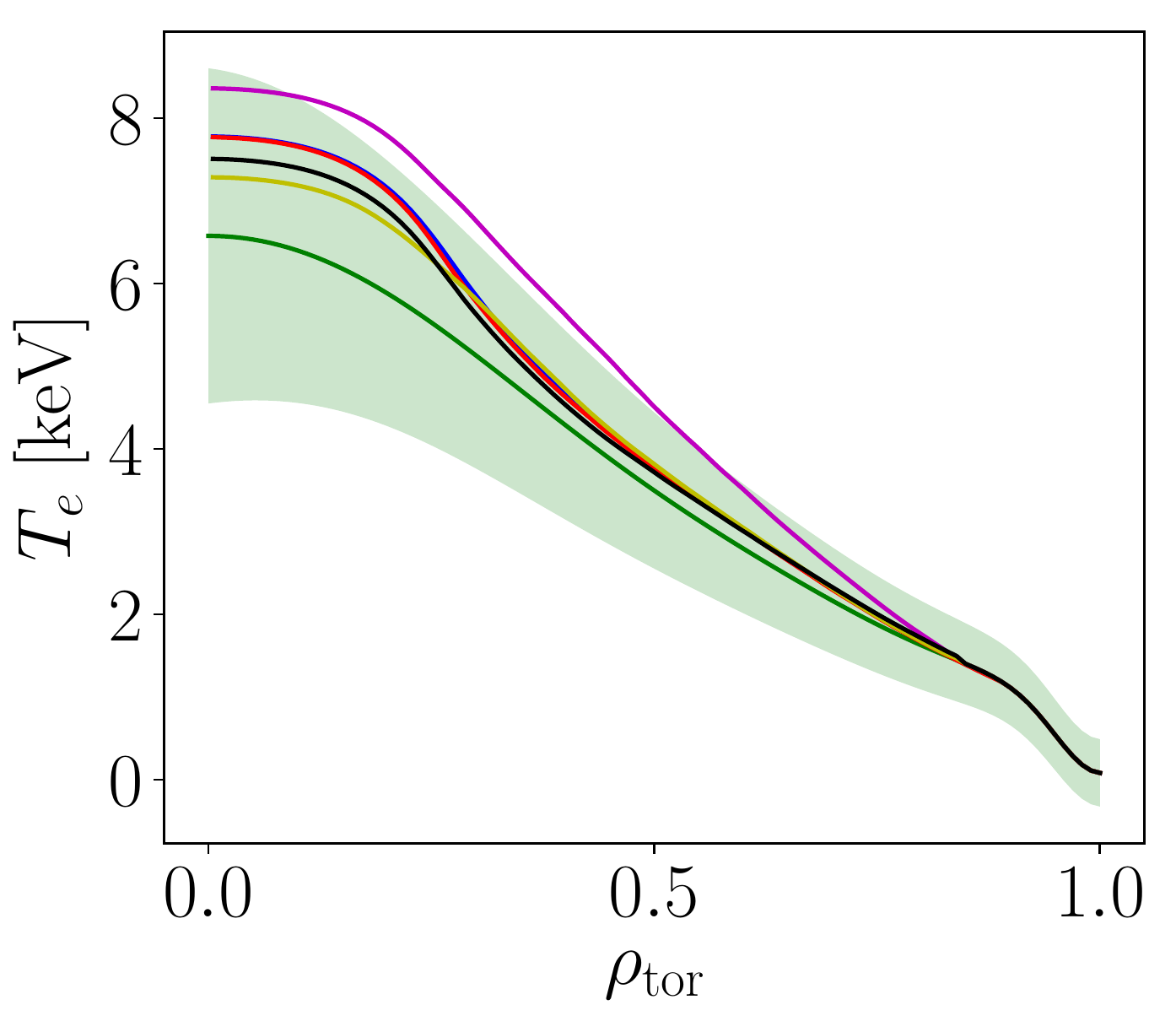}
	\includegraphics[scale=0.275]{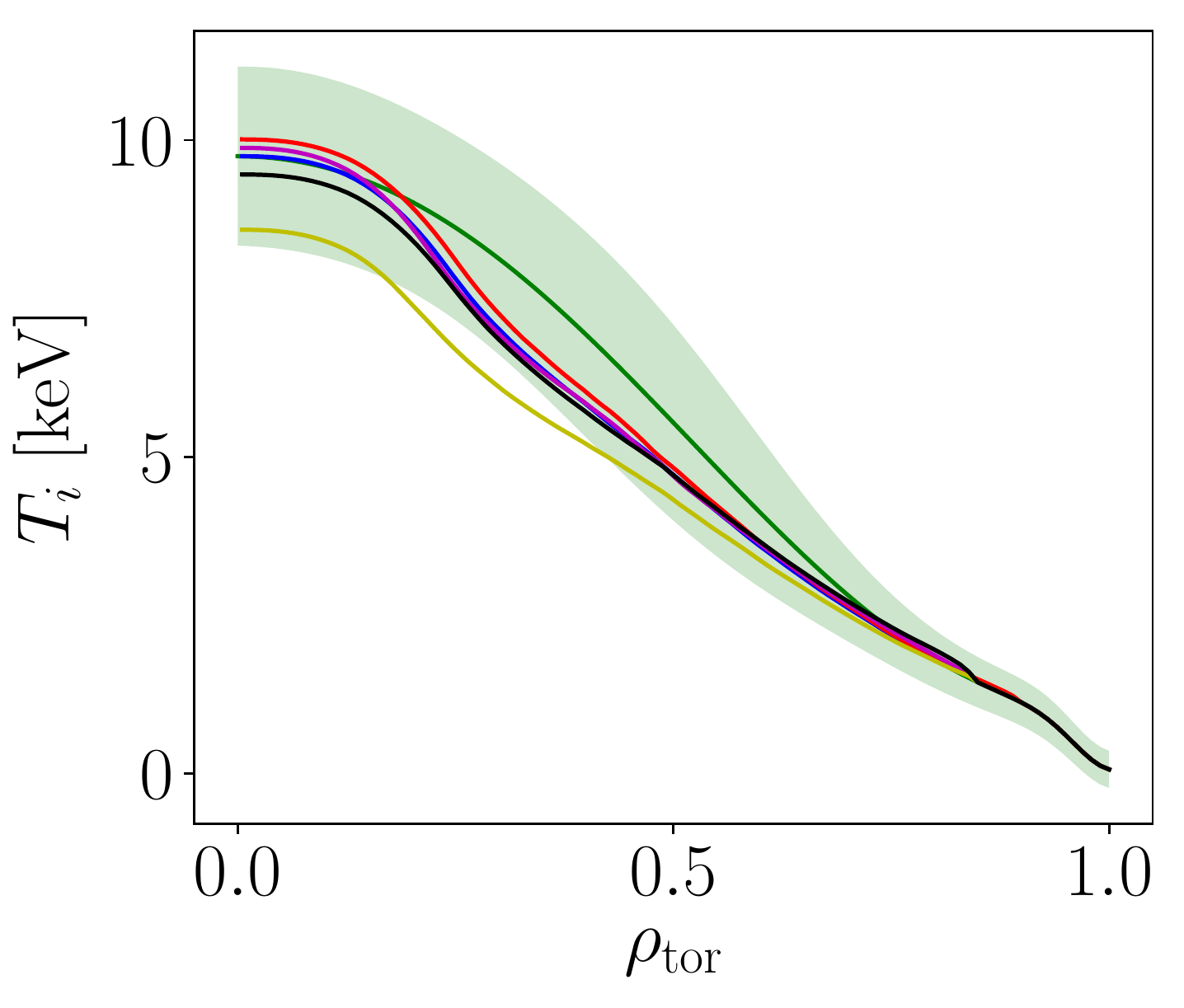}%
	\includegraphics[scale=0.275]{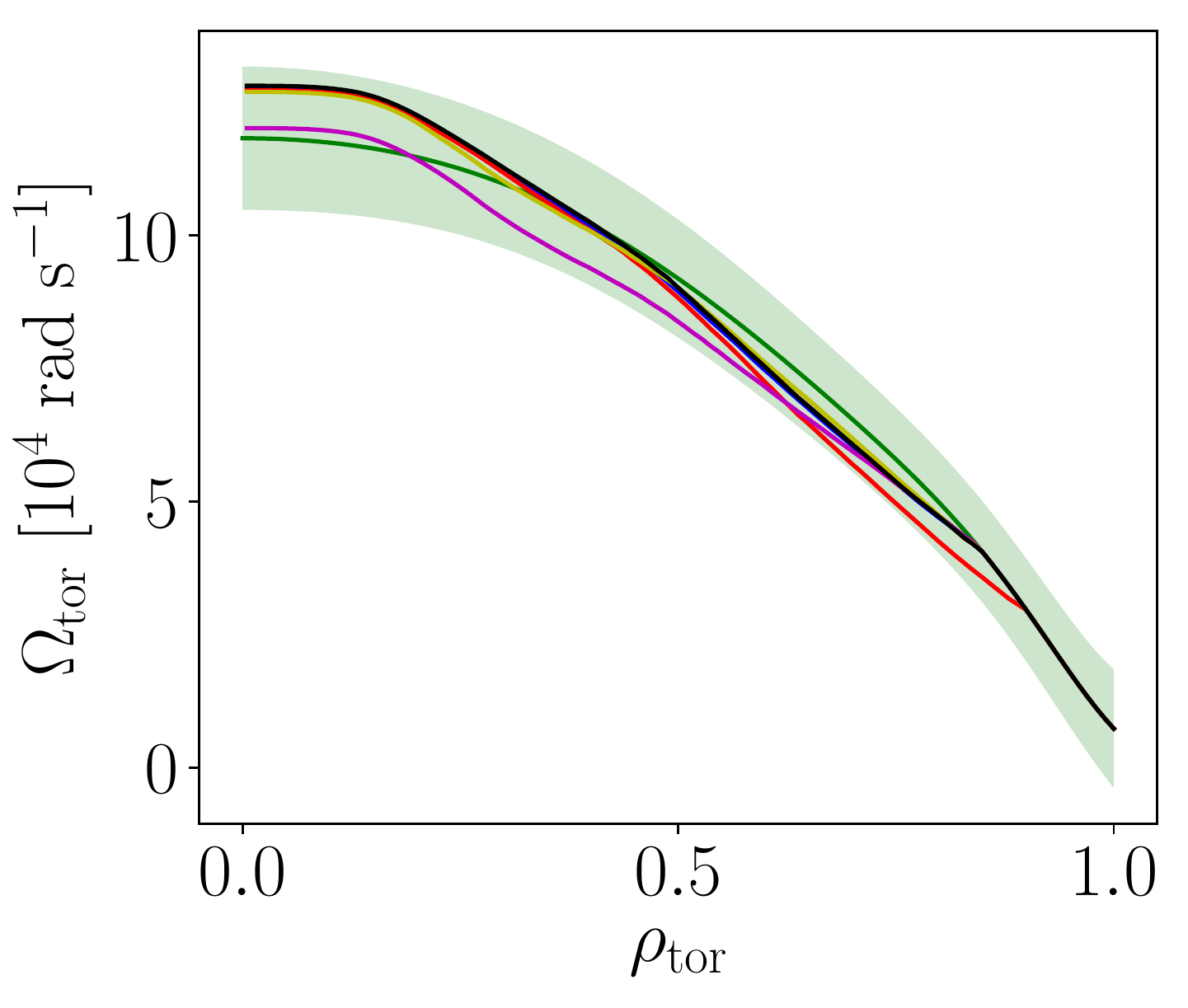}
	\caption{Results of the sensitivity studies regarding the addition or inclusion of known physical phenomena, where the input profiles (green lines) are compared against the output profiles (see legend) and the base case scenario (blue line). Upper left: Electron density profiles. Upper right: Electron temperature profiles. Lower left: Ion temperature profiles. Lower right: Toroidal angular frequency profiles.}
	\label{fig:PhysicsSensitivityResultsJET92436}
\end{figure}

In order to test the applicability of known physical phenomena for recovering the plasma conditions in JET \#92436, including the ad-hoc electromagnetic stabilisation factor, a number of sensitivity studies were performed based on the exclusion of certain physics from the JETTO + QuaLiKiz simulation. In order to increase the rigour of these tests, the results were evaluated against the fit uncertainties provided by the GPR. The sensitivities performed for this identification analysis are as follows:
\begin{itemize}
	\itemsep 0pt
	\item moving the simulation boundary condition to $\rho_{\text{tor}} = 0.9$;
	\item switching off the calculation of electron temperature gradient (ETG) scale turbulence in QuaLiKiz;
	\item removing the ad-hoc EM-stabilization based on $W_{\text{th}}/W_{\text{tot}}$ from QuaLiKiz;
	\item removing the linear contribution of the fast ion populations from QuaLiKiz.
\end{itemize}
Figure~\ref{fig:PhysicsSensitivityResultsJET92436} shows the results of the base simulation and the described sensitivity studies established earlier in this section.

From the simulation using the base settings, the blue line in Figure~\ref{fig:PhysicsSensitivityResultsJET92436}, there is a slight overprediction of $n_e$ and $T_e$ within the inner half of the core, $\rho_{\text{tor}} \le 0.5$, and an underprediction of $T_i$ near the mid-radius, $\rho_{\text{tor}} \in \left[0.3,0.6\right]$. The agreement of these profiles are considered to be good given the $2\sigma$ uncertainties of the input profiles and the complexity of the simulation undertaken. However, further discussion on the possible explanations of these discrepancies are relegated to Section~\ref{subsec:NominalSettingsValidation}, after the introduction of results from a more statistical rigourous study.

The extension of the simulation boundary to $\rho_{\text{tor}} = 0.9$ yields similar results for the $n_e$, $T_e$, and $T_i$ profiles, which indicates good performance of the turbulent transport model in H-mode baseline plasmas up until the pedestal top and potentially into the pedestal. However, as this radial location is just inside the pedestal region, as seen in the $n_e$ profile, it was chosen to forego using this extended boundary condition as the base settings for this validation exercise. The reduced agreement of the $\Omega_{\text{tor}}$ profile at the edge when using the extended boundary condition is attributed more to a poorly resolved pedestal in the angular frequency measurements, as further evidenced by the overall smoothness of the GPR fit.

The exclusion of ETG scale turbulence from QuaLiKiz yields a significantly higher $T_e$ profile, likely due to the supression of electron heat transport generated by the ETG instabilities. The QuaLiKiz ETG model contains a rudimentary multi-scale model~\cite{aNonLinearFI-Citrin} which is not fully verified against nonlinear multiscale simulations~\cite{aMultiScale-Maeyama,aMultiScaleGK-Staebler,aGKHeat-Howard}. Nevertheless, the excellent agreement for the $T_e$ profile using the QuaLiKiz predictions, with a significant ETG contribution, provides a compelling case for further nonlinear investigation of ETG impact in this discharge. Such an investigation is outside the scope of this paper.

The addition of the ad-hoc EM-stabilization factor significantly increases the density and ion temperature profile, as the reduction of the normalized ion temperature gradient input, $R/L_{T_i}$, to QuaLiKiz reduces the driving mechanism of ITG turbulence. As this instability is a significant transport channel for both particles and ion heat, it effectively causes the integrated modelling suite to drive the local density and ion temperature gradients higher, in order to achieve the fluxes required to balance the source terms in the simulation. While its removal improves the $n_e$ and $T_e$ predictions, the reduction of the $T_i$ prediction to values outside the $2\sigma$ GPR uncertainties motivated the decision to retain this ad-hoc factor in the base settings.

Finally, the contribution of the fast ion species in the simulations only had a minor impact on the results. However, the fast ion impact on turbulence in QuaLiKiz is currently limited to dilution and electrostatic kinetic effects. In this discharge, it is possible that the EM-stabilisation of ITG modes is further enhanced by sharp fast ion gradients, particularly those generated by IC heating at inner radii~\cite{aNonLinearFI-Citrin}. This effect is not captured by the simple ad-hoc EM-stabilisation model employed here and may partially explain the $T_i$ underprediction. Further investigation of this effect, which would involve nonlinear gyrokinetic simulations, is left for future work as it is outside the scope of this paper.

Due to the physical arguments for the settings chosen for this execution, they were designated as the \emph{base settings} to represent this particular time window in this discharge. This is further supported by the level of agreement between the input profiles from the GPR technique and the output profiles from the converged JETTO + QuaLiKiz execution, as shown by the blue line in Figure~\ref{fig:PhysicsSensitivityResultsJET92436}. From these base case settings, a wide variety of additional sensitivity studies were performed, as further discussed in Sections~\ref{subsec:RotationSensitivities} and \ref{subsec:ImpuritySensitivities}.

\subsection{Validation of the nominal settings}
\label{subsec:NominalSettingsValidation}

\begin{figure}[tb]
	\centering
	\includegraphics[scale=0.26]{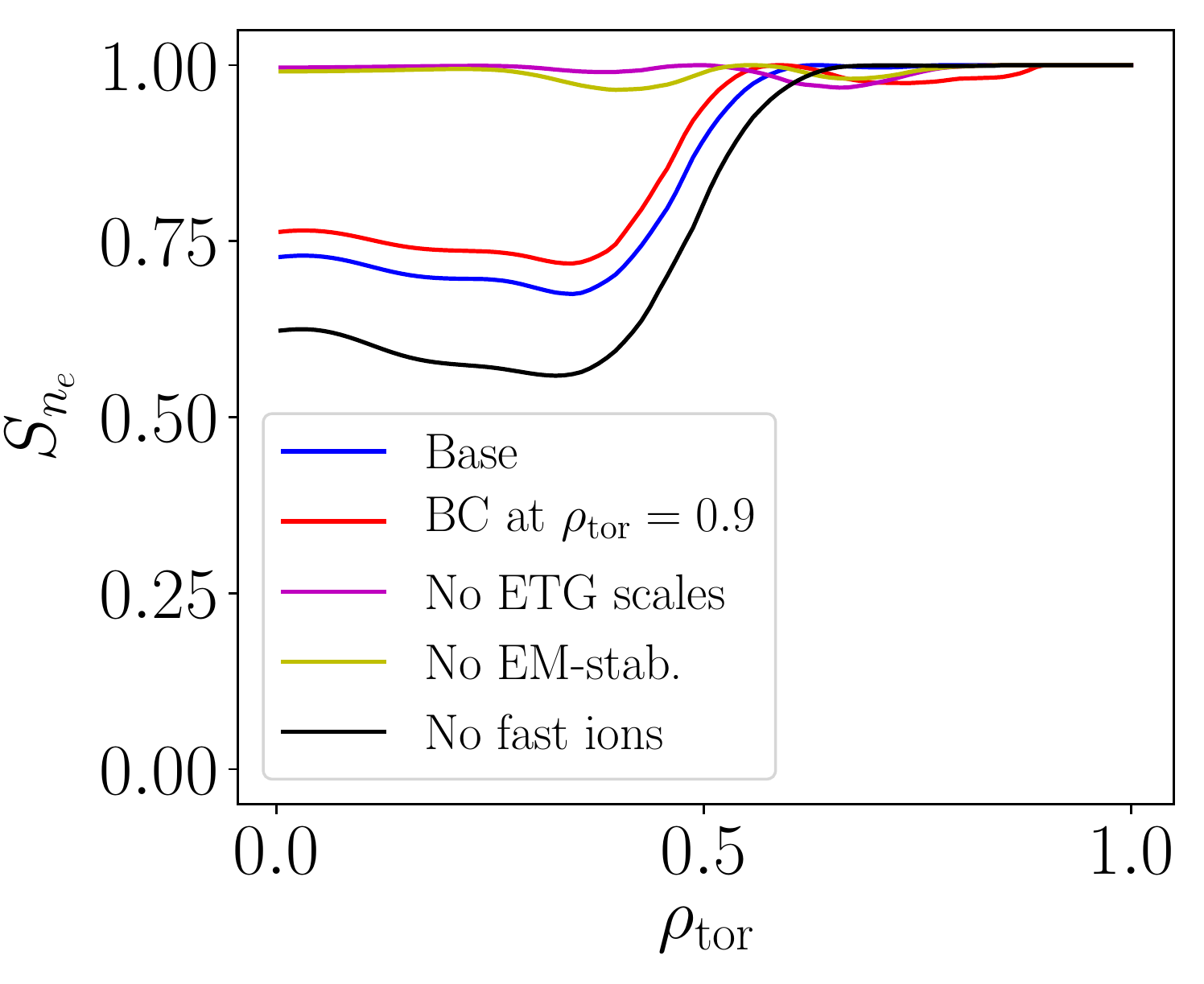}%
	\hspace{1mm}\includegraphics[scale=0.26]{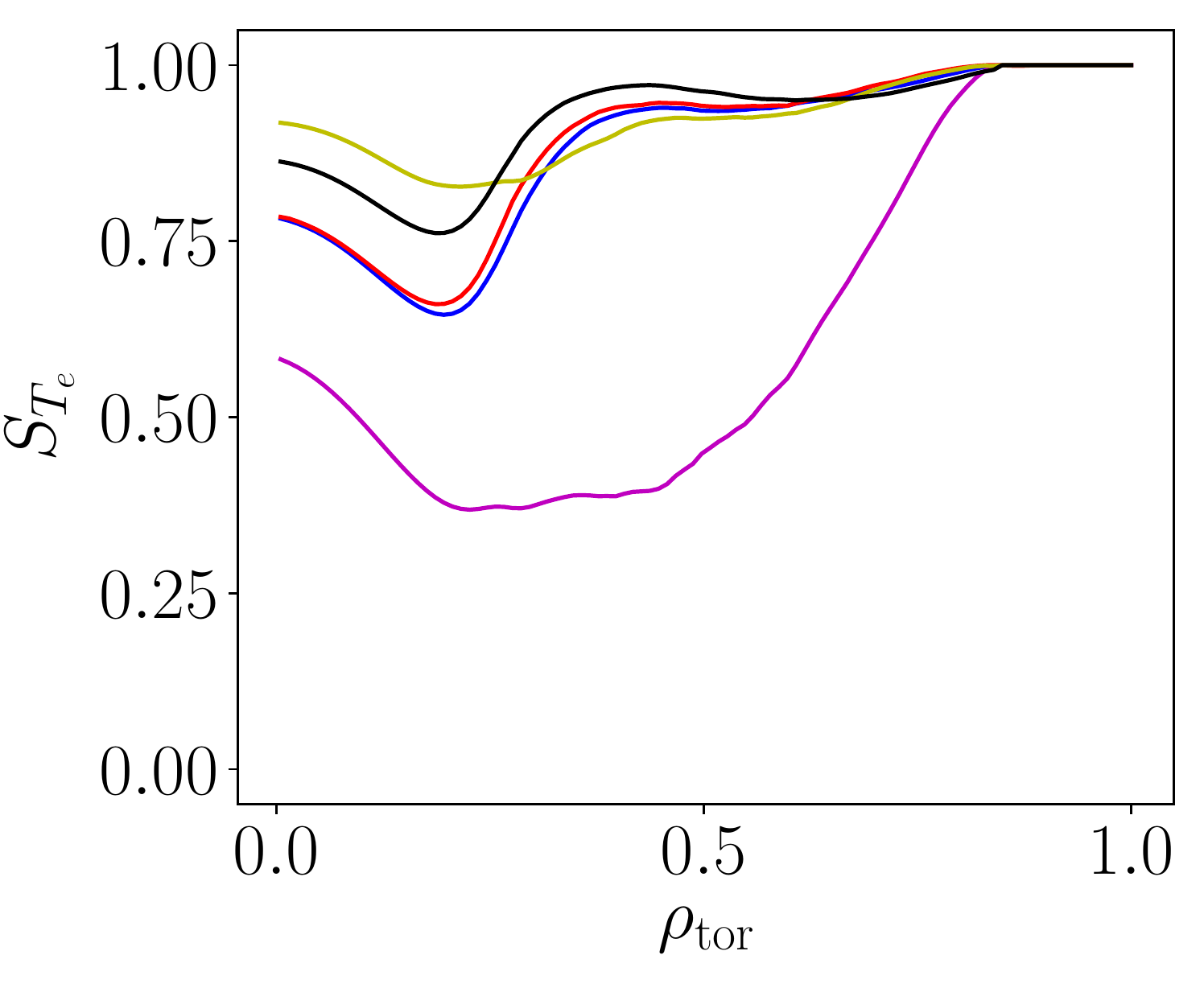}
	\includegraphics[scale=0.26]{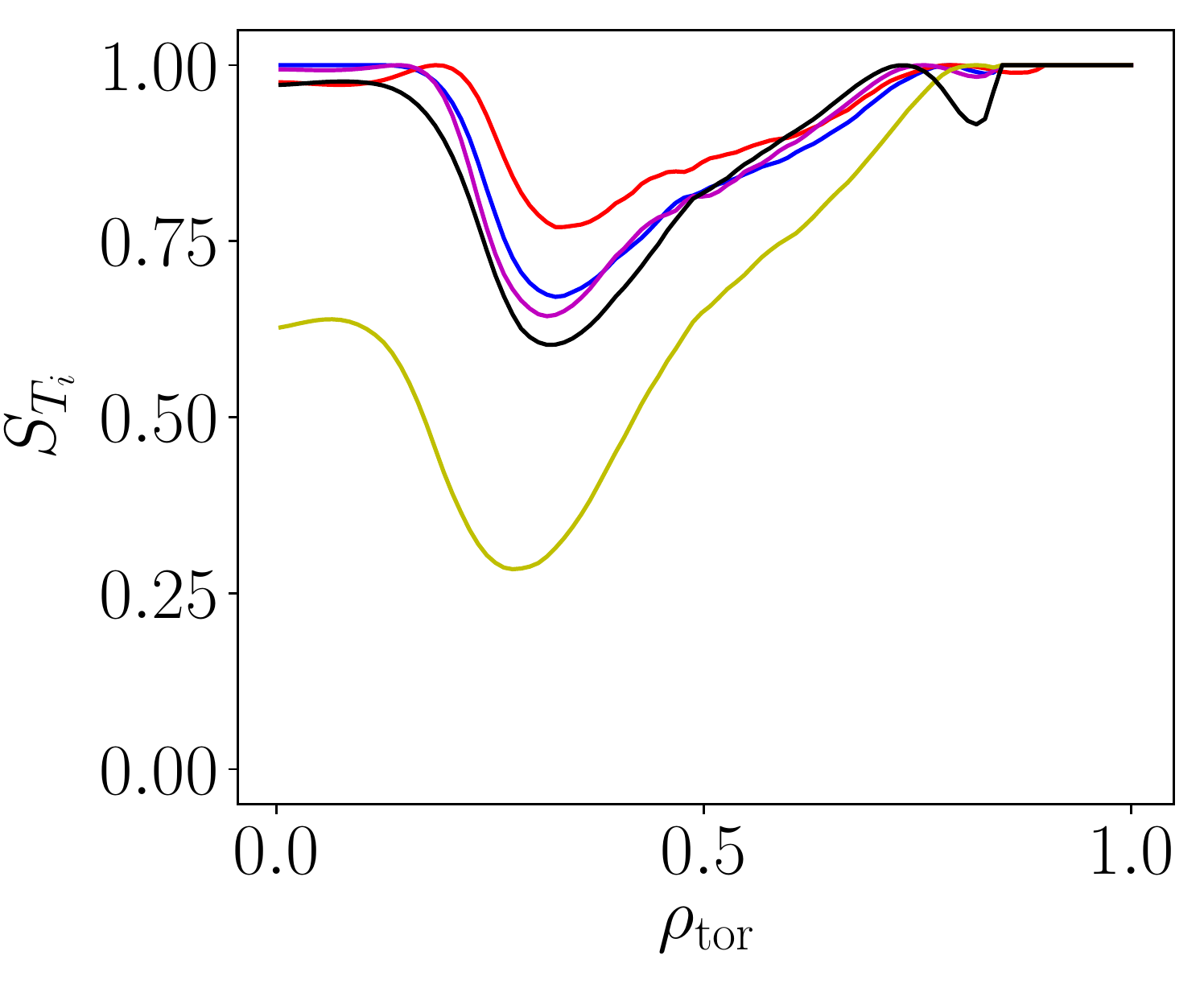}%
	\hspace{1mm}\includegraphics[scale=0.26]{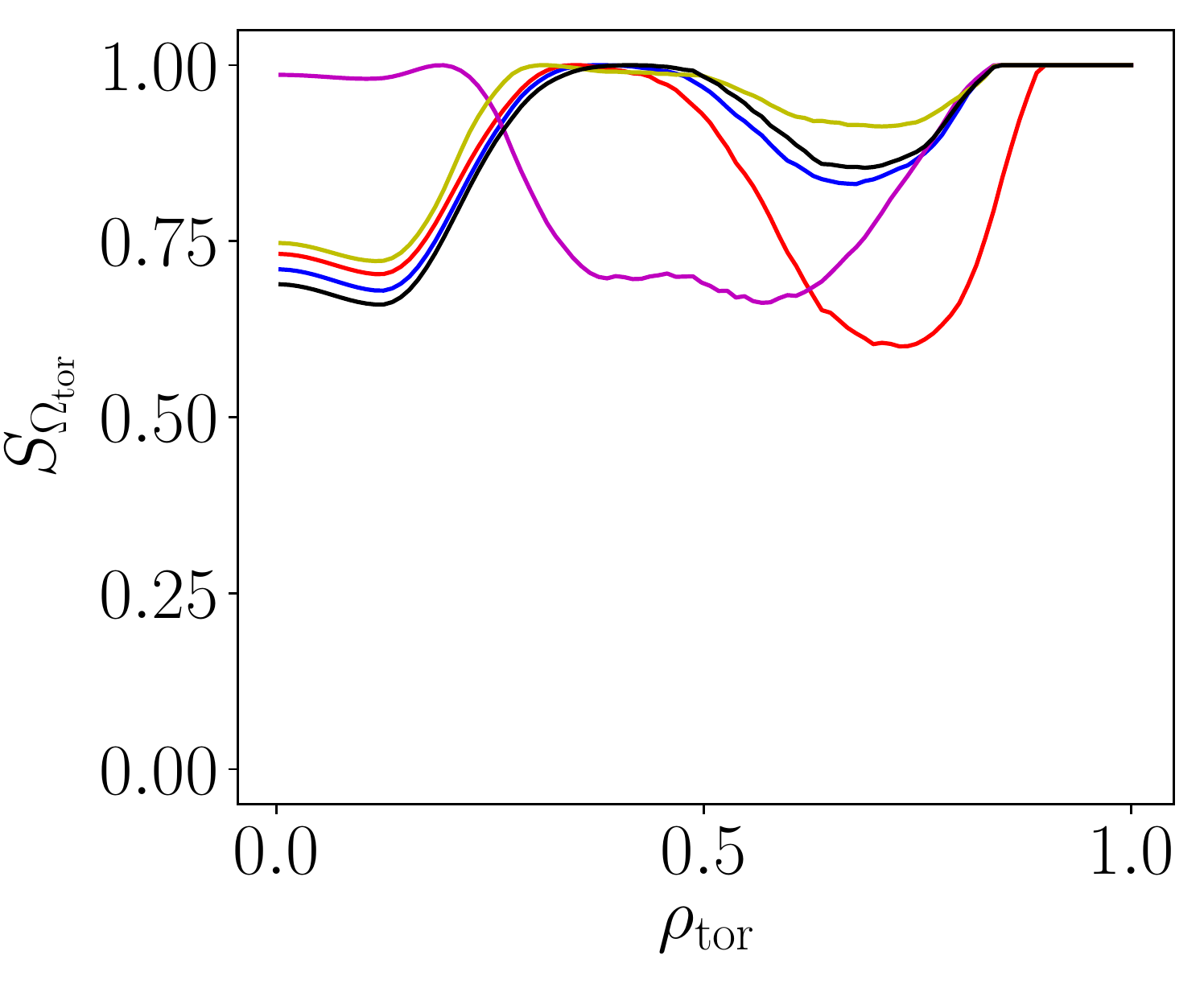}
	\caption{Point-distribution validation metric for sensitivity study results regarding the addition or inclusion of known physical phenomena (see legend). Upper left: Electron density profiles. Upper right: Electron temperature profiles. Lower left: Ion temperature profiles. Lower right: Toroidal angular frequency profiles.}
	\label{fig:PhysicsSensitivitySignificanceJET92436}
\end{figure}

Figure~\ref{fig:PhysicsSensitivitySignificanceJET92436} shows the results of applying Equation~\eqref{eq:PointDistributionSignificance} to the profiles shown in Figures~\ref{fig:PhysicsSensitivityResultsJET92436}. As shown in this figure, the level of agreement between the experimental fit and the simulation output profiles is quantitatively captured by the proposed metric, with a value of $S \ge 0.5$ indicating that the profile lies within the $\pm 2\sigma$ boundary of the experimental fit distribution. The negative impact of the removal of the ETG scale turbulence calculation in QuaLiKiz on the $T_e$ and $\Omega_{\text{tor}}$ agreement is clearly evident in Figure~\ref{fig:PhysicsSensitivitySignificanceJET92436}. In addition, the negative impact of the removal of the ad-hoc EM stabilisation factor on the $T_i$ agreement is also clearly shown.

\begin{figure}[tb]
	\centering
	\hspace{-2mm}\includegraphics[scale=0.27]{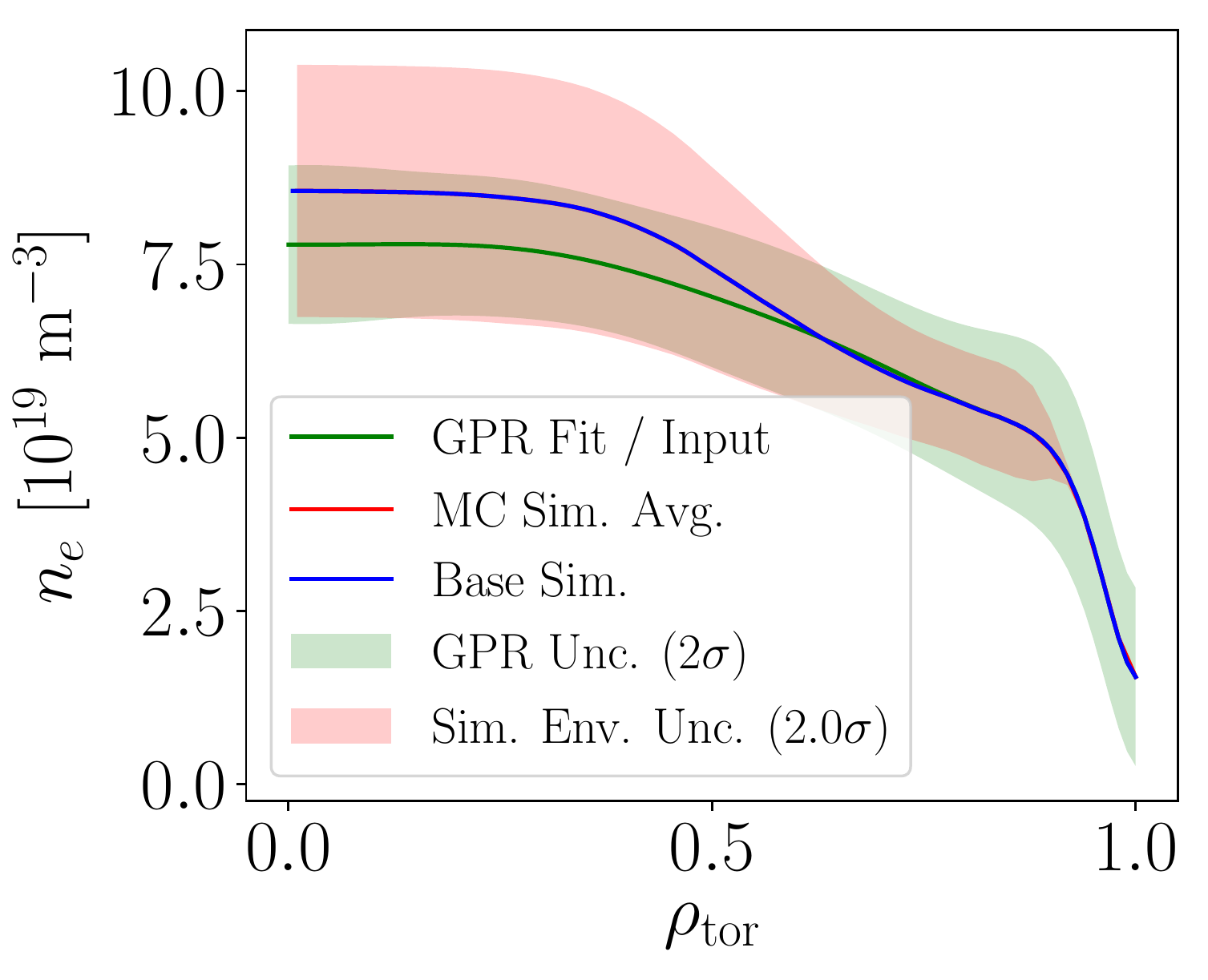}%
	\hspace{1mm}\includegraphics[scale=0.27]{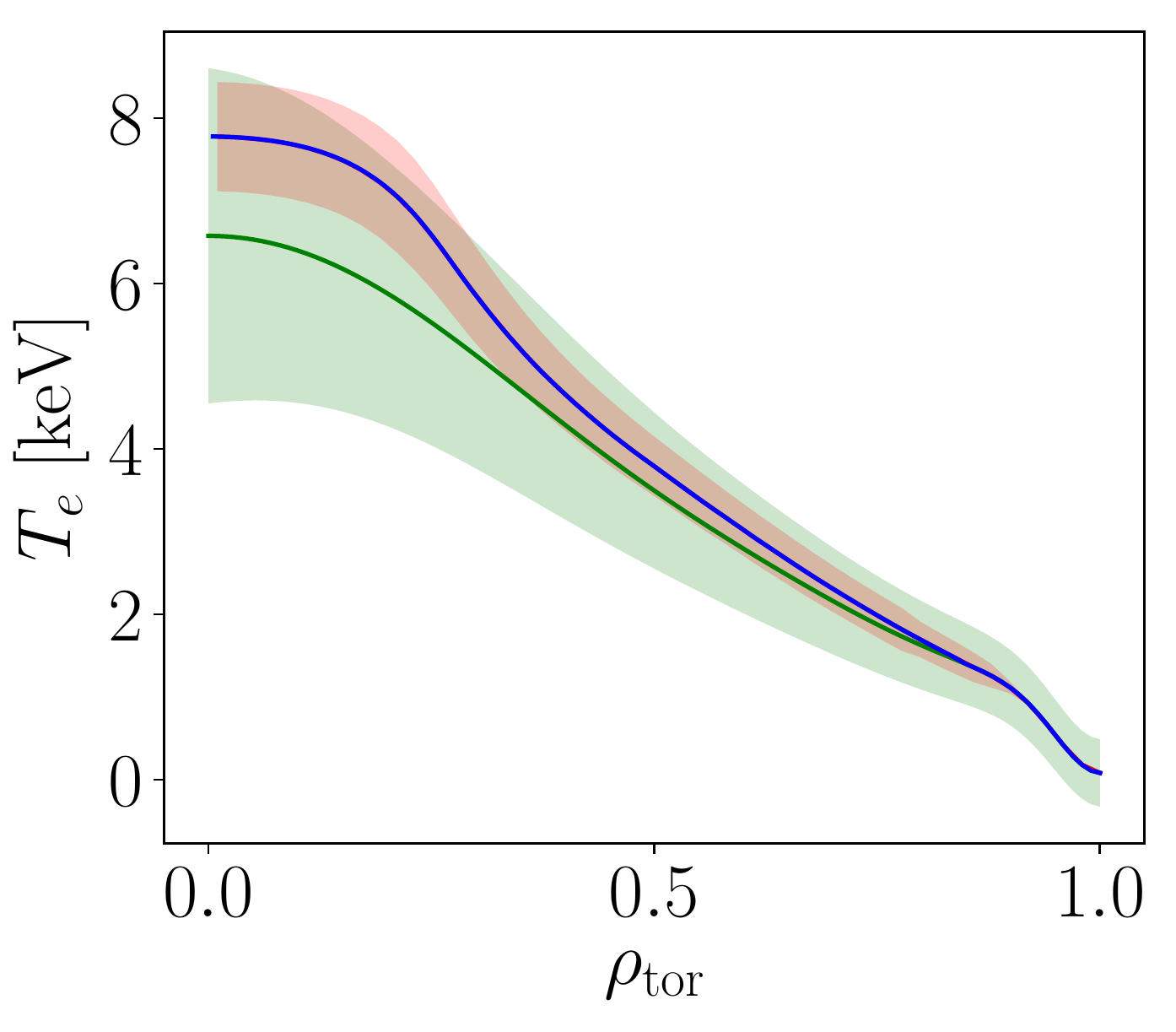}
	\hspace{4mm}\includegraphics[scale=0.27]{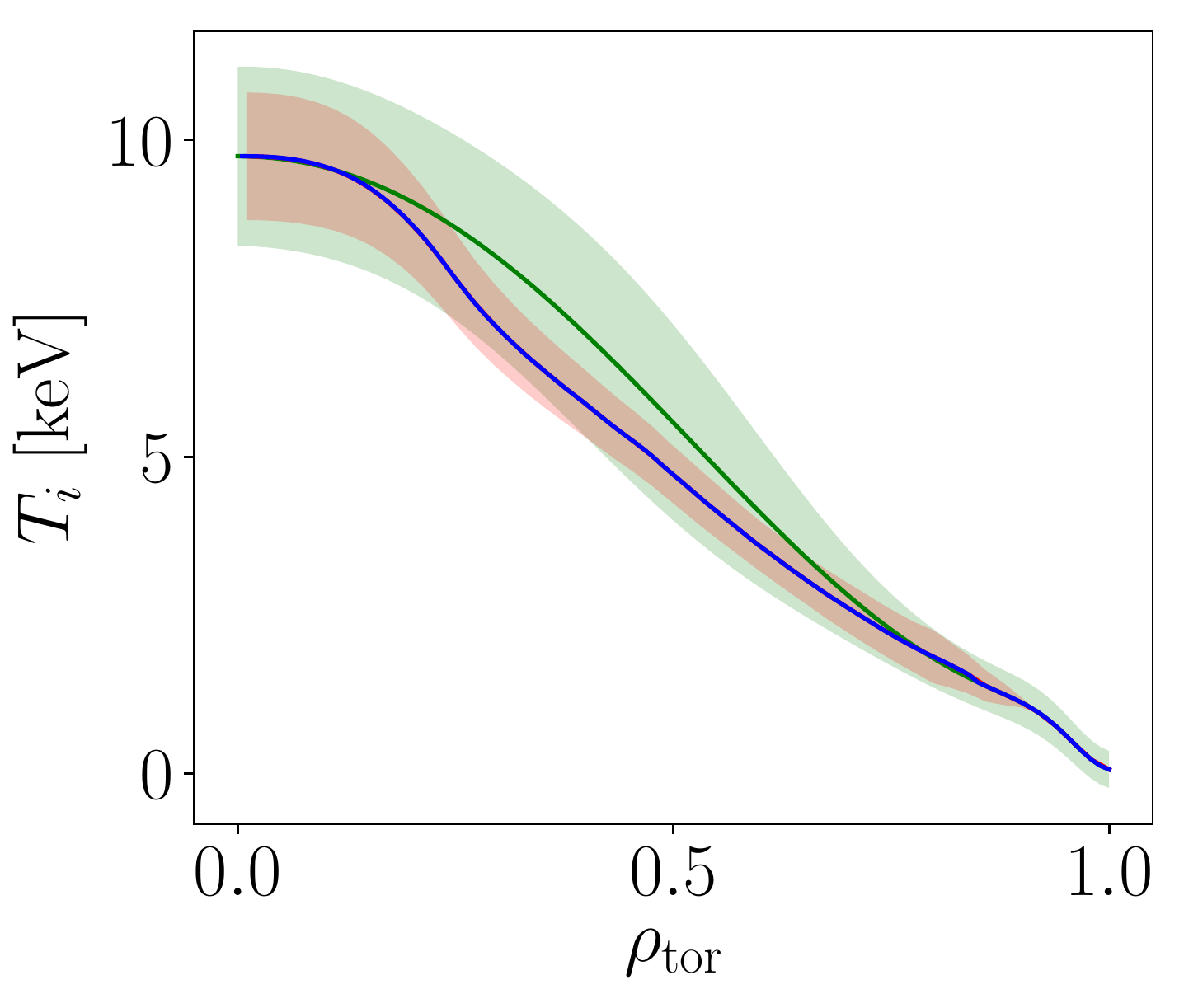}%
	\includegraphics[scale=0.27]{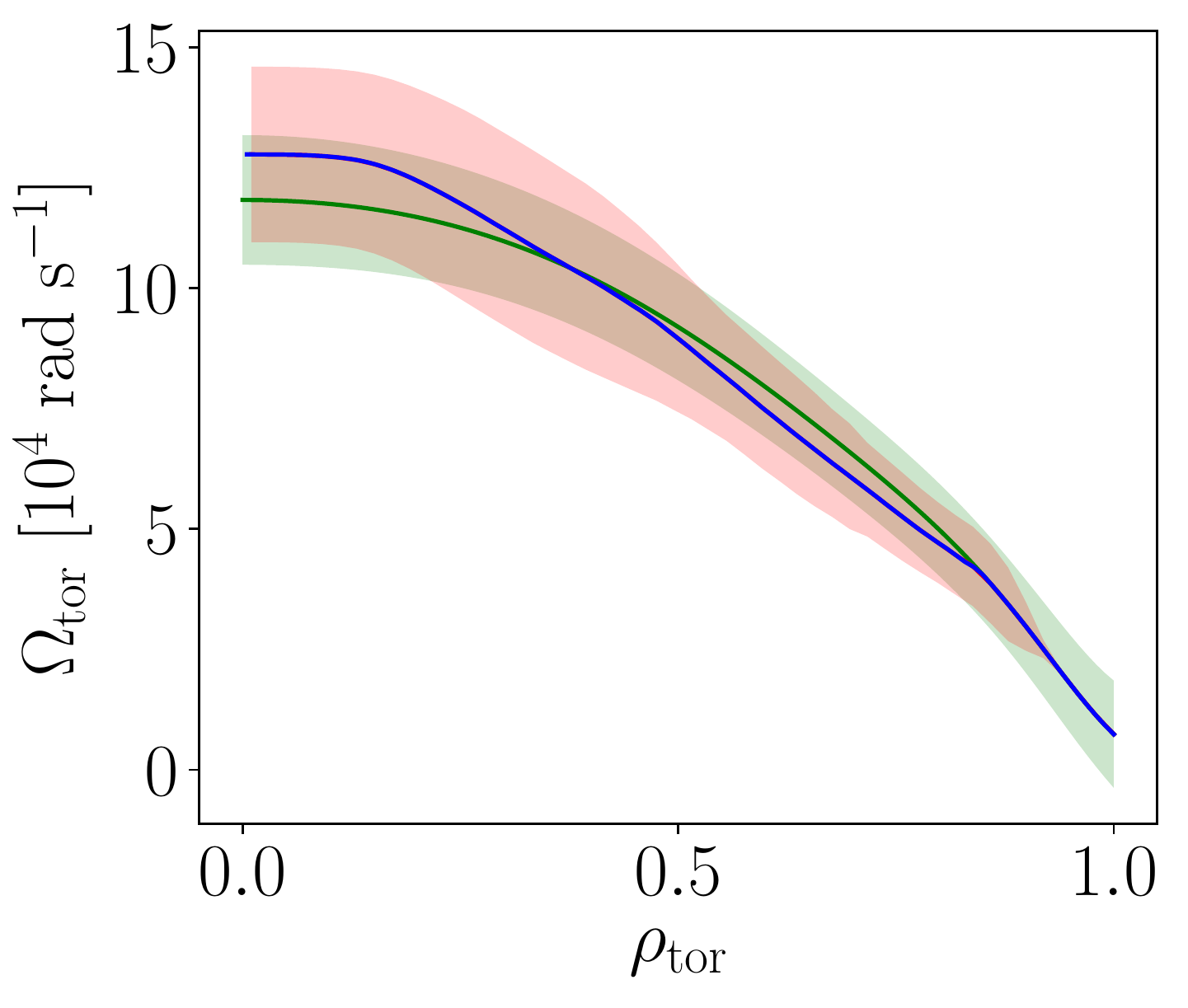}
	\caption{Comparison of GPR fit profiles (green line) and error (green shaded region) for JET \#92436 against JETTO + QuaLiKiz output (blue line), using the GPR fits as the initial / boundary conditions and the base scenario parameters. The mean output (red dashed line) and output distribution (red shaded region) was determined from a Monte Carlo sampling of the four respective initial / boundary conditions simultaneously, with 100 sample points. Upper left: Electron density profiles. Upper right: Electron temperature profiles. Lower left: Ion temperature profiles. Lower right: Toroidal angular frequency profiles.}
	\label{fig:BaseResultsJET92436}
\end{figure}

Also, since the predicted $n_e$ and $T_i$ profiles lie at the edges of the uncertainty envelopes, as shown in Figure~\ref{fig:BaseResultsJET92436}, a quantification of the agreement between prediction and experiment is desirable and will be further discussed in Section~\ref{subsec:ValidationMetricDistDist}. However, this was still chosen as the base settings as no parameter combinations were found to remedy this while simultaneously remaining strictly consistent with the experimental data.

\begin{figure}[tb]
	\centering
	\includegraphics[scale=0.255]{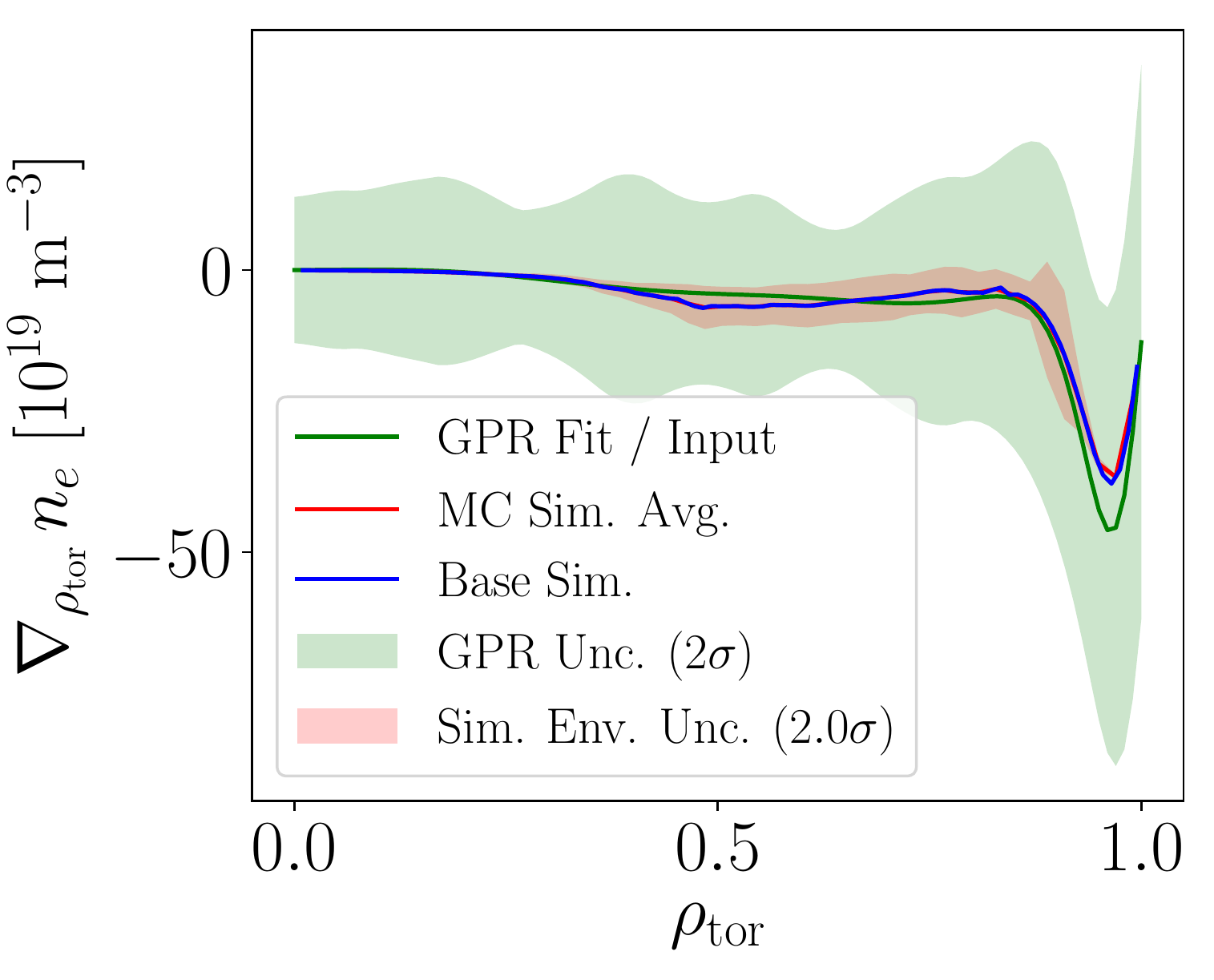}%
	\hspace{1mm}\includegraphics[scale=0.255]{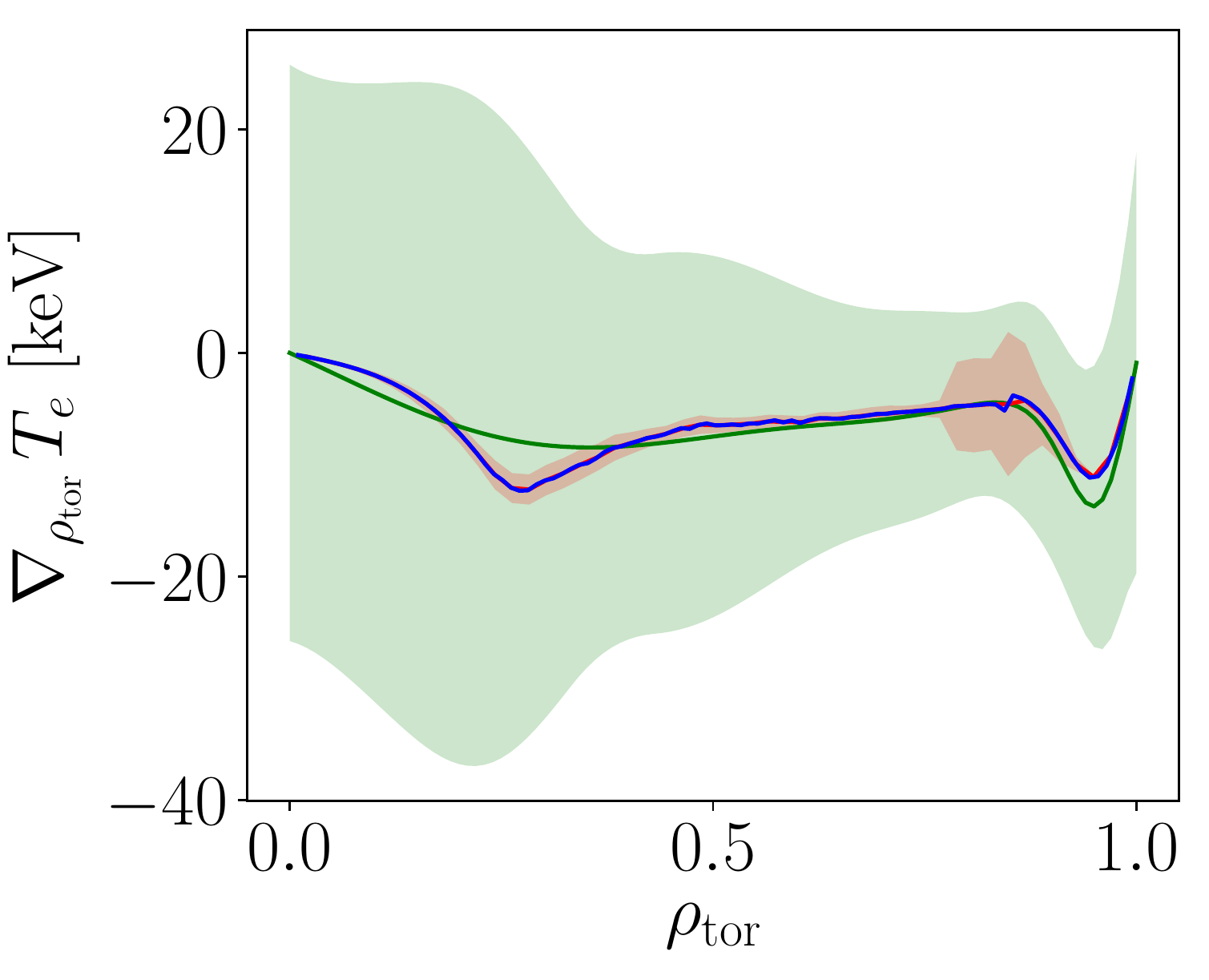}
	\includegraphics[scale=0.255]{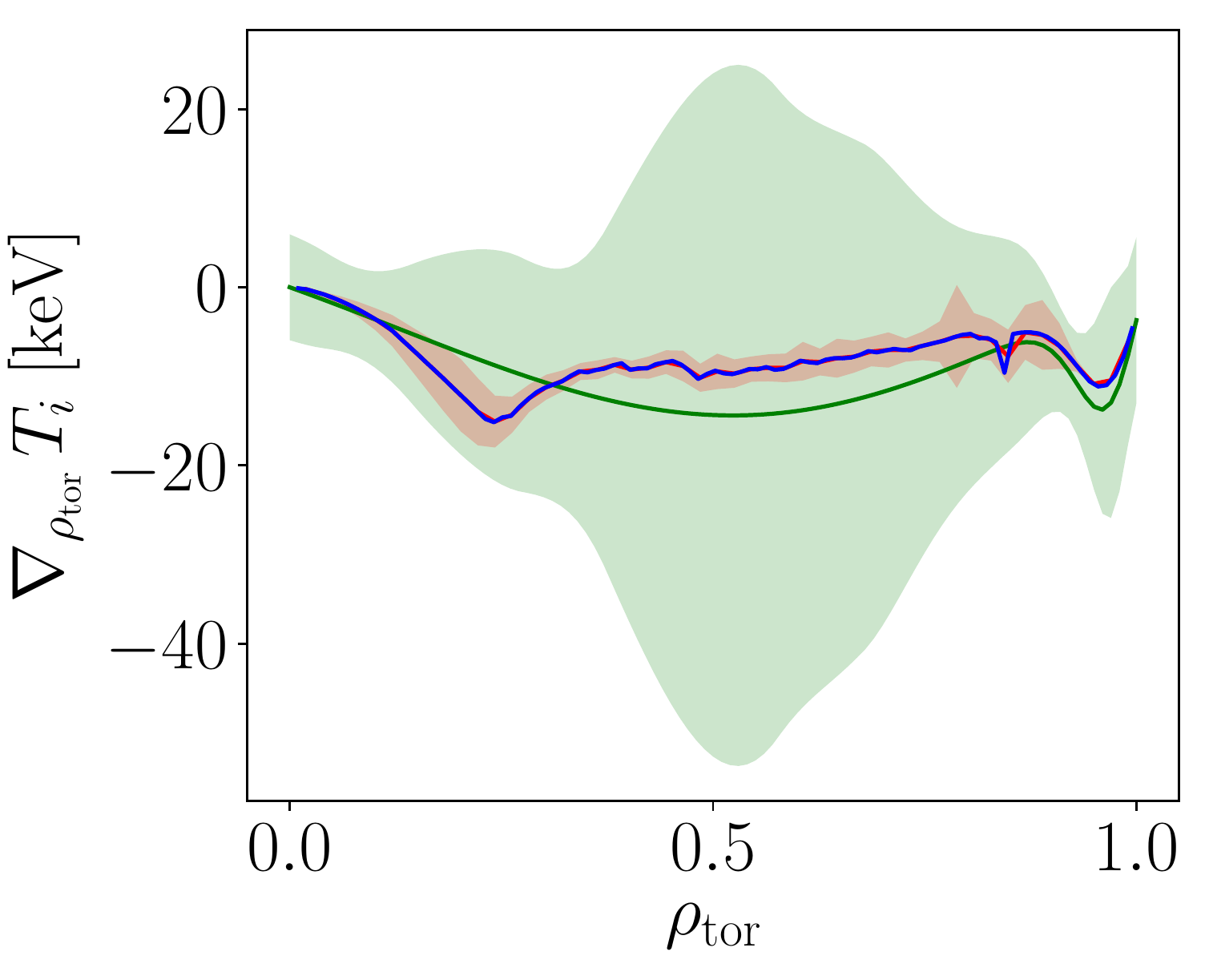}%
	\hspace{1mm}\includegraphics[scale=0.255]{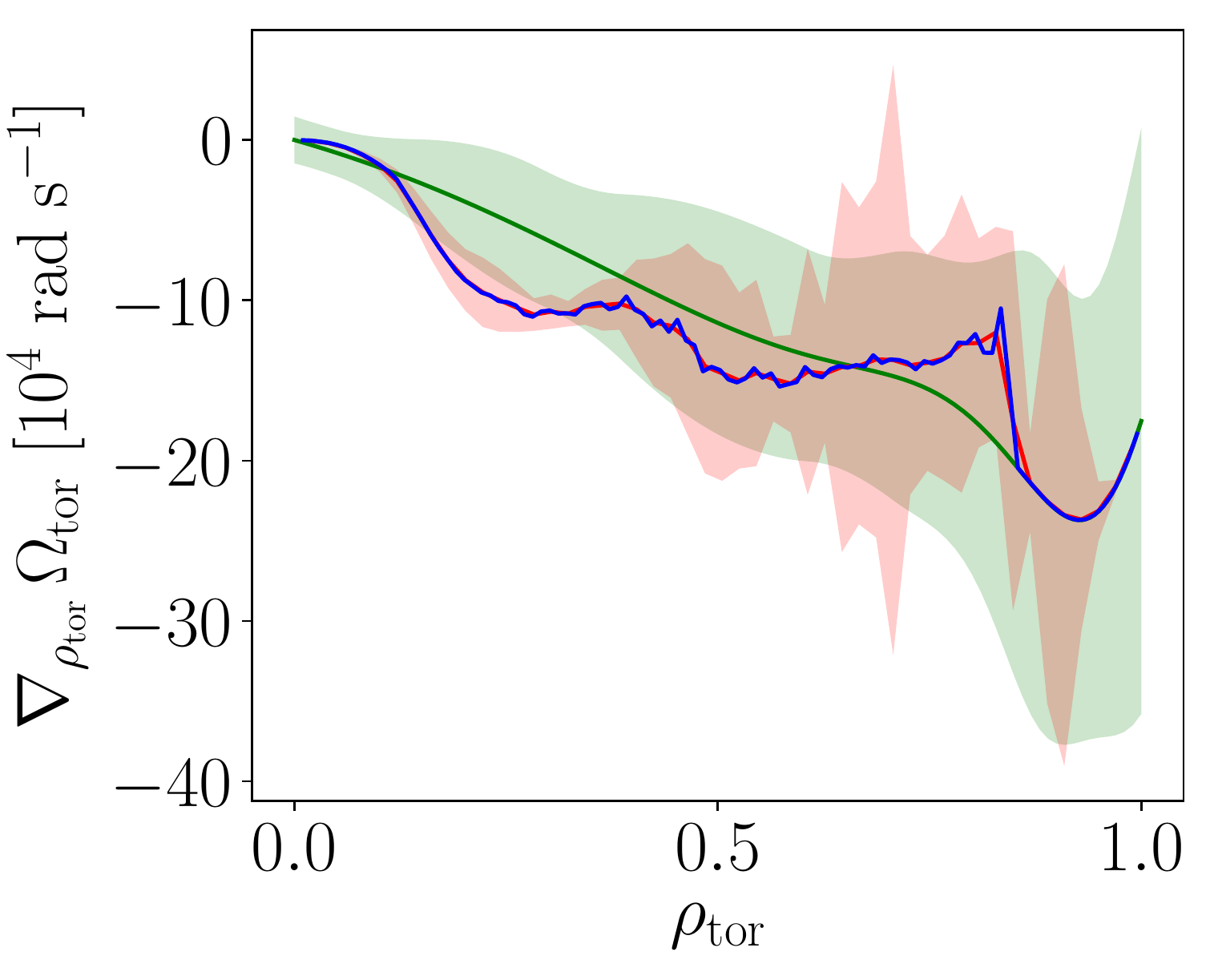}
	\caption{Comparison of GPR fit profile derivatives with respect to $\rho_{\text{tor}}$ (green line) and error (green shaded region) for JET \#92436 against JETTO + QuaLiKiz output derivatives (blue line), using the GPR fits as the initial / boundary conditions and the base scenario parameters. The mean output (red dashed line) was determined from a Monte Carlo sampling of the initial / boundary conditions, with 100 sample points. Upper left: Electron density profile derivatives. Upper right: Electron temperature profile derivatives. Lower left: Ion temperature profile derivatives. Lower right: Toroidal angular frequency profile derivatives.}
	\label{fig:BaseResultsDerivativeJET92436}
\end{figure}

Based on the uncertainty information provided by the GPR fits, the JETTO + QuaLiKiz boundary conditions for the electron density, $n_e$, electron temperature, $T_e$, ion temperature $T_i$, and angular frequency, $\Omega_{\text{tor}}$, set at $\rho_{\text{tor}} = 0.85$, were simultaneously varied within their uncertainties using a Monte Carlo approach and a normally-distributed random number generator (RNG). The red shaded regions in Figure~\ref{fig:BaseResultsJET92436} represent the results from the Monte Carlo study with 100 samples\footnote[4]{\label{foot:MC}The Monte Carlo analysis was originally performed using a slightly different boundary condition. The statistics from the original analysis was carried over to the simulations presented in this work due to limitations from its computational expense, as these statistics are not expected to change significantly as a result of the adjusted boundary condition.}, executed with the sampled profiles as both the initial and boundary condition, computing over 2~s of plasma, or $\sim10 \, \tau_E$, with simultaneous predictive updates on eight channels: the current, $j$, main ion density, $n_i$, three impurity ion densities, $n_{\text{Be}}$, $n_{\text{Ni}}$ and $n_{\text{W}}$, electron and ion temperatures, $T_e$ and $T_i$, and angular frequency, $\Omega_{\text{tor}}$. A more quantitiative statement on the level of agreement is discussed further in this section but can be considered in good agreement for the four channels for which experimental measurements exist, $n_e$, $T_e$, $T_i$ and $\Omega_{\text{tor}}$, except for an overprediction in the central $n_e$ and $T_e$ and an underprediction in the mid-radius $T_i$. Similar information about the derivatives of the profiles with respect to the radial coordinate, $\rho_{\text{tor}}$, is shown in Figure~\ref{fig:BaseResultsDerivativeJET92436}. From this plot, it becomes evident that the source of the discrepancies seen in the profiles result from differences between the fitted and simulated derivatives within $\rho_{\text{tor}}\in\left[0.4,0.7\right]$ for $n_e$, and within $\rho_{\text{tor}}\in\left[0.25,0.8\right]$ for $T_i$.

\begin{figure}[tb]
	\centering
	\includegraphics[scale=0.29]{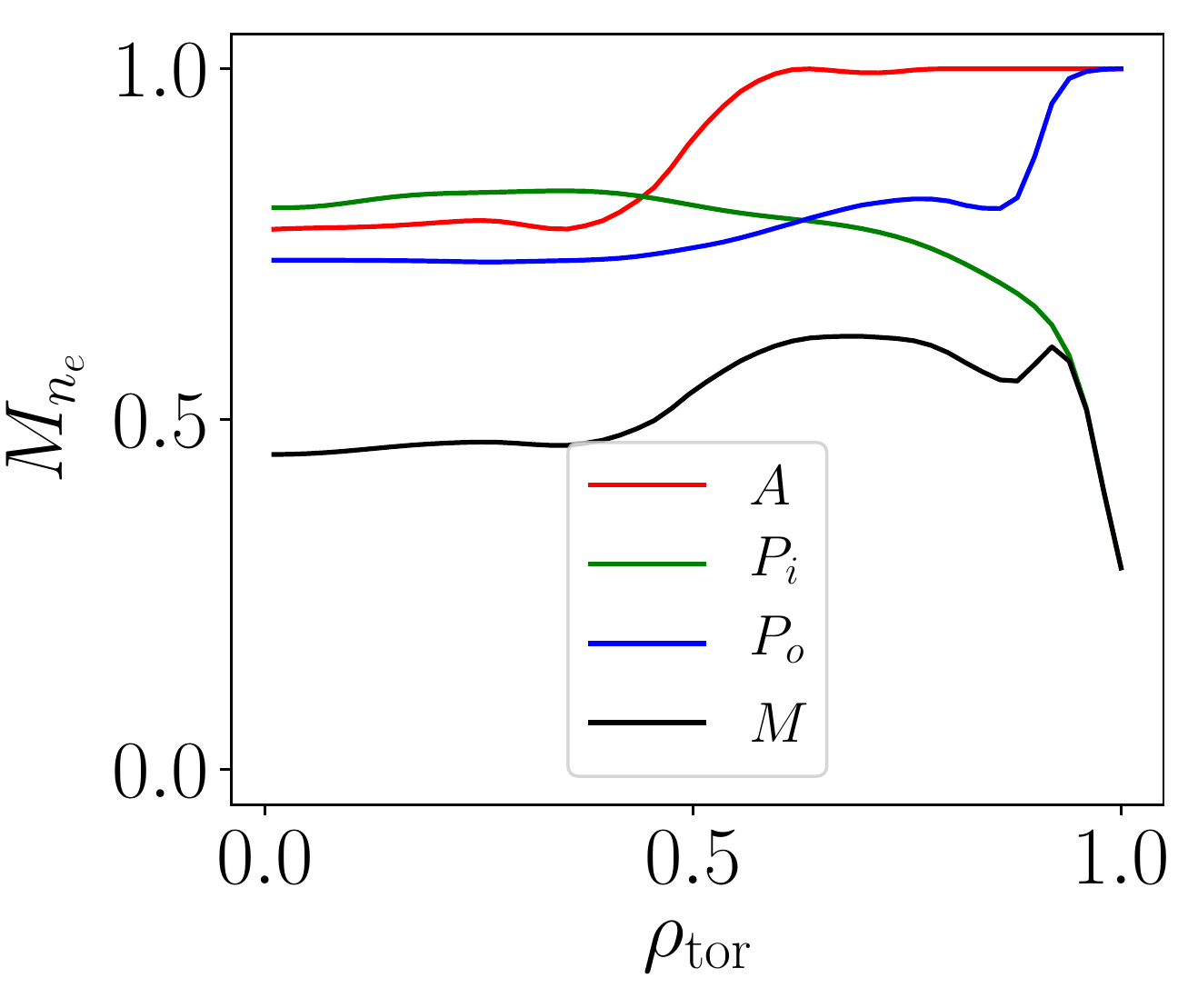}%
	\hspace{2mm}\includegraphics[scale=0.29]{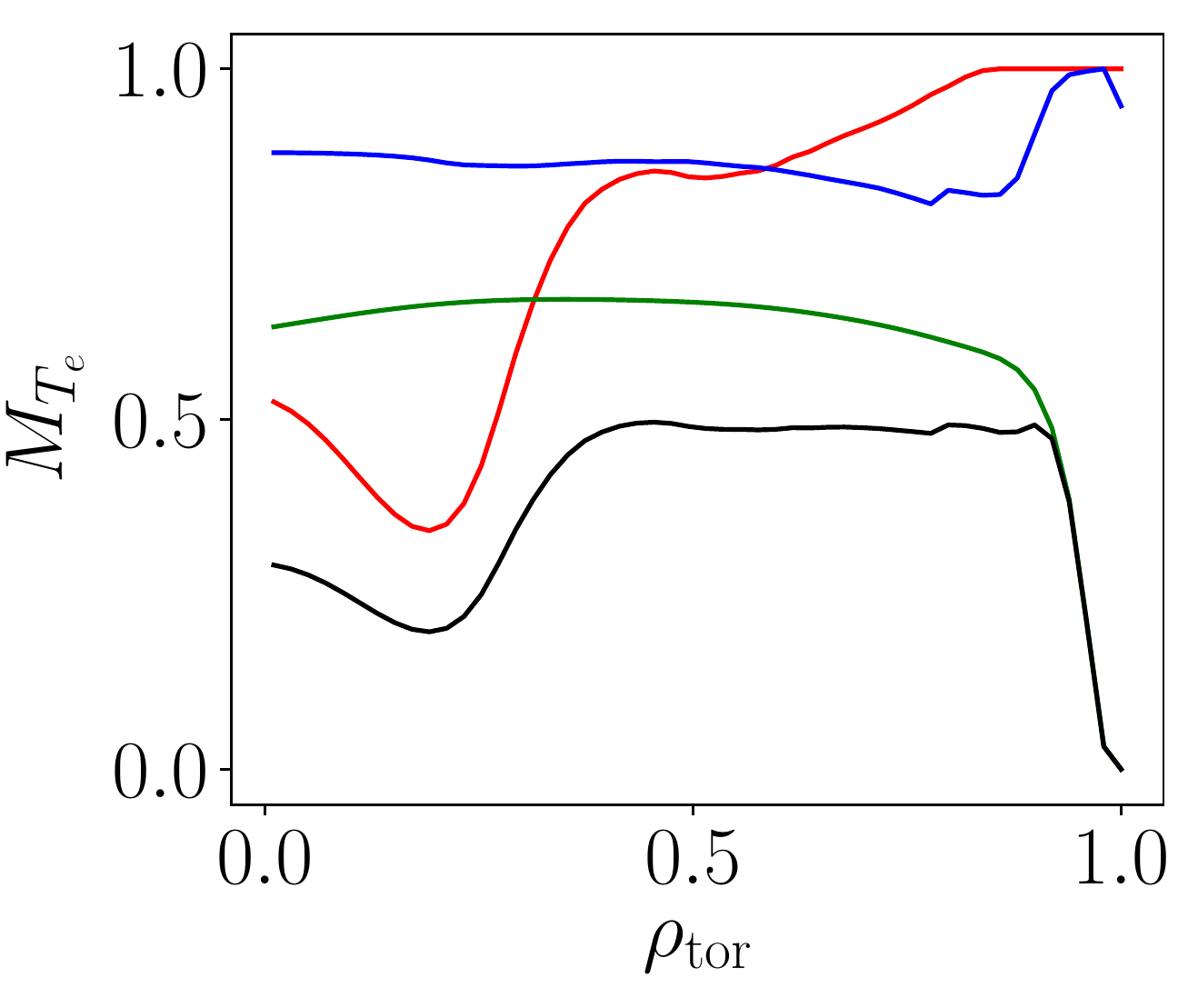}
	\includegraphics[scale=0.29]{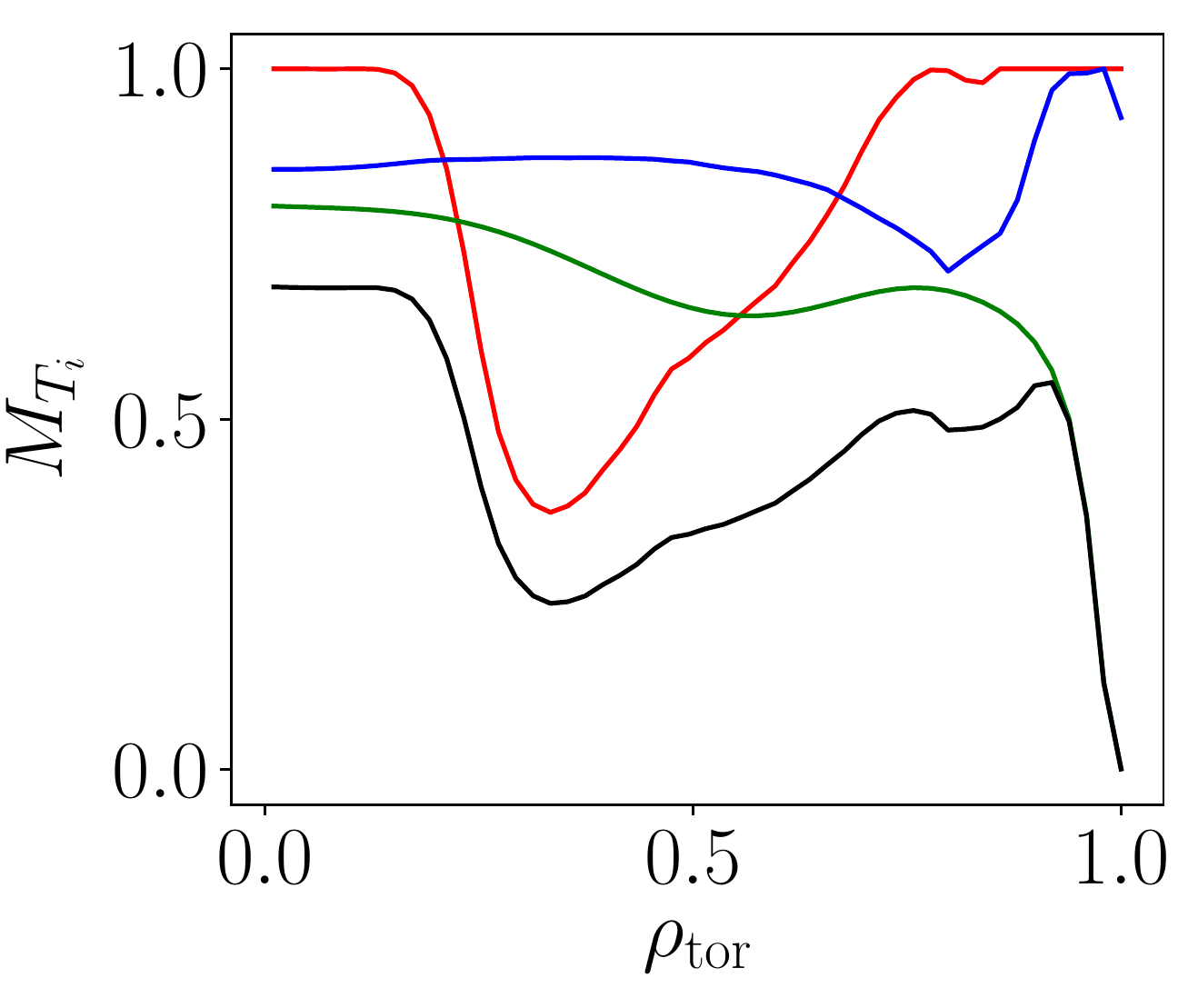}%
	\hspace{2mm}\includegraphics[scale=0.29]{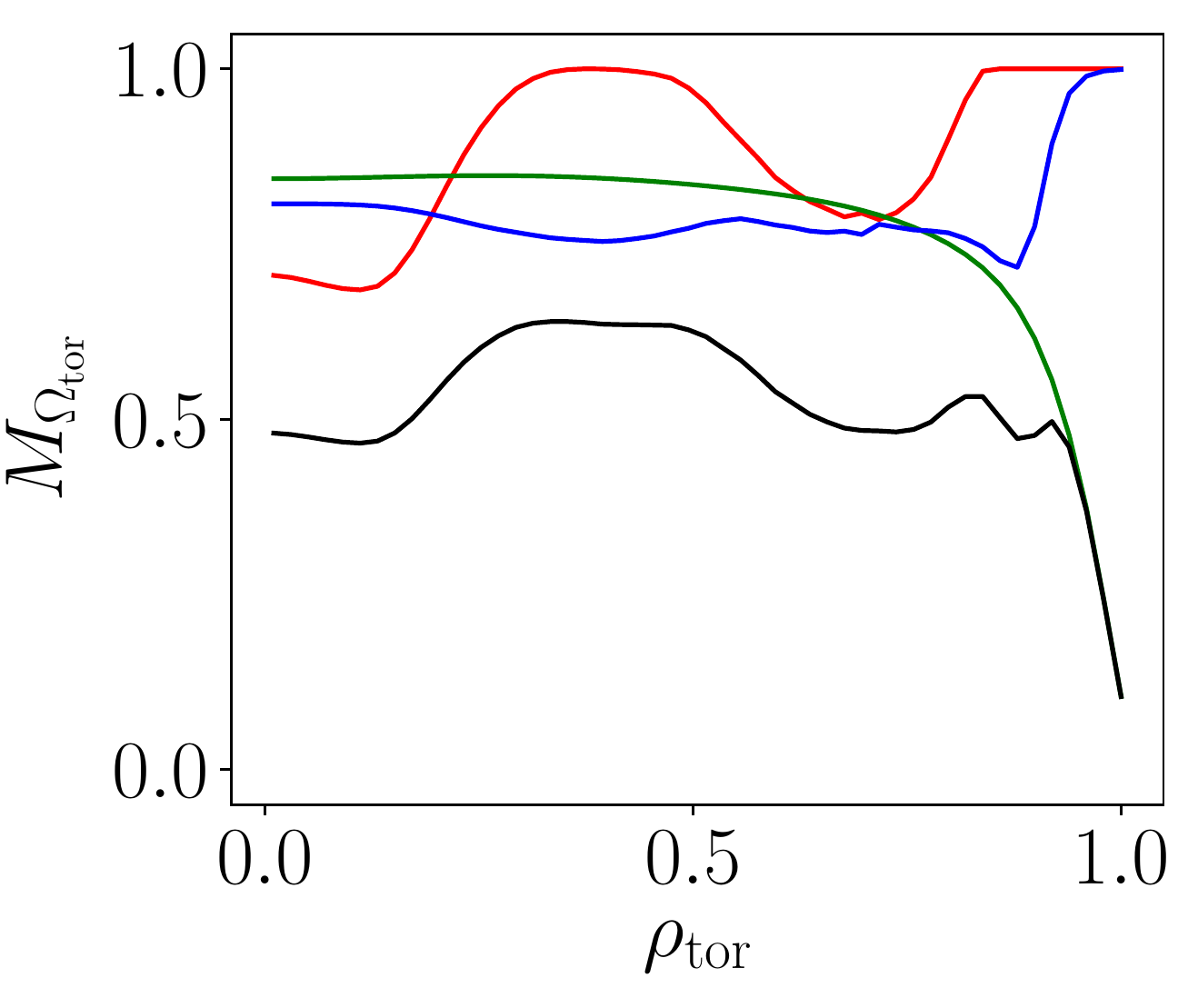}
	\caption{Distribution-distribution validation metric, $M$, for the kinetic profiles (black line) and associated components, $A$, $P_i$, $P_o$ (red, blue and green lines, respectively) calculated from the Monte Carlo sampling of the JETTO + QuaLiKiz initial / boundary conditions, based on the GPR fit uncertainties and the base scenario parameters. Only $\rho_{\text{tor}} < 0.8$ is displayed since the profiles are not evolved for $\rho_{\text{tor}} \ge 0.8$, as a result of the chosen boundary condition for the JETTO + QuaLiKiz simulations. Upper left: FOM for electron density profile. Upper right: FOM for electron temperature profile. Lower left: FOM for ion temperature profile. Lower right: FOM for toroidal angular frequency profile.}
	\label{fig:MetricResultsJET92436}
\end{figure}

Figure~\ref{fig:MetricResultsJET92436} shows the proposed FOM, given by Equation~\eqref{eq:ProposedGaussianMetric}, and its associated components, $A$, $P_i$, and $P_o$ as a function of $\rho_{\text{tor}}$, determined from the na{\"i}ve Gaussian envelope approximation calculated from the Monte Carlo boundary condition study performed on JET \#92436. The level of agreement graphically shown in Figure~\ref{fig:BaseResultsJET92436} is successfully captured by the proposed FOM for the comparisons of all profile quantities, with regions having $M \gtrsim 0.5$ indicating sufficiently good agreement between the input and output profile distributions. As expected from previous analysis, the central core $n_e$, $T_e$ and the mid-radius $T_i$ all exhibit $M < 0.5$, though it is not low enough to exclude these profiles to be untrustworthy.

\begin{figure}[tb]
	\centering
	\includegraphics[scale=0.29]{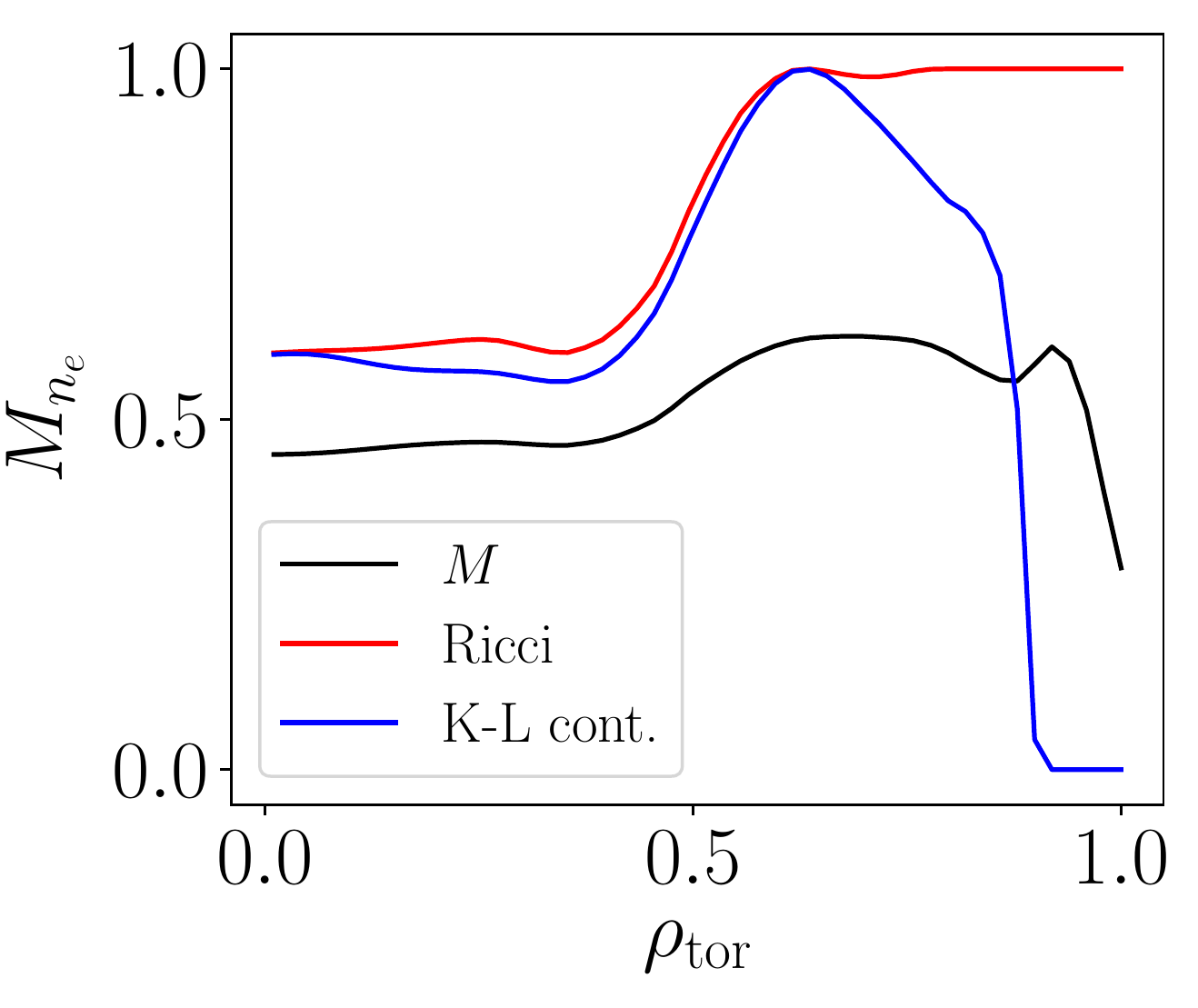}%
	\hspace{2mm}\includegraphics[scale=0.29]{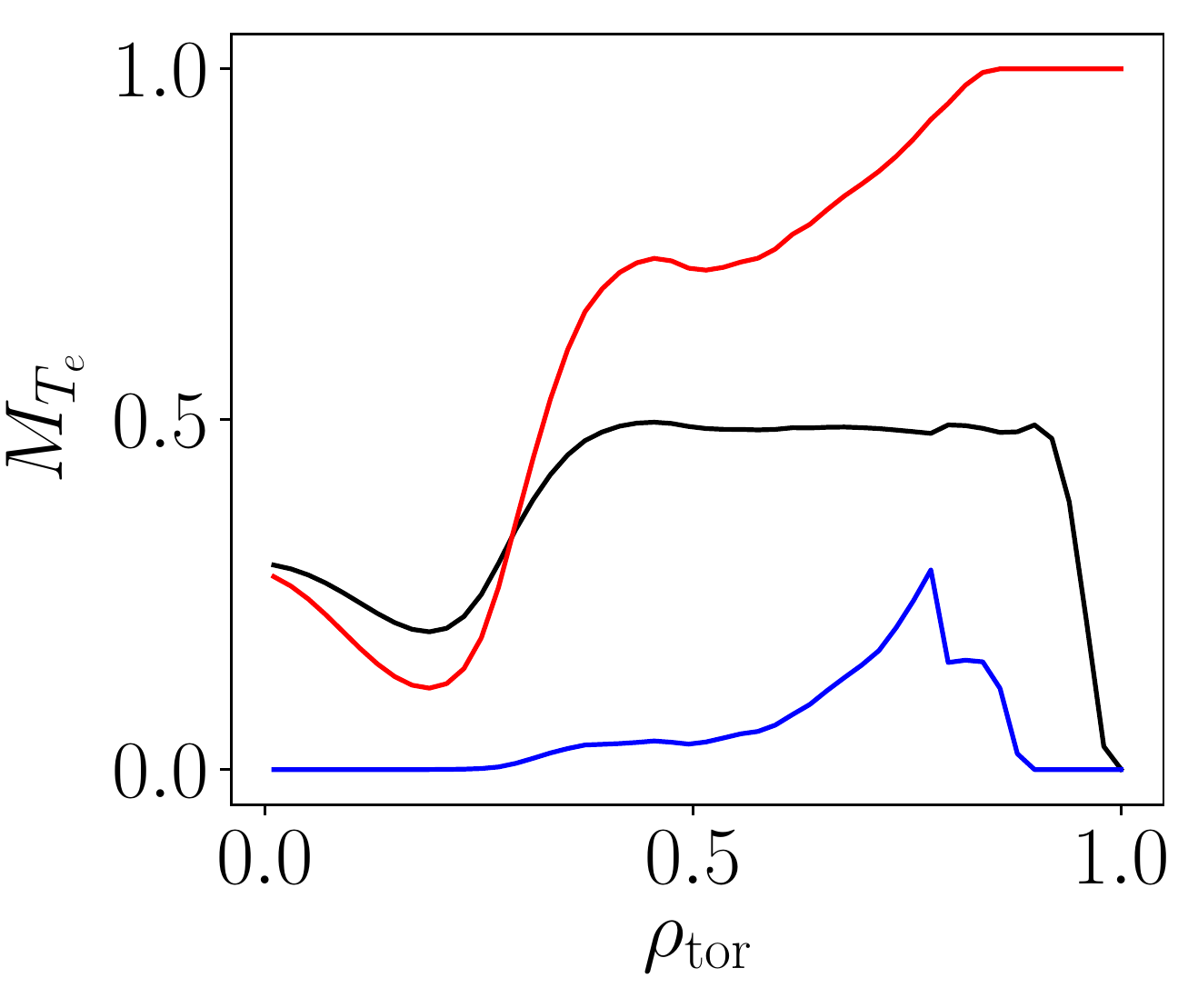}
	\includegraphics[scale=0.29]{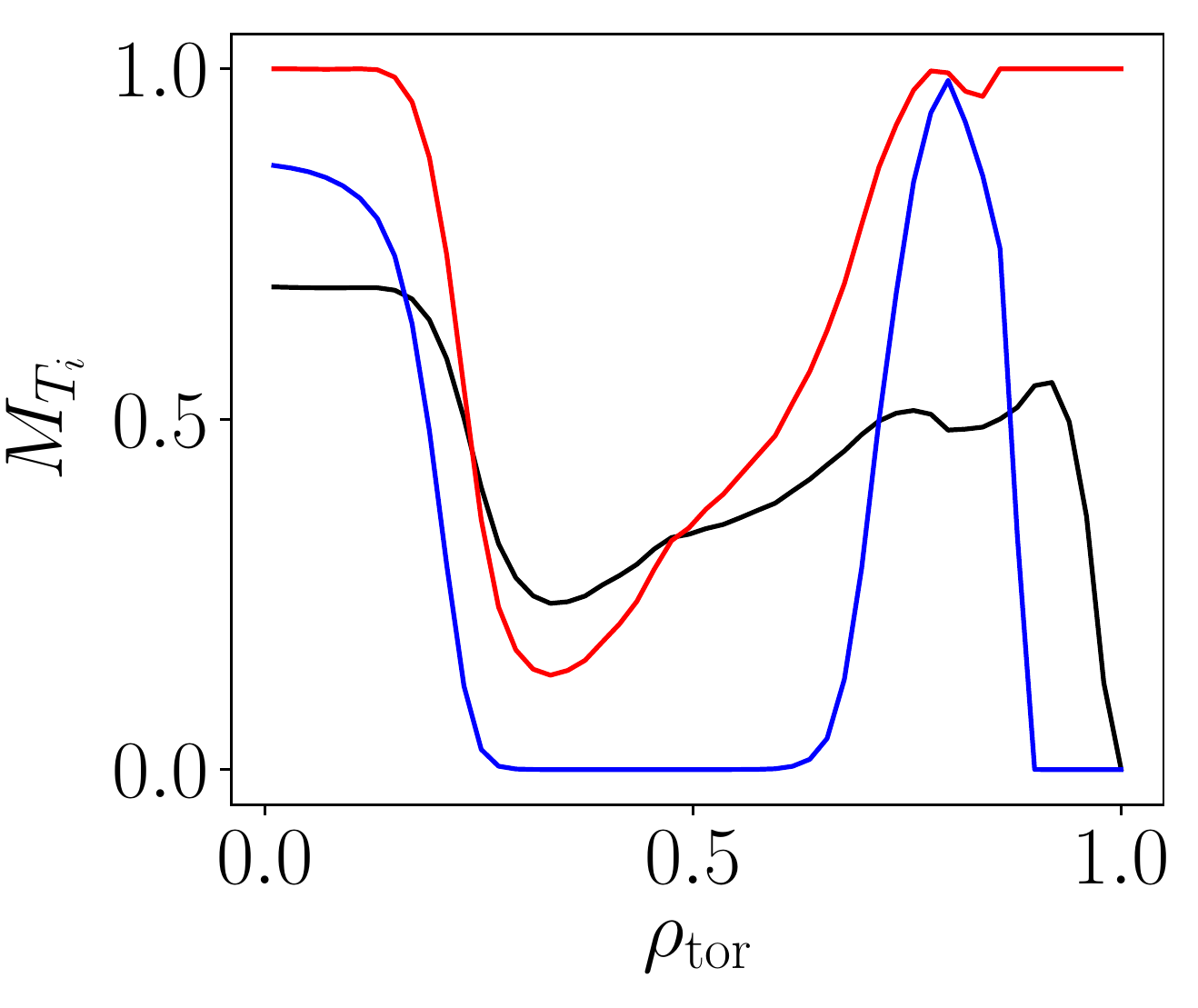}%
	\hspace{2mm}\includegraphics[scale=0.29]{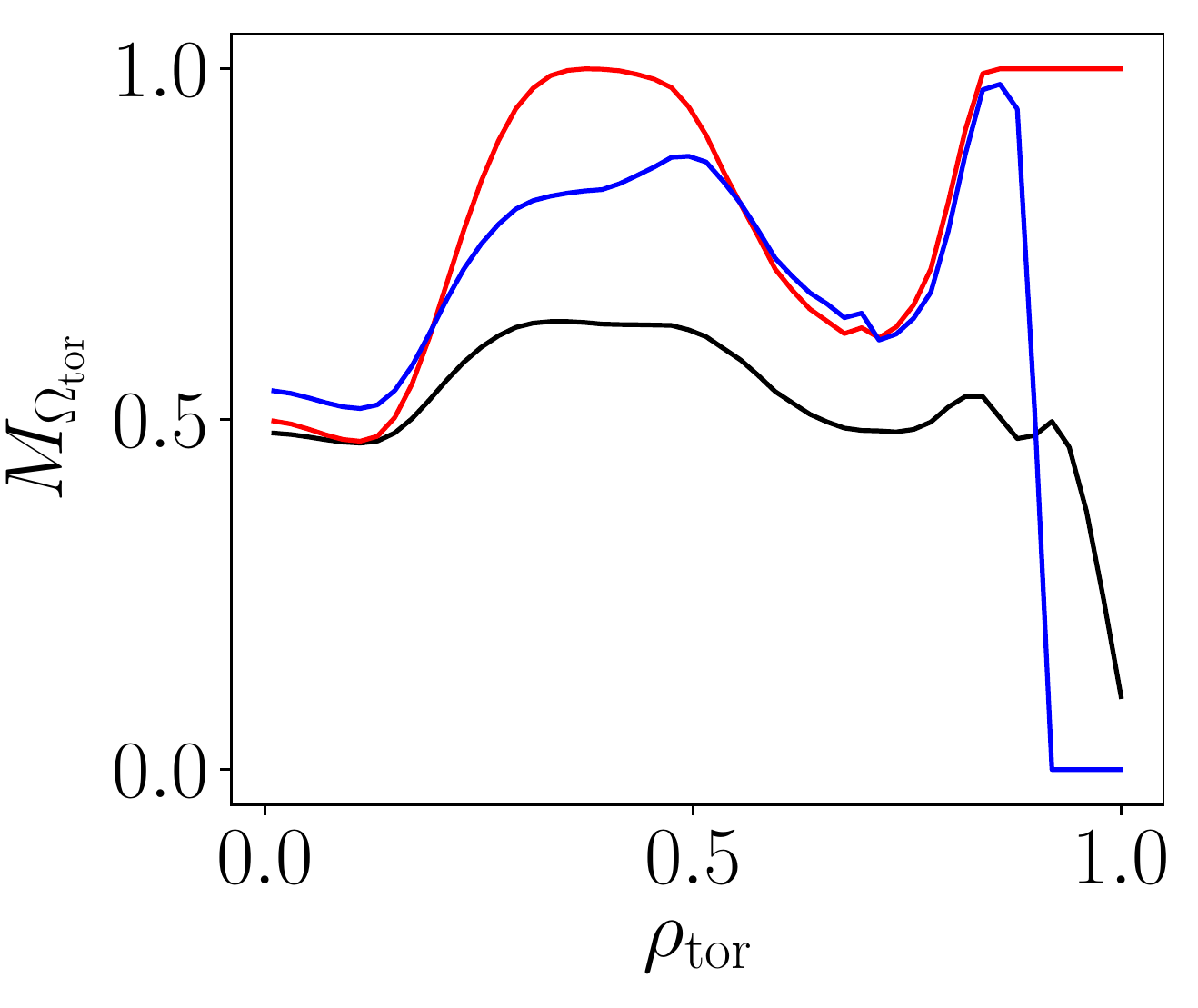}
	\caption{Comparison of the validation figure-of-merit profiles (black line), calculated from the Monte Carlo sampling of the JETTO + QuaLiKiz initial / boundary conditions based on the GPR fit uncertainties and the base scenario parameters, against other metrics, (red, blue, green, and yellow lines, respectively). Only $\rho_{\text{tor}} < 0.8$ is displayed since the profiles are not evolved for $\rho_{\text{tor}} \ge 0.8$, as a result of the chosen boundary condition for the JETTO + QuaLiKiz simulations. Upper left: Metrics for electron density profile. Upper right: Metrics for electron temperature profile. Lower left: Metrics for ion temperature profile. Lower right: Metrics for toroidal angular frequency profile.}
	\label{fig:MetricComparisonJET92436}
\end{figure}

By combining all the Monte Carlo study results into a single data set and calculating the variance, $\sigma_o^2$, of that data set using the base case result as the mean, $\mu_o$, a probability distribution can be constructed to act as the simulation output uncertainty. By using only these two statistical moments of the data set, it is na{\"i}vely assumed that its distribution can be described as Gaussian, as given by Equation~\eqref{eq:NormalDistribution}. Figure~\ref{fig:MetricComparisonJET92436} shows a comparison between the proposed FOM, given in Equation~\eqref{eq:ProposedGaussianMetric}, and the other tests which evaluate the agreement between two distributions described in Section~\ref{subsec:ValidationMetricDistDist}. The other tests provide lower scores for the simulated $n_e$ and $T_i$ profiles and provide higher scores for the simulated $T_e$ and $\Omega_{\text{tor}}$ profiles compared to the proposed FOM. This is expected behaviour as the standard tests only provide a measure of how similar the input and output distributions are to each other, in both location and shape, resulting in values of unity at the boundary condition (not shown) where the distributions are identical. The proposed FOM attempts to provide a measure of the likelihood that a profile drawn from the output distribution belongs to the input distribution and vice versa, regardless of the match in the distribution shapes, by the addition of the terms, $P_i$ and $P_o$. Since these tests do not answer the same question, the comparison shown is cannot be taken as quantitatively meaningful. However, the fact that the trends in the figure-of-merit profiles are similar provide confidence that the proposed FOM provides a valid measure of the agreement between the two distributions.

In preparation for future extrapolation exercises, Table~\ref{tbl:NeutronResultsJET92436} shows a comparison of results of the neutron rate sensitivity studies, performed with TRANSP, against the experimental total neutron rate, measured using calibrated time-resolved neutron counters with a quoted calibration error of $\sim10$\%~\cite{aNeutronCalibration-Batistoni}. All uncertainties are given with $\pm\,2 \sigma$ or at the $\sim\!95$\% confidence interval of the mean. These results provide further evidence that the $T_i$ profile from JETTO + QuaLiKiz is underpredicted, as the neutron rates using the simulated profiles consistently fall under the measured neutron rate, though still within the $\pm\,2\sigma$ uncertainty ranges for pure Ni. These results are consistent with previous works on this matter~\cite{aNeutronDeficit-Weisen} and accentuate the importance of reliable impurity composition and profile estimations for any extrapolation exercises, due to the impact of fuel dilution on the total fusion rate.

\begin{table*}[tb]
	\centering
	\caption{Results of neutron rate studies performed based on simulation data for JET \#92436, with a line-integrated $Z_{\text{eff}}$ measurement of $1.77$. Measurements taken from calibrated time-resolved neutron counters with a quoted calibration error of $\sim10$\%~\cite{aNeutronCalibration-Batistoni}}\vspace{2mm}
	\begin{tabular}{l|c|c|c}
	\toprule
	Data Source & Neutron Rate [n/s] & with $Z_{\text{eff}}$ + 20\% & Uncertainty ($\pm\,2\sigma$) \\
	\midrule
	Measured & $2.7 \times 10^{16}$ & -- & $0.2 \times 10^{16}$ \\
	GPR, only Ni & $3.4 \times 10^{16}$ & $3.3 \times 10^{16}$ & $0.7 \times 10^{16}$ \\
    GPR, Be/Ni mix & $2.8 \times 10^{16}$ & $2.5 \times 10^{16}$ & $0.5 \times 10^{16}$ \\
	JETTO + QLK, only Ni & $2.6 \times 10^{16}$ & $2.5 \times 10^{16}$ & $0.2 \times 10^{16}$ \\
	JETTO + QLK, Be/Ni mix & $2.1 \times 10^{16}$ & $1.9 \times 10^{16}$ & $0.2 \times 10^{16}$ \\
	\bottomrule
	\end{tabular}
	\label{tbl:NeutronResultsJET92436}
\end{table*}

The $T_i$ underprediction is suspected to come from the underestimation of the EM-stabilization factor provided by the ad-hoc implementation. The presence of IC auxiliary heating within the analyzed time window generates a large fast ion pressure gradient in the energy deposition region, with a maximum absolute value of $\partial p_{\text{fast,IC}} / \partial \rho_{\text{tor}} = \left(2.6 \pm 0.3\right) \times 10^5$. Due to the known dependency of this effect on large fast ion pressure gradients~\cite{aNonLinearFI-Citrin,aNonMaxwellian-DiSiena}, generated in this discharge by IC auxiliary heating, the ad-hoc EM-stabilization factor, based on $W_{\text{th}} / W_{\text{tot}}$, may underestimate the magnitude of the stabilization effect. However, the quantification of this shortcoming is outside the scope of this study and deeper investigations are left as future work.

\subsection{Impact analysis of rotation profiles}
\label{subsec:RotationSensitivities}

\begin{figure}[tb]
	\centering
	\includegraphics[scale=0.27]{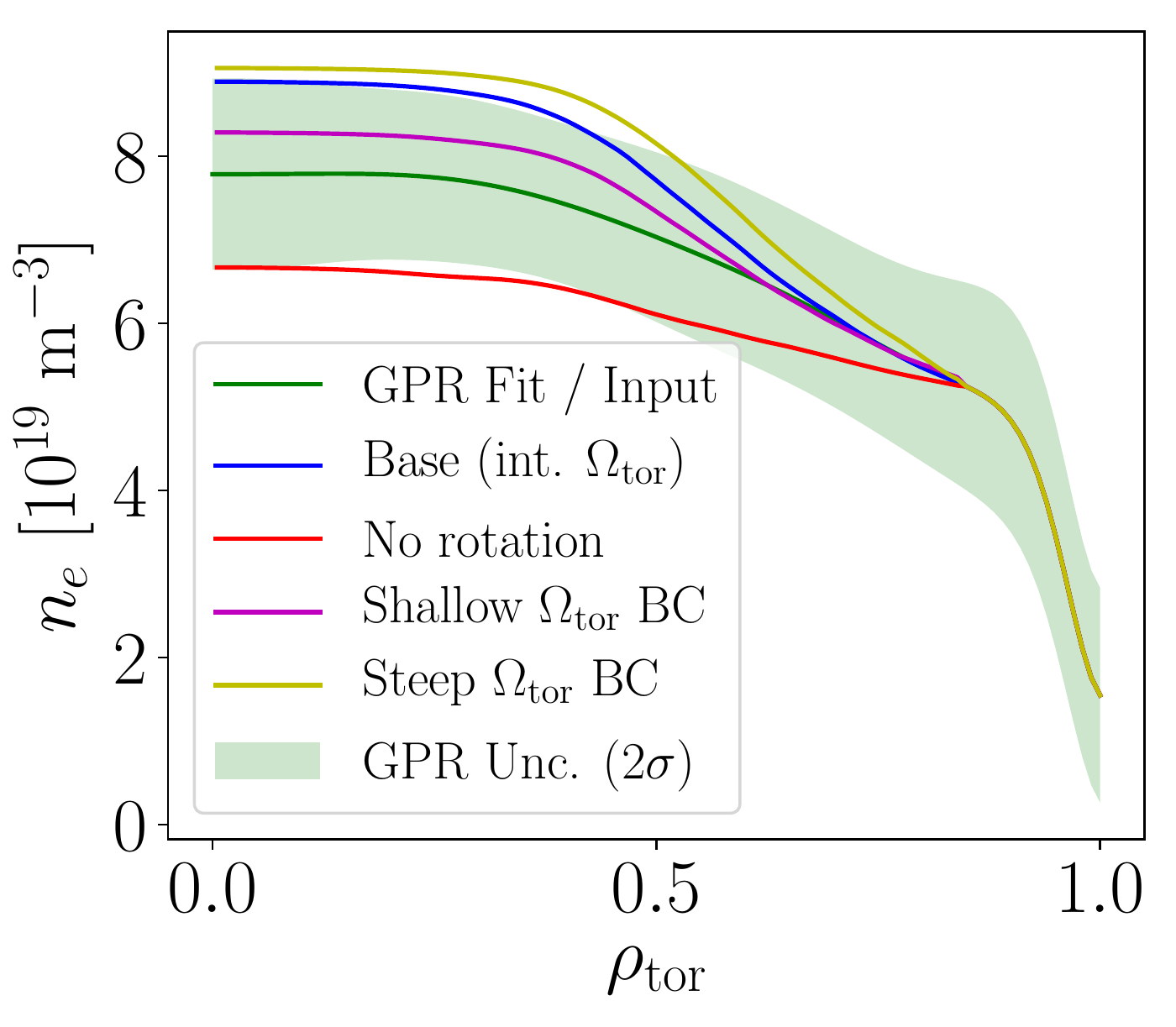}%
	\hspace{1mm}\includegraphics[scale=0.27]{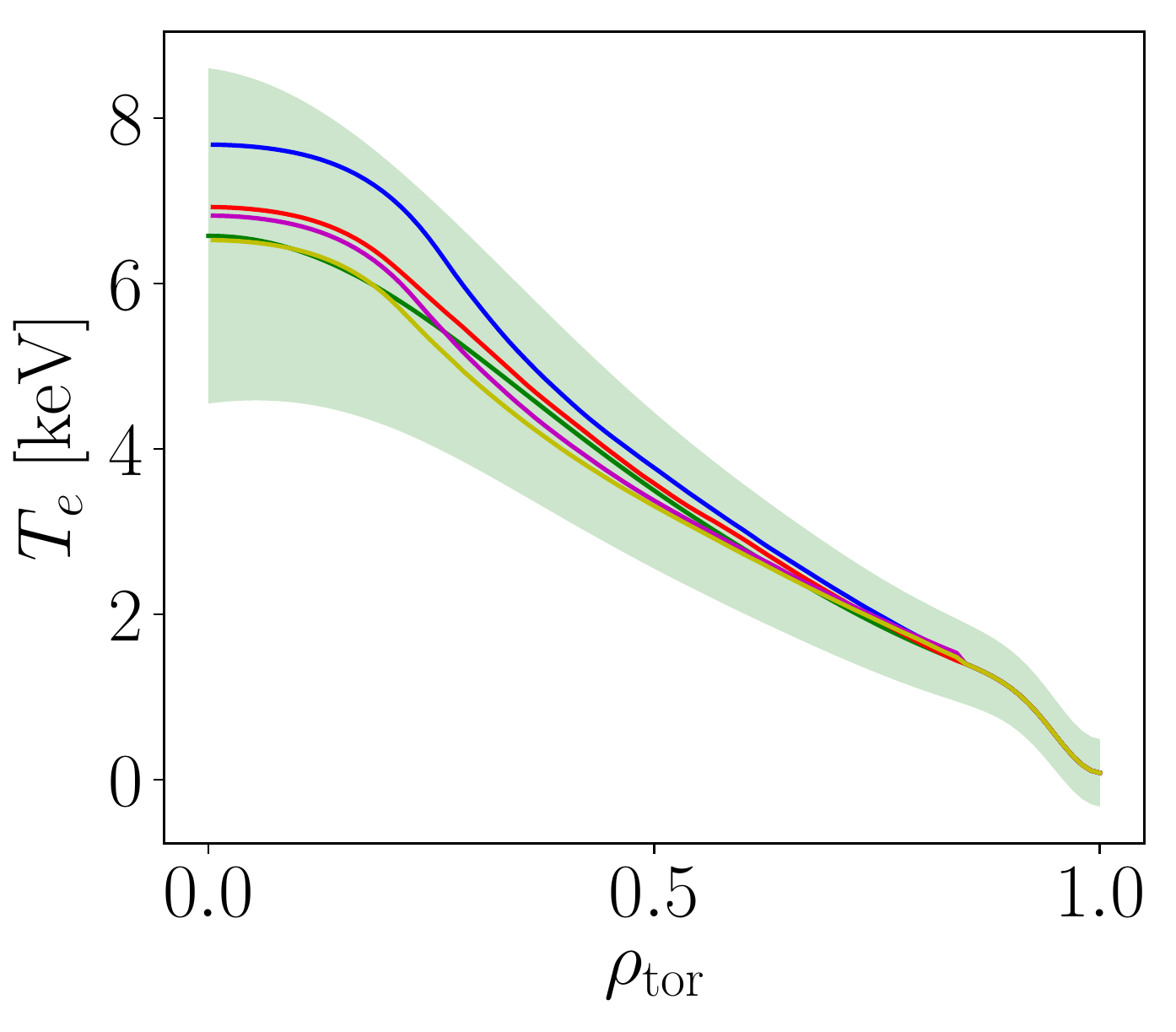}
	\hspace{-0.5mm}\includegraphics[scale=0.27]{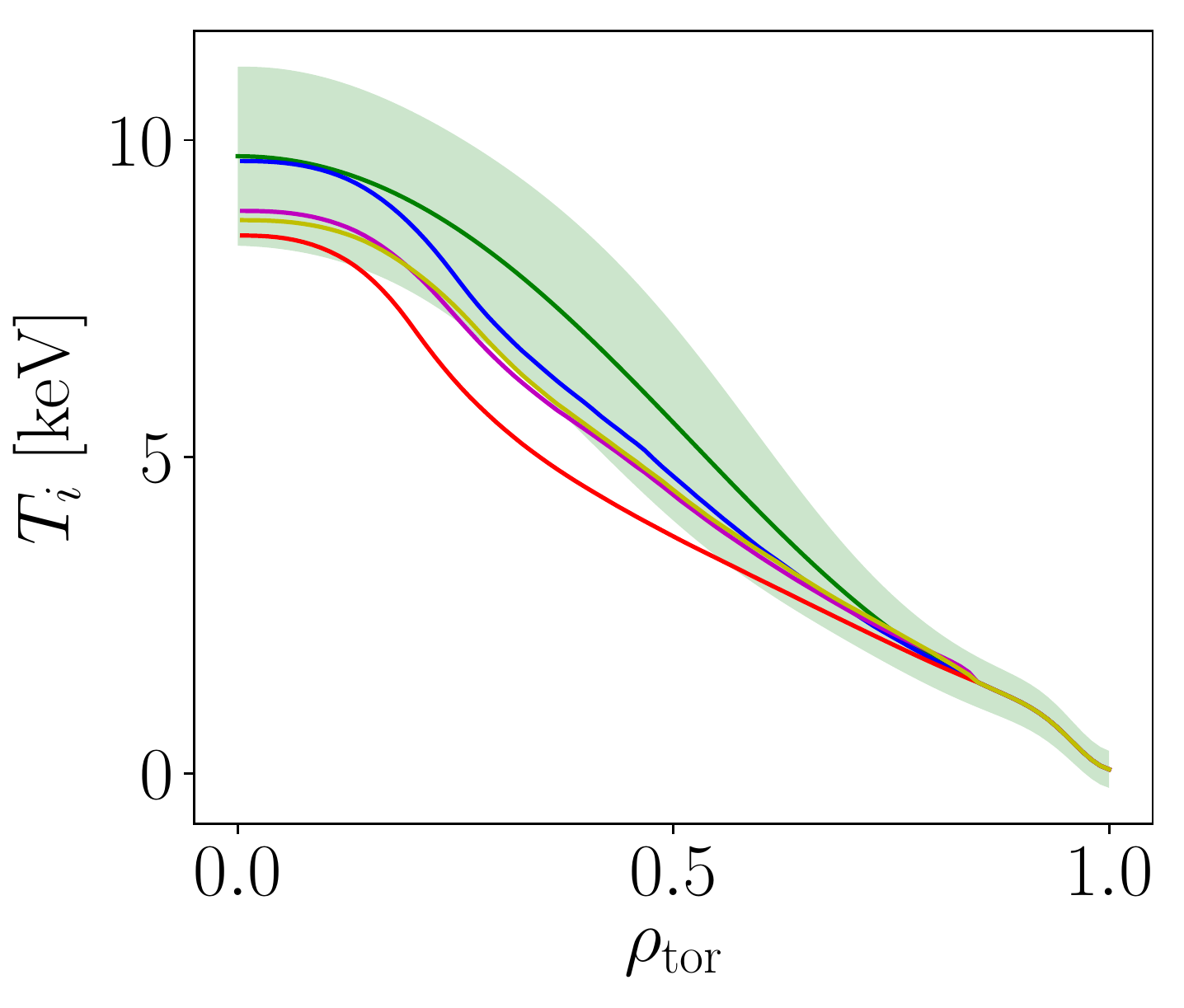}%
	\includegraphics[scale=0.27]{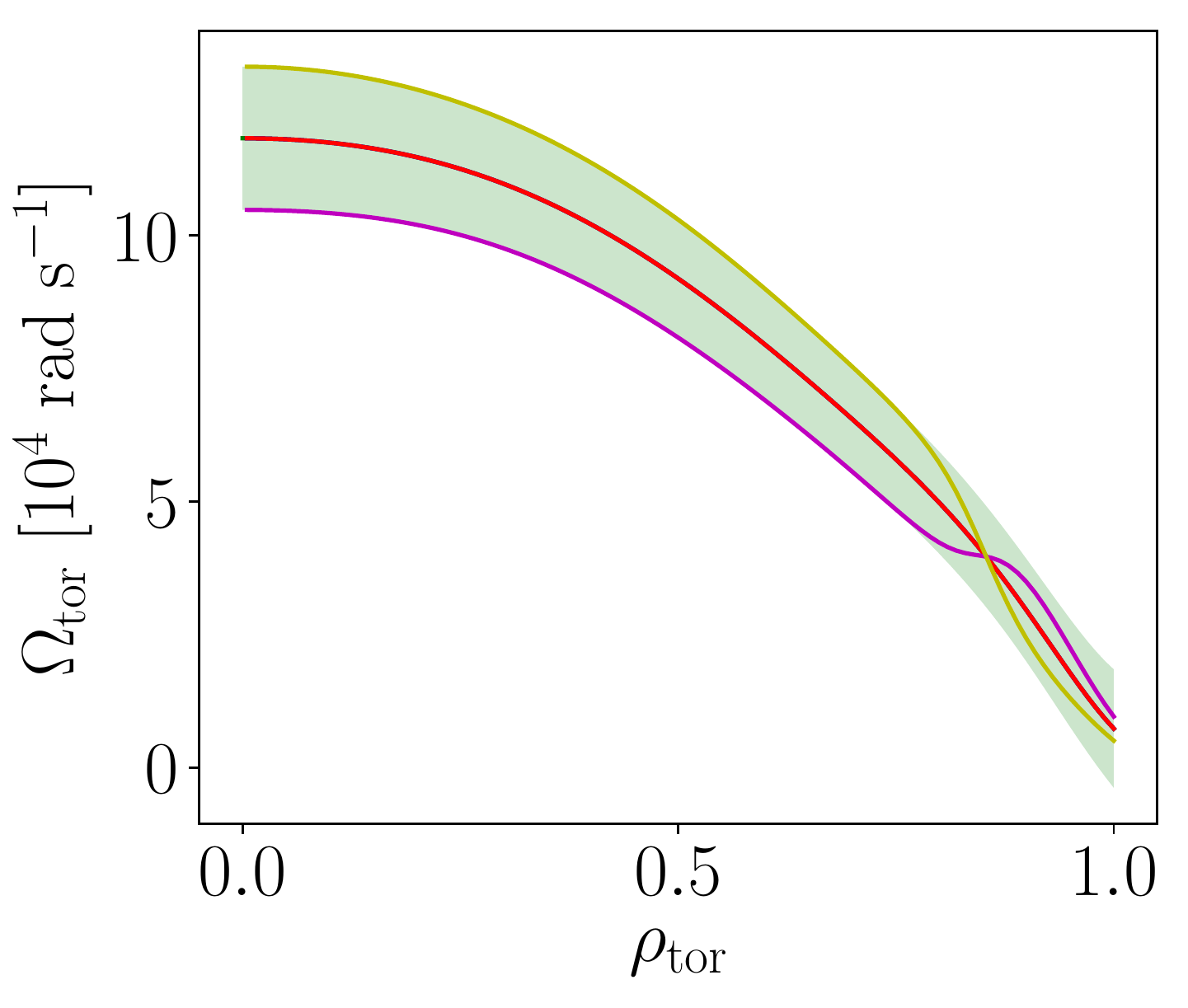}
	\caption{Results of the sensitivity studies regarding the toroidal rotation profile modifications, where the input profiles (green lines) are compared against the output profiles (see legend) and the base case scenario (blue line). These simulations were performed using interpretive momentum transport. Upper left: Electron density profiles. Upper right: Electron temperature profiles. Lower left: Ion temperature profiles. Lower right: Toroidal angular frequency profiles.}
	\label{fig:RotationSensitivityResultsJET92436}
\end{figure}

In addition to the Monte Carlo analysis of the model sensitivity to boundary conditions, the GPR fit uncertainties also allow for greater statistical rigour in the modification of prescribed input profiles for the integrated model. A typical example of such a prescribed input profile is the toroidal angular frequency, $\Omega_{\text{tor}}$, within simulations using interpretative momentum transport, meaning that the $\Omega_{\text{tor}}$ profile remains fixed to its initial condition. To demonstrate this capability and the effects of this modification within this discharge, the following sensitivity studies were performed:
\begin{itemize}
	\itemsep 0pt
	\item switching off the rotational contributions to the fluxes calculated by QuaLiKiz, applied only to $\rho_{\text{tor}} \ge 0.5$ in the base settings~\cite{aQLK-Citrin};
	\item adjusting the angular rotation, $\Omega_{\text{tor}}$, profile within $\pm\,2\sigma$ to have a steeper gradient at the simulation boundary;
	\item adjusting the angular rotation, $\Omega_{\text{tor}}$, profile within $\pm\,2\sigma$ to have a shallower gradient at the simulation boundary.
\end{itemize}
Figure~\ref{fig:RotationSensitivityResultsJET92436} shows the results of these sensitivity studies.

The impact of the rotation shear in QuaLiKiz appears primarily on the density profile, which is attributed to a strong $E \times B$ shear stabilisation effect on ITG instabilities within QuaLiKiz. Despite the fact that ITG instabilities also drive ion heat transport, this effect is not as prevalent in the $T_i$ profile. It is likely that the increasing density gradient prevents further increase of $T_i$ with increasing $E \times B$ shear, as this simulation performs both predictive heat and particle transport. This increasing density gradient increases the turbulence drive and compensates the stabilising effect of the $E \times B$ shear on the ion heat flux. The sensitivity of density peaking to the rotational shear is consistent with previous works~\cite{aCoreEdge-Garcia,aEMStabilisation-Citrin}, although a more detailed transport analysis of this effect is recommended and left for future work.

\begin{figure}[tb]
	\centering
	\includegraphics[scale=0.26]{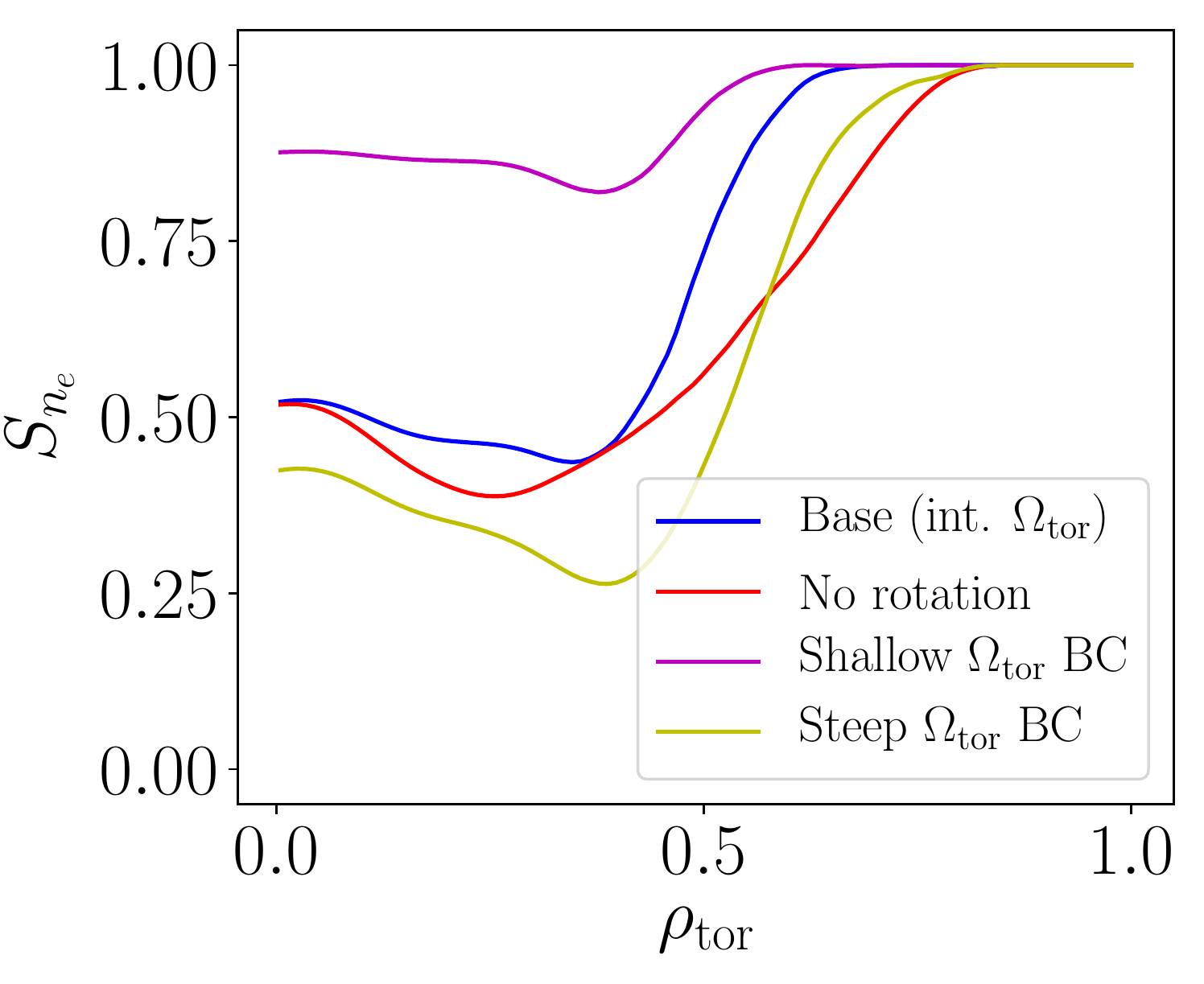}%
	\hspace{1mm}\includegraphics[scale=0.26]{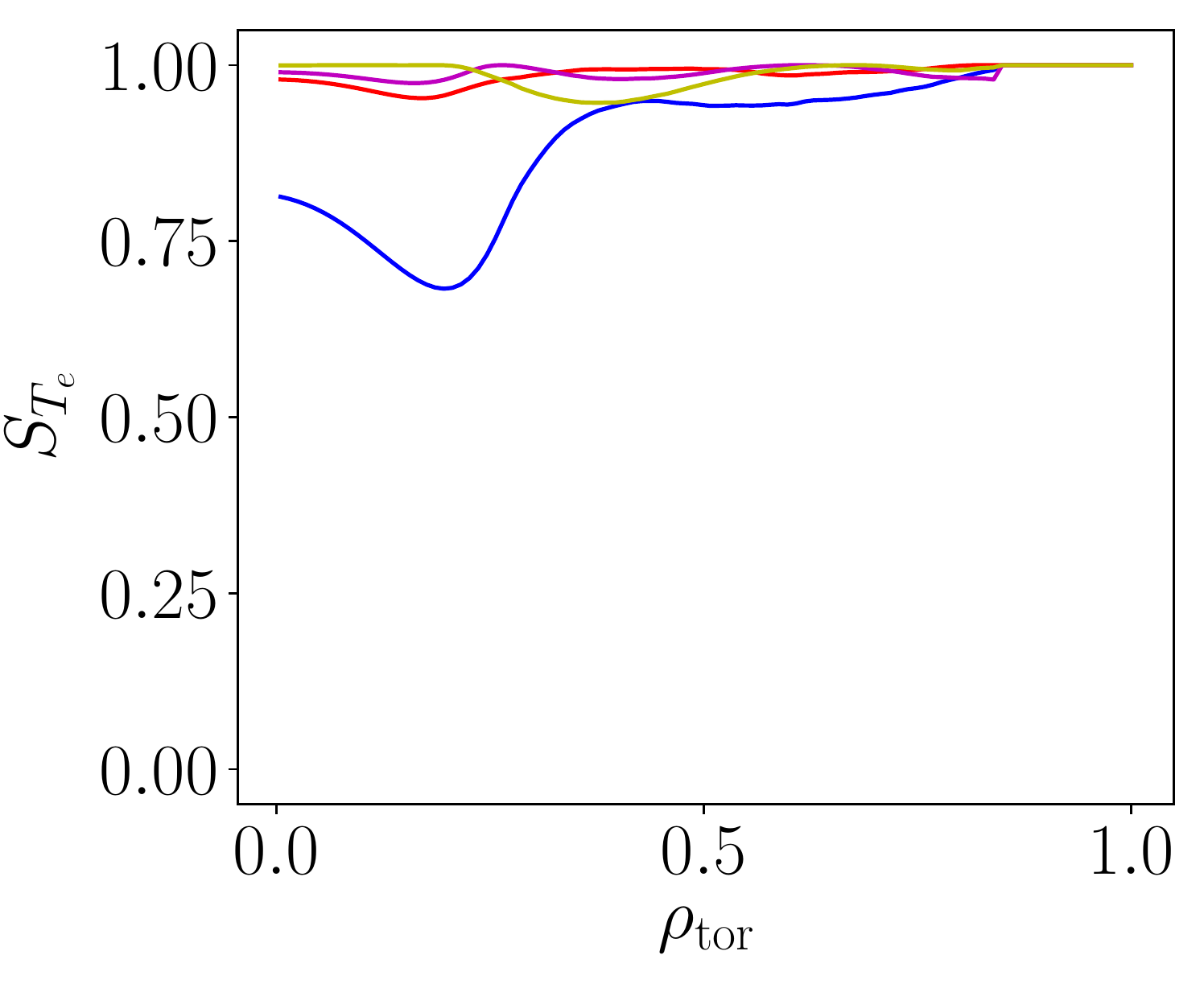}
	\includegraphics[scale=0.26]{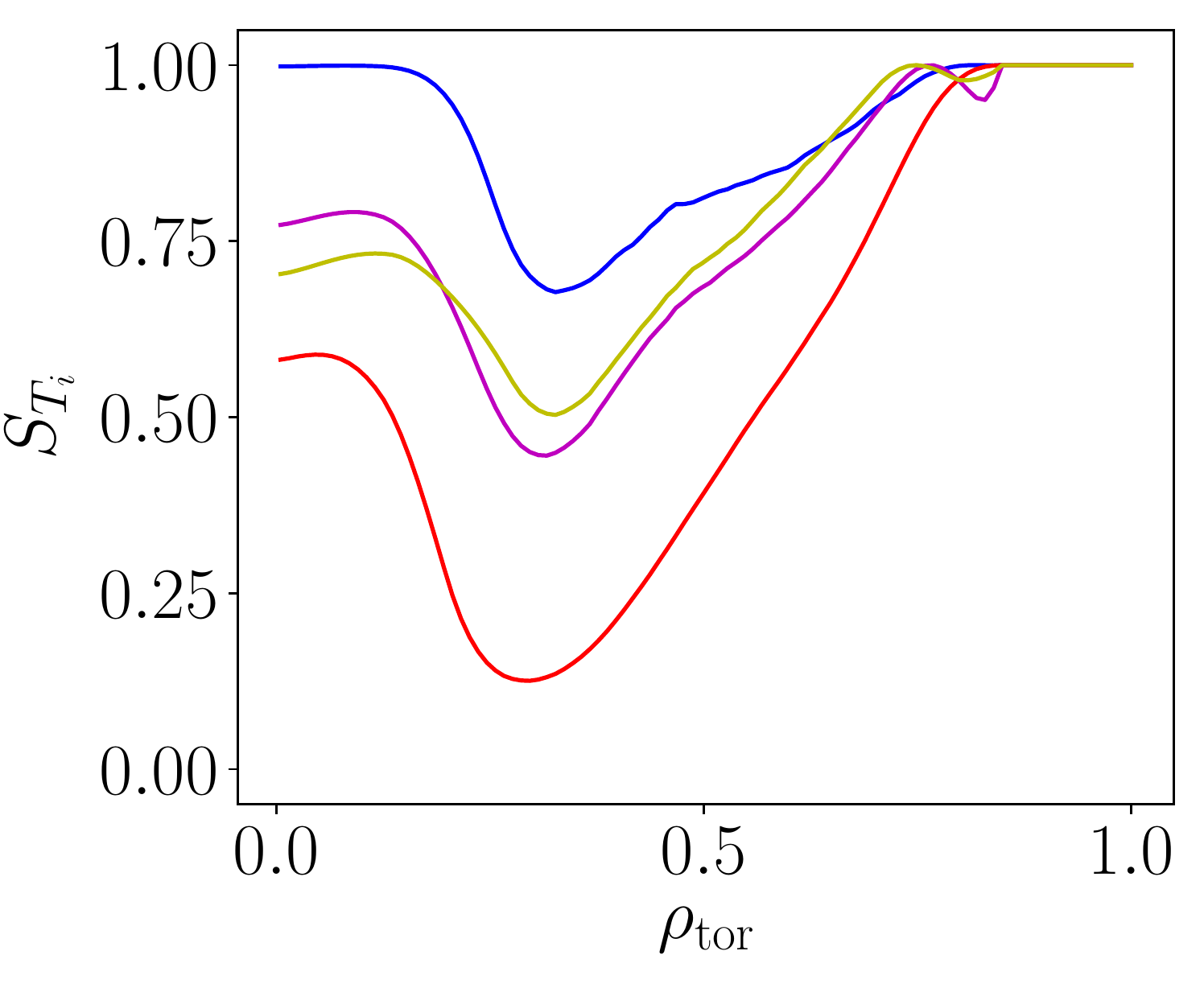}%
	\hspace{1mm}\includegraphics[scale=0.26]{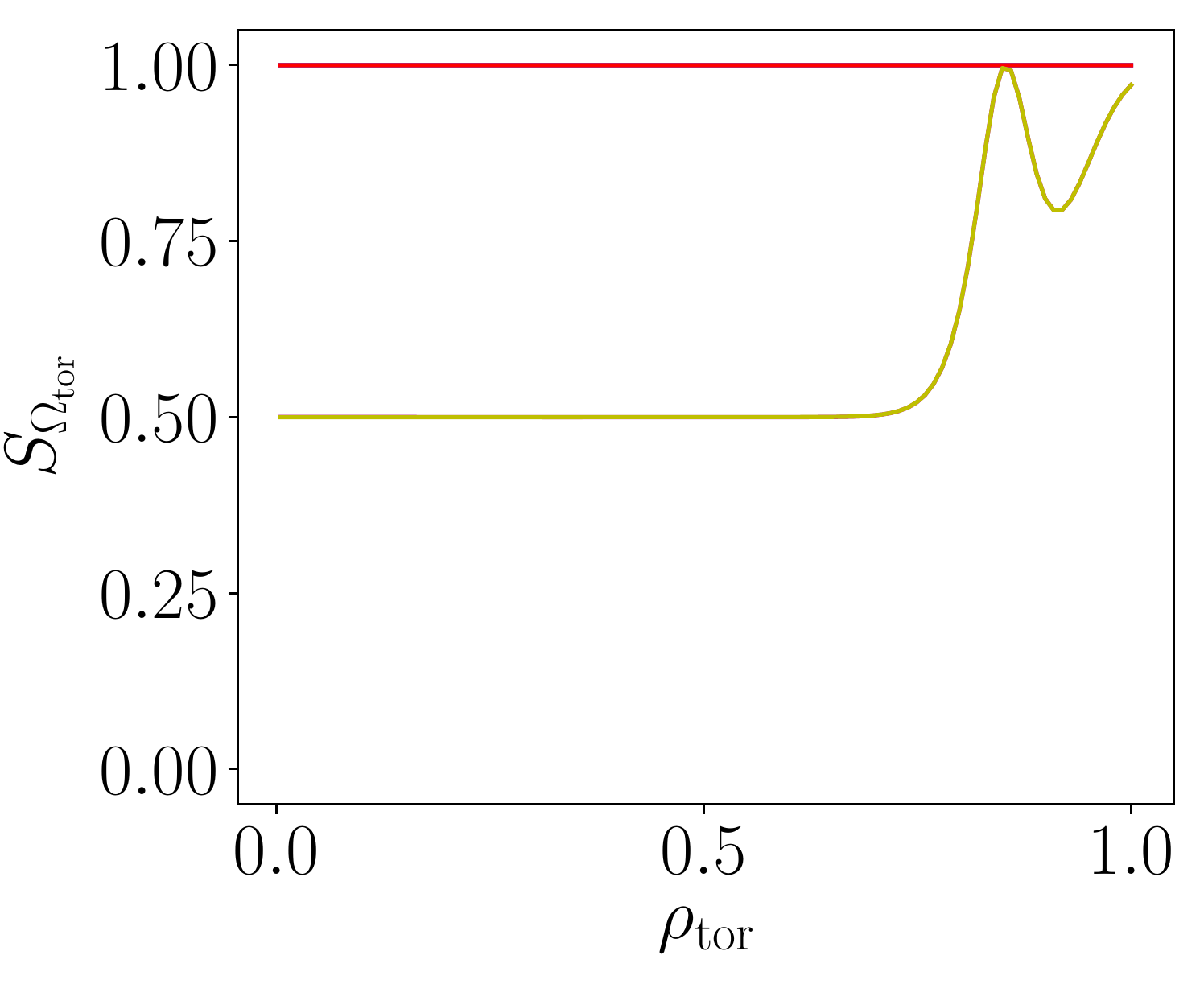}
	\caption{Point-distribution validation metric for sensitivity study results regarding the toroidal rotation profile modifications (see legend). These simulations were performed using interpretive momentum transport, thus the $\Omega_{\text{tor}}$ figure-of-merit calculation is not meaningful here. Upper left: Electron density profiles. Upper right: Electron temperature profiles. Lower left: Ion temperature profiles. Lower right: Toroidal angular frequency profiles.}
	\label{fig:RotationSensitivitySignificanceJET92436}
\end{figure}

Figure~\ref{fig:RotationSensitivitySignificanceJET92436} shows the analysis of these sensitivities according to the FOM given in Equation~\eqref{eq:PointDistributionSignificance}. The negative impact of the removal of the rotation contributions within QuaLiKiz on the $n_e$ and $T_i$ agreement can be seen.

\subsection{Impact analysis of impurity concentration and composition}
\label{subsec:ImpuritySensitivities}

\begin{figure}[tb]
	\centering
	\includegraphics[scale=0.27]{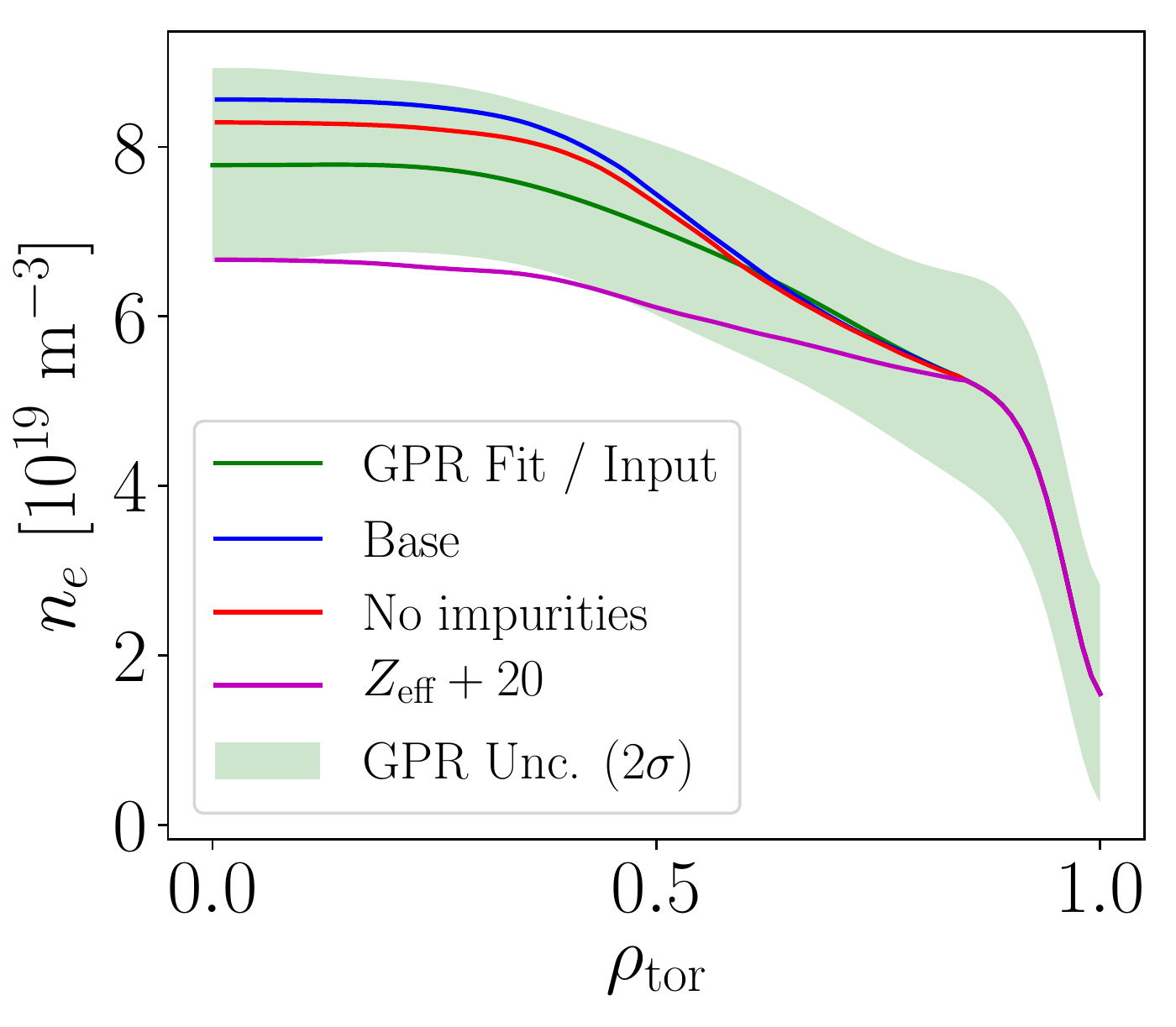}%
	\hspace{1mm}\includegraphics[scale=0.27]{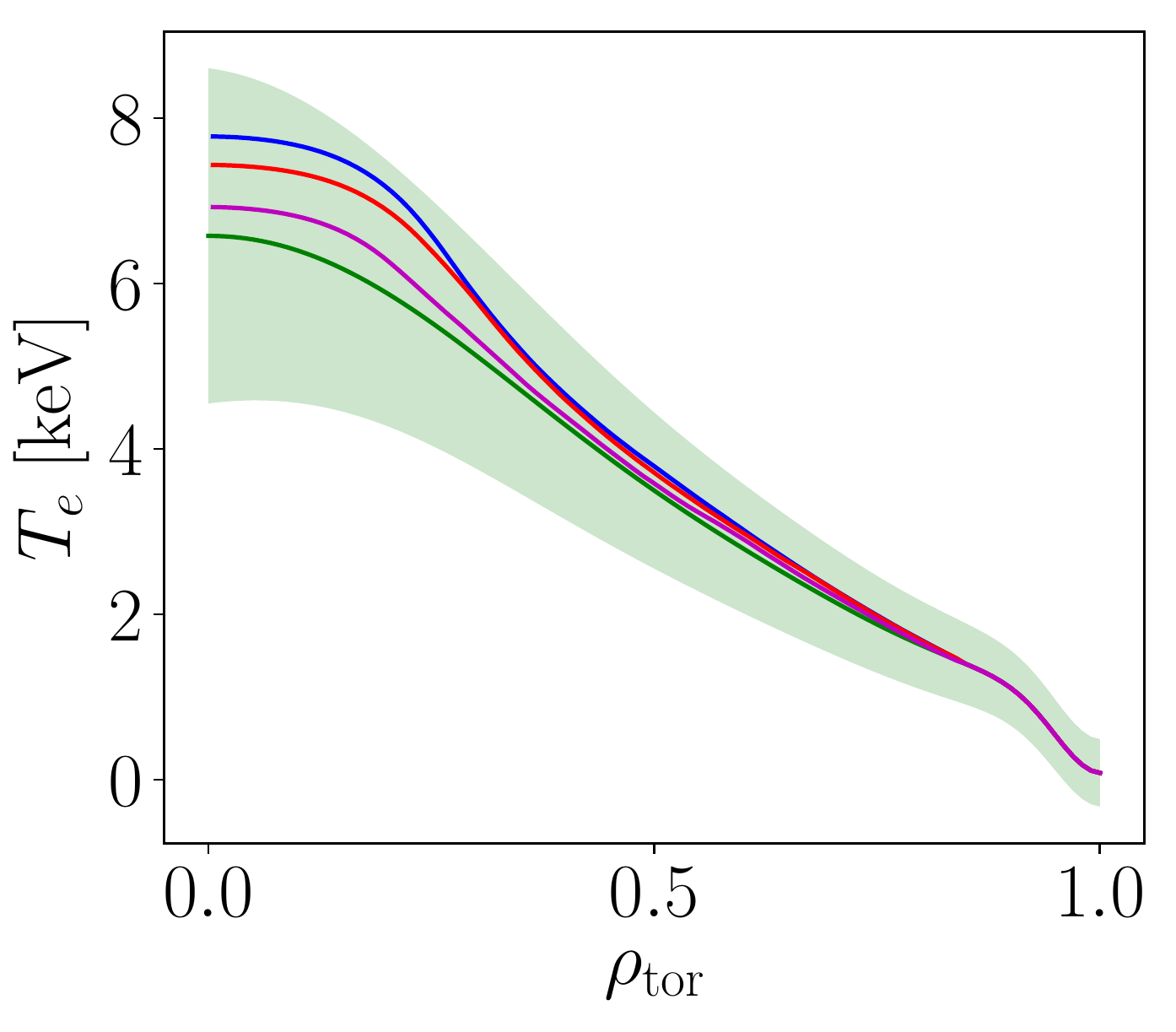}
	\includegraphics[scale=0.27]{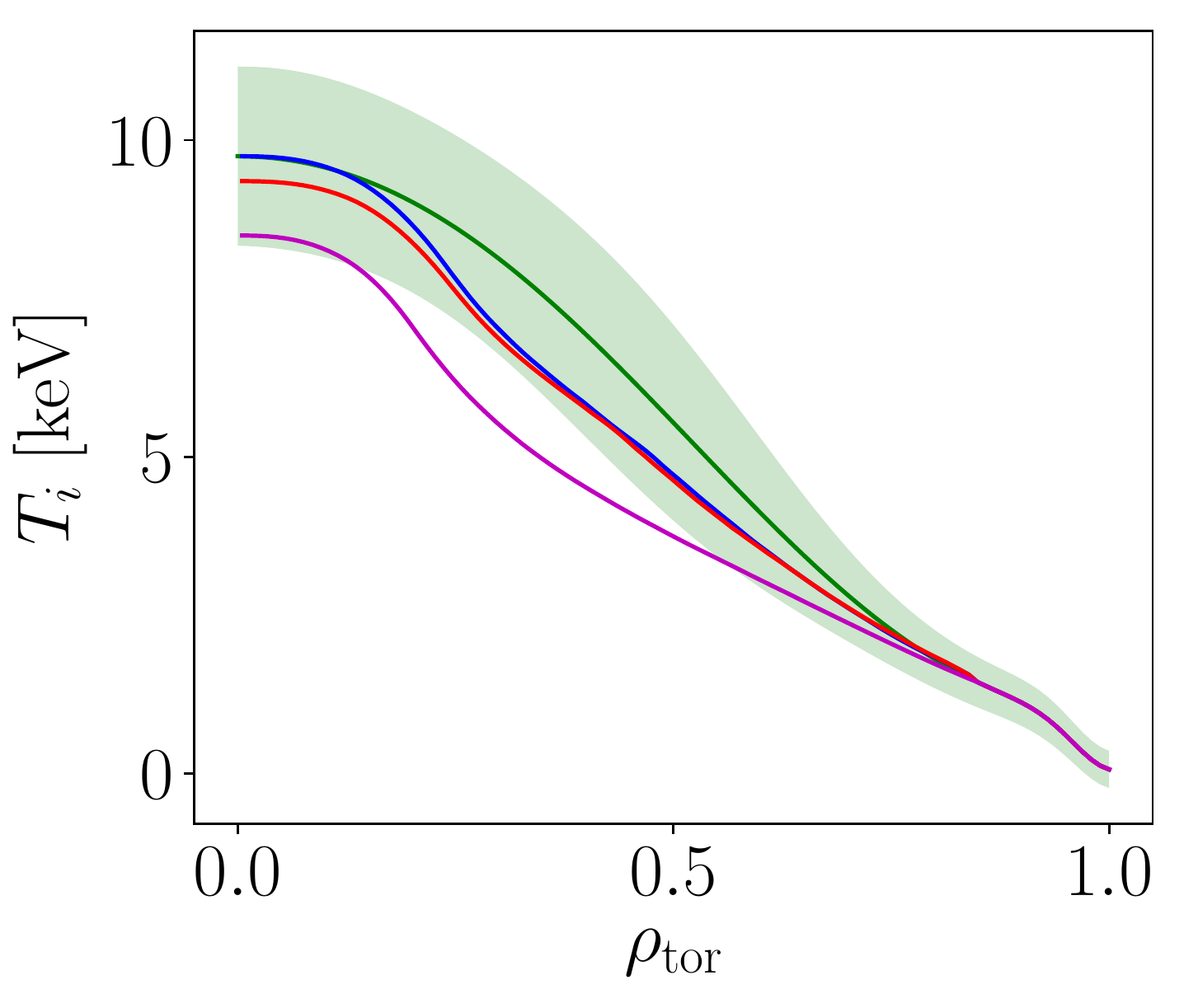}%
	\includegraphics[scale=0.27]{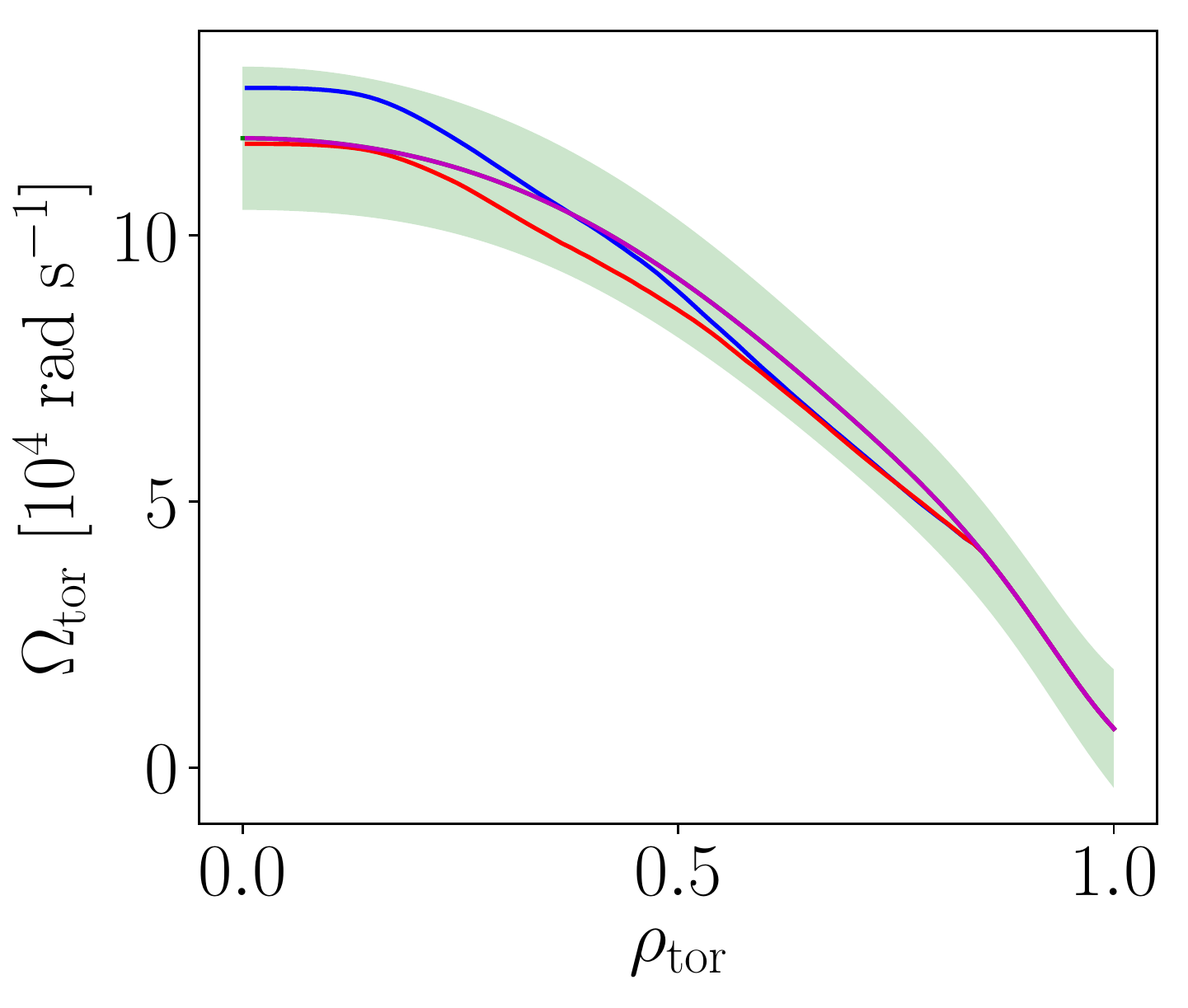}
	\caption{Results of the sensitivity studies regarding impurity concentration and composition modifications, where the input profiles (green lines) are compared against the output profiles (see legend) and the base case scenario (blue line). These simulations were performed using interpretive momentum transport. Upper left: Electron density profiles. Upper right: Electron temperature profiles. Lower left: Ion temperature profiles. Lower right: Toroidal angular frequency profiles.}
	\label{fig:ImpuritySensitivityResultsJET92436}
\end{figure}

Due to the inclusion of predictive impurity transport calculations via SANCO, sensitivity tests were also performed regarding the impurity concentration and composition in order to assess the validity of the chosen base settings. The sensitivities performed for this analysis are as follows:
\begin{itemize}
	\itemsep 0pt
	\item increasing the initial $Z_{\text{eff}}$ condition by 20\%;
	\item setting $Z_{\text{eff}}=1$ in the JETTO simulation by disabling the impurity transport module.
\end{itemize}
Figure~\ref{fig:ImpuritySensitivityResultsJET92436} shows the results of these sensitivity studies.

The dilution is expected to strongly stabilize ITG turbulence~\cite{aDilutionITG-Migliuolo,aDilutionITG-Ennever}, which is reflected in the increase of $T_i$ with increasing $Z_{\text{eff}}$. However, the trend is also noticed in the $T_e$ and an inverse trend is seen in the $\Omega_{\text{tor}}$ profile, where increased $Z_{\text{eff}}$ slows down the plasma rotation. An analysis of the QuaLiKiz growth rates show that the ETG instabilities become less dominant as the $Z_{\text{eff}}$ increases, possibly explaining the rise in the $T_e$. This is consistent with known ETG critical threshold dependencies~\cite{aCritETG-Jenko}. It is uncertain whether the impurities have a direct impact on the momentum transport or the observations are a result of an indirect effect through the modification of other profiles, particularly the $T_e$ profile in this case. However, the clarification of the underlying mechanisms and their relative strengths is left for future work.

Overall, all of the performed physics studies appear to agree with current literature, with the largest impact on the formation of these particular profiles being the stabilizing effect of flow shear. This highlights the importance of ensuring self-consistent profiles through the predictive calculation of the $\Omega_{\text{tor}}$ profile simultaneously with the other quantities, especially for extrapolation to D-T plasmas.

The application of the point-distribution metric described in Equation~\eqref{eq:PointDistributionSignificance} to the data presented in Figure~\ref{fig:ImpuritySensitivityResultsJET92436} were omitted from this paper as they did not reveal additional insights.

\section{Conclusions}
\label{sec:Conclusions}

A novel implementation of model validation, incorporating the use of Gaussian process regression techniques in profile fitting, has been proposed and demonstrated in JETTO integrated modelling of JET-ILW discharge \#92436 with the QuaLiKiz quasilinear turbulence model. A comparison between the fitted and simulated profiles showed that an excellent level of agreement was achieved in all channels, with discrepancies in both the core $n_e$ and $T_e$ profiles as well as in the mid-radius $T_i$. However, due to the high sensitivity of the $n_e$ prediction in the model to the simulation boundary conditions, only the temperature profile discrepancies are considered to warrant more in-depth studies. Two figures-of-merit were proposed, one for comparing a point and a distribution and one for comparing two distributions, which were shown to sufficiently quantify the level of agreement between the experimental profiles and the simulated profiles, with figure-of-merit scores $\gtrsim 0.5$ indicating the two profiles fall within $\pm 2 \sigma$ uncertainty bounds of each other. All four of the major predictive channels in the base simulation have figure-of-merit scores $> 0.5$ for the point-distribution metric and $\sim0.5$ for the distribution-distribution metric, effectively and quantitatively evaluating the degree of trust that can be assigned to these simulation results.

Additionally, the neutron rates calculated from the fitted and simulated profiles, using interpretive TRANSP, were compared against the measured neutron rate, agreeing with the experimental value within the $\pm\,2\sigma$ uncertainties. This showed the sensitivity of the neutron rate prediction to the various impurity composition estimates and relative concentration estimates, made via $Z_{\text{eff}}$. The discrepancy in the simulated $T_i$ profile is suspected to be the result of an incomplete description of the fast ion contributions to the instabilities driving turbulent transport. It was also shown that the $\Omega_{\text{tor}}$ profile is crucial to the accuracy of integrated modelling results of JET discharges and that QuaLiKiz is capable of providing reasonable momentum transport predictions within the studied plasma regime. This capability is expected to be important for extrapolating to future scenarios, such as deuterium-tritium plasmas.

As the proposed data processing and fitting procedure lends itself well to automatization, provided that the hyperparameter optimisation settings have been properly tuned, it opens the possiblity of large-scale model verification and validation through the comparison of thousands of different discharges. However, a bottleneck remains in performing the integrated modelling executions due to the complexity of setting up the simulations and ensuring self-consistency between any additional inputs to the model. Future work is foreseen in applying this procedure to similar data from other tokamak devices, such as ASDEX-Upgrade, Alcator C-Mod, and WEST, with the aim of developing a large database of discharges suitable for performing model verification and validation studies on a wide variety of gyrokinetic codes. Furthermore, when combined with their corresponding model outputs, this large database can be used to generate training sets for neural network emulations of the target model. Such emulations not only reduce the computational resources required for Monte Carlo uncertainty quantification studies, similar to the one presented in this paper, but also allow for the development of extremely quick and reliable model emulators for use in scenario optimization and tokamak controller design.

\section{Acknowledgements}
\label{sec:Acknowledgements}

This work has been carried out within the framework of the EUROfusion Consortium and has received funding from the Euratom research and training programme 2014--2018 under grant agreement No 633053. The views and opinions expressed herein do not necessarily reflect those of the European Commission.

\printbibliography

\vfill
\pagebreak
\appendix

\section{Gaussian processes}
\label{app:GaussianProcess}

A \emph{Gaussian process (GP)}~\cite{bGP-Rasmussen,bGP-Bishop} is strictly defined a collection of random variables whose joint distribution, along with the joint distribution of any finite set of them, is Gaussian. As such, a GP is completely specified by its mean function, $m\!\left(x\right)$, and its covariance function, $k\!\left(x_1,x_2\right)$, also known as \emph{kernel}, where $x$, $x_1$ and $x_2$ all represent the same coordinate space. In general, the theory assumes that $m\!\left(x\right) = 0$, as it is simple to devise a transformation which makes this true and is easy to apply to the data before and reverse after the completion of the algorithm.

Given a $N$-dimensional set of data, $\left(X,Y\right)$, a kernel function describing the covariance between the data points, $k\!\left(x_1,x_2,\theta\right)$, and a measurement error function, $r\!\left(x\right)$, a \emph{Gaussian process regression (GPR)} attempts to fit the data in a statistically rigourous manner. One advantage of this approach lies in the replacement of pre-defined basis functions with the kernel, allowing for a more generalized fit. However, the disadvantage is that the fitted function cannot be expressed in an analytical form, excluding further analysis of the fits in terms of comparing the mathematical model selection against equations derived from theoretical interpretations of the underlying process producing the data. All algorithms mentioned in this section are implemented within the ``GPR1D" tool, written in the Python programming language for applying the GPR technique to one-dimensional data.

\subsection{Mathematical Overview}
\label{subsec:GPRMathematics}

As mentioned in Section~\ref{subsec:DataProcessing}, the GPR applies Bayesian statistical principles to general regression fitting. This section provides a brief synopsis of the mathematical concepts used in the derivation of the GPR algorithm.
	
The generalized regression model is represented as follows:
\begingroup\makeatletter\def\f@size{9}\check@mathfonts
\def\maketag@@@#1{\hbox{\m@th\normalsize\normalfont#1}}%
\begin{equation}
\label{eq:StandardRegressionModel}
	y = \mathbf{\Phi}\!\left(x,\beta\right) + \varepsilon
\end{equation}\endgroup
where $\mathbf{\Phi}$ represent the set of basis functions in the regression analysis, $\beta$ represents the set of fit coefficients or free parameters to be adjusted by the fitting routine, and $\varepsilon$ represents the residuals of the model.

The derivation starts with the assumption that the optimal value of $\beta$ lies within a probability distribution, $p\!\left(\beta\right)$, known as the \emph{prior}. Then, as data points, $\left(X,Y\right)$, are added to test this hypothesis, the Bayesian inference framework can be applied to updating the prior to form the \emph{posterior} distribution as follows:
\begingroup\makeatletter\def\f@size{9}\check@mathfonts
\def\maketag@@@#1{\hbox{\m@th\normalsize\normalfont#1}}%
\begin{equation}
\label{eq:RegressionBayesTheorem}
	p\!\left(\beta|X,Y\right) = \frac{p\!\left(Y|X,\beta\right) p\!\left(\beta\right)}{p\!\left(Y|X\right)}
\end{equation}\endgroup
where the denominator, $p\!\left(Y|X\right)$, is known as the \emph{marginal likelihood} which is a normalization factor determined by the combined likelihood over all possible models. The posterior distribution effectively describes all the possible models within the confines of the pre-defined basis functions, $\mathbf{\Phi}$, and their respective probabilities of being the correct model given the available data, $\left(X,Y\right)$.

However, the purpose of a regression algorithm is usually to make predictions of points, $\left(X_*,Y_*\right)$, that are not explicitly provided as input data, as a means of interpolating or extrapolating to unexplored territory. The predictive distribution, $p\!\left(Y_*|X_*,X,Y\right)$, determines the probability distribution of predictions across all the regression models described by the posterior distribution, as follows:
\begingroup\makeatletter\def\f@size{9}\check@mathfonts
\def\maketag@@@#1{\hbox{\m@th\normalsize\normalfont#1}}%
\begin{equation}
\label{eq:GPRPredictiveDistribution}
	p\!\left(Y_*|X_*,X,Y\right) = \int_{-\infty}^{\infty} p\!\left(Y_*|X_*,\beta\right) p\!\left(\beta|X,Y\right) \text{d}\beta
\end{equation}\endgroup
The normalization factor of the predictive distribution is omitted from the previous equation as it is often unnecessary in practice, ie. only the moments of the predictive distribution are regularly calculated. However, for completeness, this normalization factor is identical to the marginal likelihood from the posterior distribution.

Furthermore, if all the probability distributions in this framework are assumed to be \emph{Gaussian} or \emph{normally-distributed}, then these equations can be analytically solved and simplified, resulting in the equations of the GPR predictive algorithm. This analytical solution provides an added advantage that the explicit definition of the basis functions, $\mathbf{\Phi}$, can be replaced by a more generic concept, the model covariance function, $k\!\left(x_1,x_2\right)$. This replacement, known as the \emph{kernel trick}, effectively allows for the use of an infinite set of basis functions through a clever selection of the model covariance function, making the GPR technique more similar to an \emph{universal function approximator}.

\subsection{Predictive algorithm}
\label{subsec:PredictiveAlgorithm}

Firstly, the contribution of the measurement noise, or the \emph{output noise}, to the kernel must be defined. In order to account for the possibility of spatially-varying noise, it was decided to implement this noise as such:
\begingroup\makeatletter\def\f@size{9}\check@mathfonts
\def\maketag@@@#1{\hbox{\m@th\normalsize\normalfont#1}}%
\begin{equation}
\label{eq:NoiseFunction}
	R\!\left(x_1,x_2\right) = r\!\left(x_1\right) r\!\left(x_2\right) \delta\!\left(x_1 - x_2\right)
\end{equation}\endgroup
Then, after selecting the \emph{hyperparameters}, denoted as a set with $\theta$, for the kernel, a prediction of the fit and its confidence interval, evaluated at the points, $X_*$, can be made using the following set of equations~\cite{bGP-Rasmussen,ahGP-Kersting}:
\begingroup\makeatletter\def\f@size{9}\check@mathfonts
\def\maketag@@@#1{\hbox{\m@th\normalsize\normalfont#1}}%
\begin{equation}
\label{eq:PredictedModel}
	\begin{aligned}
	Y_* &= K\!\left(X_*,X\right) L^{-1} \, Y \\
	\sigma_{Y_*}^2 &= L_* - K\!\left(X_*,X\right) L^{-1} K\!\left(X,X_*\right)
	\end{aligned}
\end{equation}\endgroup
where the short-hand $K=K\!\left(X,X\right)$, $R=R\!\left(X,X\right)$, $K_*=K\!\left(X_*,X_*\right)$, $R_*=R\!\left(X_*,X_*\right)$, $L=K+R$, and $L_*=K_*+R_*$ was used to improve the readability. Most GPR implementations, including this one, modify the $Y$-values such that $\bar{Y} = 0$ and $Y \in \left[-1,1\right]$ and reverse these changes afterwards, in order to improve the numerical stability of the algorithm.

Additionally, provided that the derivatives of the kernel can be calculated, the derivatives of the fit and its confidence intervals can also be predicted directly from the data using the following equations~\cite{adGP-McHutchon}:
\begingroup\makeatletter\def\f@size{9}\check@mathfonts
\def\maketag@@@#1{\hbox{\m@th\normalsize\normalfont#1}}%
\begin{equation}
\label{eq:PredictedDerivatives}
	\begin{aligned}
	Y_*' &= \frac{\partial K\!\left(X_*,X\right)}{\partial X_*} L^{-1} \, Y \\
	\sigma_{Y_*'}^2 &= \frac{\partial^2 L_*}{\partial X_* \, \partial X_*} - \frac{\partial K\!\left(X_*,X\right)}{\partial X_*} L^{-1} \frac{\partial K\!\left(X,X_*\right)}{\partial X_*}
	\end{aligned}
\end{equation}\endgroup

If the error function, $r\!\left(x\right)$, is not known, it can be estimated using a separate GPR on the data set, $\left(X,\Sigma_Y\right)$, where $\Sigma_Y$ represents the standard deviation of the $Y$-values~\cite{ahGP-Kersting}. For this GPR, it is recommended to assume a constant error function with a normalized value of $\sim10^{-3}$.

\subsection{Hyperparameter optimization}
\label{subsec:HypOptimization}

Within the GPR framework, the hyperparameters of the chosen kernel, $\theta$, act as the free variables which can be adjusted to fine-tune the fit. The optimal value for these hyperparameters can be obtained by maximizing the \emph{log-marginal-likelihood (LML)}, given as follows~\cite{bGP-Rasmussen}:
\begingroup\makeatletter\def\f@size{9}\check@mathfonts
\def\maketag@@@#1{\hbox{\m@th\normalsize\normalfont#1}}%
\begin{equation}
\label{eq:LogMarginalLikelihood}
	\ln{p\!\left(Y|X\right)} = -\frac{1}{2} Y^T L^{-1} Y - \frac{\lambda}{2} \ln{\left|L\right|} - \frac{N}{2} \ln{2 \pi}
\end{equation}\endgroup
where the hyperparameter-dependence is given by $L \equiv L\!\left(\theta\right)$, the vertical brackets represent the determinant of the enclosed matrix, $\lambda$ is the \emph{regularization} parameter, used to control the degree of complexity in the model, and $N$ is the number of data points to be fit. By maximizing this value, the chosen model is the most probable match to the input data but provides no guarantee that the physical process behind the data is modelled correctly. Note that this is only one optimization criterion that can be applied to the hyperparameters and that other criteria may provide models that have different relations to the data.

Most maximization algorithms require the derivative of Equation~\eqref{eq:LogMarginalLikelihood} with respect to each of the hyperparameters, $\theta_j$, which can be calculated directly, provided that an analytical form exists for the derivative of the kernel function with respect to these hyperparameters, as such:
\begingroup\makeatletter\def\f@size{9}\check@mathfonts
\def\maketag@@@#1{\hbox{\m@th\normalsize\normalfont#1}}%
\begin{equation}
\label{eq:LogMarginalLikelihoodDerivativeApp}
	\frac{\partial \ln{p\!\left(Y|X\right)}}{\partial \theta_j} = \frac{1}{2} Y^T L^{-1} \frac{\partial K}{\partial \theta_j} L^{-1} Y - \frac{\lambda}{2} \text{tr}\!\left(L^{-1} \frac{\partial K}{\partial \theta_j}\right)
\end{equation}\endgroup
where $\text{tr}\!\left(...\right)$ represents the trace, the sum of all the entries along the main diagonal, of the enclosed matrix. The desired solution will then be the combination of hyperparameters, $\theta$, which satisfy the following criteria:
\begingroup\makeatletter\def\f@size{9}\check@mathfonts
\def\maketag@@@#1{\hbox{\m@th\normalsize\normalfont#1}}%
\begin{equation}
\label{eq:MaximizationCriteriaApp}
	\nabla_{\theta} \ln{p\!\left(Y|X\right)} = \mathbf{0}
\end{equation}\endgroup
However, since Equation~\eqref{eq:LogMarginalLikelihoodDerivativeApp} typically forms a non-linear system of equations for the set of $\theta$, it becomes difficult to calculate the solution explicitly. Thus, an iterative method, such as a gradient-based optimization algorithm, is used to find the desired solution. Of the optimization methods discussed below, the nominal implementation uses the \emph{Adam} method for fitting both the plasma profiles and the associated error function.

\subsubsection{Gradient ascent method}
\label{subsubsec:OptimizerGradAscent}

The most basic gradient-based optimization algorithm for maximization problems is known as the \emph{gradient ascent} method. It starts with an initial guess, $\theta_0$, and iteratively updates that guess in increments, labelled with index $i$, as follows:
\begingroup\makeatletter\def\f@size{9}\check@mathfonts
\def\maketag@@@#1{\hbox{\m@th\normalsize\normalfont#1}}%
\begin{equation}
\label{eq:StepUpdate}
	\theta_{i+1} = \theta_i + \Delta \theta_i
\end{equation}\endgroup
where the \emph{step}, $\Delta \theta_i$, is calculated according to a \emph{step estimator}.

The step estimator used in this method is as follows:
\begingroup\makeatletter\def\f@size{9}\check@mathfonts
\def\maketag@@@#1{\hbox{\m@th\normalsize\normalfont#1}}%
\begin{equation}
\label{eq:SE_Grad}
	\Delta \theta_i = \gamma \, \mathbf{G}_i
\end{equation}\endgroup
where $\gamma$, called the \emph{learning rate}, was set to a value of $10^{-5}$ and
\begingroup\makeatletter\def\f@size{9}\check@mathfonts
\def\maketag@@@#1{\hbox{\m@th\normalsize\normalfont#1}}%
\begin{equation}
\label{eq:GradientVector}
	\mathbf{G}_i \equiv \nabla_{\theta} \ln{p\!\left(Y|X\right)} |_{\theta=\theta_i}
\end{equation}\endgroup

This simple method is considered to be the most robust out of all the gradient-based optimization algorithms but it also suffers from a slow convergence rate. Thus, a number of additional methods were implemented in an attempt to improve the performance of the algorithm, though only the ones relevant to this application will be discussed here.

\subsubsection{Adaptive moment estimation method (Adam)}
\label{subsubsec:OptimizerAdam}

This method introduces a way for the algorithm to autonomously adjust the learning rate for each hyperparameter individually, such that a more intelligent approach path to the optimal solution can be determined. This is done by including some memory of both the gradient with respect to the hyerparameters and the square of the gradient to the step estimator. This is done as follows~\cite{aAdam-Kingma}:
\begingroup\makeatletter\def\f@size{9}\check@mathfonts
\def\maketag@@@#1{\hbox{\m@th\normalsize\normalfont#1}}%
\begin{equation}
\label{eq:SE_Adam}
	\Delta \theta_i = \gamma \left[\widehat{\mathbf{V}}_i^{\sfrac{1}{2}} + \varepsilon\right]^{-1} \widehat{\mathbf{M}}_i
\end{equation}\endgroup
where all operations are done element-wise, $\gamma$ was set to $10^{-2}$, $\varepsilon \sim 10^{-8}$ is provided to avoid division-by-zero errors in the algorithm and
\begingroup\makeatletter\def\f@size{9}\check@mathfonts
\def\maketag@@@#1{\hbox{\m@th\normalsize\normalfont#1}}%
\begin{equation}
\label{eq:NormalizedGradientMoments}
	\begin{gathered}
	\widehat{\mathbf{M}}_i = \frac{1}{1 - \beta_1^i} \mathbf{M}_i \;, \quad \mathbf{M}_i = \beta_1 \mathbf{M}_{i-1} + \left(1 - \beta_1\right) \mathbf{G}_i \\
	\widehat{\mathbf{V}}_i = \frac{1}{1 - \beta_2^i} \mathbf{V}_i \;, \quad \mathbf{V}_i = \beta_2 \mathbf{V}_{i-1} + \left(1 - \beta_2\right) \mathbf{G}_i^2
	\end{gathered}
\end{equation}\endgroup
where $\beta_1 \in \left[0,1\right]$, $\beta_2 \in \left[0,1\right]$ and $\mathbf{G}_i$ is given by Equation~\eqref{eq:GradientVector}. For fusion data, $\beta_1 = 0.4$ and $\beta_2 = 0.8$ were found to adequate choices for these memory factors.

Mathematically, this algorithm can be seen as attempting to pick steps which minimize a weighted \emph{$l_2$-norm} of the gradient, represented by $\mathbf{V}_i$, and a strong penalty is applied to steps which dramatically increase this value. Thus, the algorithm tends to move towards the regions where the gradients are zero with fewer iterations. This conveniently turns out to be the desired behaviour as the solutions to the maximization problem have a gradient of zero.

\section{Detailed base simulation settings}
\label{app:BaseSimulationSettings}

The JETTO + QLK settings used as the definition of the base settings in this paper are detailed in Tables~\ref{tbl:DetailedBaseSettingsJETTO}, \ref{tbl:DetailedBaseSettingsSANCO}, and \ref{tbl:DetailedBaseSettingsQuaLiKiz}. 

\begin{table*}[tb]
	\centering
	\begin{threeparttable}
	\caption{Summary table of most pertinent JETTO settings of the base case simulation, discussed in Section~\ref{subsec:NominalSettings}.}
		\begin{tabular}{l|c}
			\toprule[1.5pt]
			Field Name / Option & Value / Setting \\
			\midrule
			JAMS Version & v080817 \\
			Shot number & 92436 \\
			Number of grid points & 101 \\
			Start time\tnote{1}  (s) & 50 \\
			End time\tnote{1}  (s) & 52 \\
			Min. time step (s) & $10^{-13}$ \\
			Max. time step (s) & $10^{-3}$ \\
			Ion (1) mass (u) & 2 \\
			Simulation boundary & $\rho_{\text{tor}} = 0.85$ \\
			Equilibrium & ESCO \\
			Equilibrium boundary & 0.998 \\
			Toroidal Field & 2.8~T \\
			Plasma Current & 2.9~MA \\
			Neoclassical transport model & NCLASS \\
			\midrule
			Additional transport\tnote{2} &
			\begin{tabular}{C{3cm} C{3cm} C{3cm}}
				Electron heat & Ion heat & Particle
			\end{tabular} \\
			Shape & \begin{tabular}{C{3cm} C{3cm} C{3cm}}
				Gaussian & Gaussian & Gaussian
			\end{tabular} \\
			Centre & \begin{tabular}{C{3cm} C{3cm} C{3cm}}
				0.0 & 0.0 & 0.0
			\end{tabular} \\
			Height (cm$^2$ s$^{-1}$) & \begin{tabular}{C{3cm} C{3cm} C{3cm}}
				$10^4$ & $10^4$ & $2 \times 10^4$
			\end{tabular} \\
			Width & \begin{tabular}{C{3cm} C{3cm} C{3cm}}
				0.212 & 0.212 & 0.212
			\end{tabular} \\
			\bottomrule
		\end{tabular}
		\begin{tablenotes}
			\item[1] The reference time, $t=0$, in the PPF data system at JET is set as the time when the magnetic coils start ramping up, instead of the time of plasma breakdown as is usual in the fusion physics community. The time interval between these two events is typically $40$~s at JET. This interval is not subtracted here such that the entries in this table represent the exact inputs to the simulation suite for replication purposes.
			\item[2] These transport coefficients were added to the computed ones before advancing the transport calculation iteration, and were used here to emulate the MHD-induced transport within the central core for reasons of simplicity.
		\end{tablenotes}
		\label{tbl:DetailedBaseSettingsJETTO}
	\end{threeparttable}
\end{table*}

\begin{table*}[tb]
	\centering
	\begin{threeparttable}
		\caption{Summary table of most pertinent SANCO settings of the base case simulation, discussed in Section~\ref{subsec:NominalSettings}.}
		\begin{tabular}{l|c}
			\toprule[1.5pt]
			Field Name / Option & Value / Setting \\
			\midrule
			Impurity &
			\begin{tabular}{C{2cm} C{2cm} C{2cm}}
				Be & Ni & W
			\end{tabular} \\
			Index &
			\begin{tabular}{C{2cm} C{2cm} C{2cm}}
				1 & 2 & 3
			\end{tabular} \\
			Mass (u) &
			\begin{tabular}{C{2cm} C{2cm} C{2cm}}
				9 & 58 & 184
			\end{tabular} \\
			Charge (e) &
			\begin{tabular}{C{2cm} C{2cm} C{2cm}}
				4 & 28 & 74
			\end{tabular} \\
			ADAS Year &
			\begin{tabular}{C{2cm} C{2cm} C{2cm}}
				96 & 89 & 50
			\end{tabular} \\
			Initial condition & Coronal \\
			Boundary condition & $\rho_{\text{tor}} = 0.85$ \\
			Effective charge, $Z_{\text{eff}}$ (e) & 1.76 \\
			Time steps per JETTO step & 25 \\
			\bottomrule
		\end{tabular}
	\label{tbl:DetailedBaseSettingsSANCO}
	\end{threeparttable}
\end{table*}

\begin{table*}[tb]
	\centering
	\begin{threeparttable}
		\caption{Summary table of most pertinent JETTO--QuaLiKiz settings of the base case simulation, discussed in Section~\ref{subsec:NominalSettings}.}
		\begin{tabular}{l|c}
			\toprule
			Field Name / Option & Value / Setting \\
			\midrule
			JETTO--QuaLiKiz Version & Git hash: 8c97aca0c0 \\
			Inner simulation boundary & $\rho_{\text{tor,min}} = 0.15$ \\
			Outer simulation boundary & $\rho_{\text{tor,max}} = 0.85$ \\
			Particle diffusion multiplier & 1.0 \\
			Bohm electron diffusion multiplier & 0.08 \\
			Bohm ion diffusion multiplier & 0.08 \\
			Bohm momentum diffusion Prandtl number & 1.25 \\
			\bottomrule[1.5pt]
		\end{tabular}
		\label{tbl:DetailedBaseSettingsQuaLiKiz}
	\end{threeparttable}
\end{table*}

\end{document}